\documentstyle[12pt,epsf,amsfonts]{article}

\def\eps{\epsilon}

\def\A#1{{\cal A}^{#1}}

\def\ib{{\bar\imath}}

\def\tr{\mathop{\rm tr}\nolimits}

\def\spa#1.#2{\left\langle#1\,#2\right\rangle}
\def\spb#1.#2{\left[#1\,#2\right]}

\def\eqn#1{Eq.~(\ref{#1})}
\def\eqns#1#2{Eqs.~(\ref{#1}) and~(\ref{#2})}
\def\eqnss#1#2{Eqs.~(\ref{#1})-(\ref{#2})}
\def\fig#1{Fig.~{\ref{#1}}}
\def\sec#1{Section~{\ref{#1}}}
\def\app#1{Appendix~\ref{#1}}

\def\feynsl#1{
  \setbox0=\hbox{/} \setbox1=\hbox{$#1$}
  \dimen0=\wd0 \advance\dimen0 by -\wd1 \divide\dimen0 by 2
  \ifdim\wd0>\wd1 \raise.15ex\copy0\kern-\wd0\kern\dimen0\copy1\kern\dimen0
  \else \kern-\dimen0\raise.15ex\copy0\kern-\dimen0\kern-\wd1\copy1\fi}

\newskip\humongous \humongous=0pt plus 1000pt minus 100pt

\newif\ifdtup

\makeatletter
\def\@eqnnum{\hbox{\reset@font\rm(\theequation)}}
\let\make@eqnnum=\@eqnnum %
\def\eqnum#1{\dec@eqnnum \global\def\make@eqnnum{\reset@font\rm(#1)}%
\def\@currentlabel{#1}%
}
\def\inc@eqnnum{\addtocounter{equation}{1}}
\def\dec@eqnnum{\addtocounter{equation}{-1}}
\@definecounter{equation}%
\@addtoreset{equation}{section} %
\def\theequation@prefix{{\thesection}.} %
\def\theequation{\theequation@prefix\arabic{equation}}%
\makeatother

\def    \br#1#2          {\mbox{$\langle #1 \, #2 \rangle$}}
\def    \sq#1#2          {\mbox{$\left[  #1 \, #2 \right]$}}
\def    \sap#1#2#3       {\mbox{$\langle #1 | #2 |#3  \rangle$}}
\def    \t#1#2#3         {\mbox{$s_{#1 #2 #3}$}}
\def    \s#1#2           {\mbox{$s_{#1 #2}$}}
\def    \sapp#1#2#3#4    {\mbox{$\langle #1 | (#2+#3) |#4  \rangle$}}
\def    \sep             {\mbox{$\,|\;$}}
\def    \br(#1,#2)          {\mbox{$\langle #1 \, #2 \rangle$}}
\def    \sq(#1,#2)          {\mbox{$\left[  #1 \, #2 \right]$}}
\def    \t(#1,#2,#3)        {\mbox{$s_{#1 #2 #3} $}}
\def    \s(#1,#2)           {\mbox{$s_{#1 #2}$ }}
\def    \Pperp(#1)          {\mbox{$k_{#1 \perp}$ }}
\def    \PperpStar(#1)      {\mbox{$k_{#1 \perp}^*$ }}
\def    \Pperpt(#1)          {\mbox{$|k_{#1 \perp}|^2$ }}
\def    \x(#1)              {\mbox{$x_{#1}$ }}
\def    \qoperp              {\mbox{$ q_{1\perp}$ }}
\def    \qtperp              {\mbox{$ q_{2\perp}$ }}
\def    \qoperpStar          {\mbox{$ q_{1\perp}^*$ }}
\def    \qtperpStar          {\mbox{$ q_{2\perp}^*$ }}
\def    \qoperpt              {\mbox{$ |q_{1\perp}|^2$ }}
\def    \qtperpt              {\mbox{$ |q_{2\perp}|^2$ }}
\def    \qomqtperpt           {\mbox{$ |q_{1\perp}-q_{2\perp}|^2$ }}

\begin{document}

\begin{titlepage}

\hspace*{\fill}\parbox[t]{4cm}{
DFTT 45/99\\
\today}

\begin{center}
{\Large\bf Factorization of tree QCD amplitudes in
the high-energy limit and in the collinear limit}\\
\vspace{1.cm}

{Vittorio Del Duca}\\
\vspace{.2cm}
{\sl Istituto Nazionale di Fisica Nucleare\\ Sezione di Torino\\
via P. Giuria, 1\\ 10125 - Torino, Italy}\\

\vspace{.5cm}
and \\
\vspace{.5cm}

{Alberto Frizzo and Fabio Maltoni}\\
\vspace{.2cm}
{\sl Dipartimento di Fisica Teorica\\
Universit\`a di Torino\\
via P. Giuria, 1\\ 10125 - Torino, Italy}\\

\vspace{.5cm}

\begin{abstract}
In the high-energy limit, we compute the gauge-invariant three-parton 
forward clusters, which in the BFKL theory constitute the tree parts of the 
NNLO impact factors. In the triple collinear limit, we obtain the polarized 
double-splitting functions. For the unpolarized and the spin-correlated
double-splitting functions, our results agree with the ones obtained
by Campbell-Glover and Catani-Grazzini, respectively.
In addition, we compute the four-gluon forward cluster, which in the BFKL 
theory
forms the tree part of the NNNLO gluonic impact factor. In the quadruple 
collinear limit we obtain the unpolarized triple-splitting functions, while in
the limit of a three-parton central cluster we derive the Lipatov vertex
for the production of three gluons, relevant for the calculation of 
a BFKL ladder at NNLL accuracy.
Finally, motivated by the reorganization of the color in the
high-energy limit, we introduce a color decomposition 
of the purely gluonic tree amplitudes in terms of the
linearly independent subamplitudes only.
\end{abstract}

\end{center}
 \vfil

\end{titlepage}

\section{Introduction}

QCD calculations of multijet rates beyond the leading order (LO) in
the strong coupling constant $\alpha_s$ are generally quite involved.
However, in recent years it has become clear how to construct 
general-purpose algorithms for the calculation of multijet rates
at next-to-leading order (NLO) accuracy~\cite{gg,slicing,ks,subtr,dipole}.
The crucial point is to organise the cancellation of the infrared
(i.e. collinear and soft) singularities in a universal, i.e.
process-independent, way. The universal pieces in a NLO calculation
are given by the tree-level splitting~\cite{ap} and eikonal~\cite{bcm,bg} 
functions, and by the universal structure of the poles of the one-loop
amplitudes~\cite{gg,ks,kst}.

Eventually, the same procedure will permit the construction of
general-purpose algorithms at next-to-next-to-leading order (NNLO) accuracy.
It is mandatory then to fully investigate the infrared structure of the
phase space at NNLO. The universal pieces needed to organise the cancellation
of the infrared singularities are given by the tree-level 
double-splitting~\cite{glover,cat}, double-eikonal~\cite{bg,catgra} and 
splitting-eikonal~\cite{glover,catgra} functions, by the one-loop 
splitting~\cite{bds,dk} and eikonal~\cite{bds} functions,
and by the universal structure of the poles of the two-loop
amplitudes~\cite{catani}.

Another outstanding issue in QCD, at first sight unrelated to the topics 
discussed above, is the calculation of the higher-order corrections
to the BFKL equation~\cite{fkl,bfkl}. In scattering processes characterised
by two large and disparate scales, like $s$, the squared parton center-of-mass
energy, and $t$, a typical momentum transfer, the BFKL equation resums the
large logarithms of type $\ln(s/t)$. The LO term of the resummation
requires gluon exchange in the cross channel, which for a given scattering
occurs at ${\cal O}(\alpha_s^2)$. The corresponding QCD amplitude
factorizes then into a gauge-invariant effective amplitude formed by 
two scattering centers, the LO impact factors, connected by the gluon 
exchanged in the cross channel. The LO impact factors are characteristic
of the scattering process at hand. The BFKL equation resums then the universal
leading-logarithmic (LL) corrections, of ${\cal O}(\alpha_s^n\ln^n(s/t))$,
to the gluon exchange in the cross channel. 
The building blocks of the BFKL resummation are the Lipatov 
vertex~\cite{lipat}, i.e. the effective gauge-invariant emission
of a gluon along the gluon ladder in the cross channel, and the gluon
reggeization~\cite{fkl}, i.e. the LL part of the one-loop corrections
to the gluon exchange in the cross channel.

The accuracy of the BFKL equation is improved by computing the
next-to-leading logarithmic (NLL) corrections~\cite{ff,cc}, i.e.
the corrections of ${\cal O}(\alpha_s^n\ln^{n-1}(s/t))$,
to the gluon exchange in the cross channel.
In order to do that, the universal building blocks of the BFKL ladder must be
computed to NLL accuracy. These are given by the tree
corrections to the Lipatov vertex, i.e. the emission of two 
gluons~\cite{fl,ptlipnl,flqq} or of a $\bar{q} q$ pair~\cite{flqq,ptlipqq}
along the gluon ladder, by the one-loop corrections to the Lipatov 
vertex~\cite{1loop,1loopeps,1loopds}, and finally by the NLL gluon
reggeization~\cite{2loop}, i.e. the NLL part of the two-loop corrections
to the gluon 
ladder. However, to compute jet production rates at NLL accuracy, the
impact factors must be computed at NLO~\cite{dsif,colf}. For jet 
production at large
rapidity intervals, they are given by the one-loop corrections~\cite{dsif} 
to the LO impact factors, and by the tree 
corrections~\cite{fl,ptlipnl,ptlipqq,thuile}, i.e. the emission
of two partons in the forward-rapidity region.
In the collinear or soft limits, the latter reduce to the tree
splitting or eikonal functions~\cite{rio}.

To further improve the accuracy of the BFKL ladder one needs to compute
the next-to-next-to-leading logarithmic (NNLL) corrections, i.e.
the corrections of ${\cal O}(\alpha_s^n\ln^{n-2}(s/t))$, to the gluon
ladder. At present it is not known whether such corrections can be
resummed. If that is the case,
the universal building blocks of a BFKL ladder at NNLL would be:
the emission of three partons along the gluon ladder, the one-loop
corrections to the emission of two partons along the ladder,
the two-loop corrections to the Lipatov vertex, and the gluon
reggeization at NNLL accuracy. None of them is known at present.
In this paper we compute the gluonic NNLO Lipatov vertex, i.e.
the emission of three gluons along the ladder. 

In addition,
to compute jet production rates at NNLL accuracy, the BFKL ladder should
be supplemented by impact factors at NNLO. They are not known either.
In this paper we compute their tree components, i.e. the emission of
three partons in the forward-rapidity region. By taking then the triple 
collinear limit of the tree NNLO impact factors, we obtain the polarized 
double-splitting functions. Summing over the parton polarizations,
we obtain the unpolarized and the spin-correlated double-splitting 
functions, previously computed in Ref.~\cite{glover} and \cite{cat},
respectively, in the
conventional dimensional regularization (CDR) scheme. Since we sum over
two helicity states of the external partons, as it is done in the
dimensional reduction (DR) scheme~\cite{fdh,dimred},
our results agree with the ones in the CDR scheme by setting there
the dimensional regularization scheme (RS) parameter $\epsilon=0$.

For a scattering with production of $m$ partons,
we define the $n$-parton cluster, with $m > n$, as the set of $n$ final-state 
partons where the distance in rapidity between any two partons in the 
cluster is much smaller than the rapidity distance between a parton 
inside the cluster and a parton outside. 
In the BFKL theory, $(n+1)$-parton forward clusters 
provide the tree parts of ${\rm N}^n{\rm LO}$ impact factors, while
$(n+1)$-parton central clusters provide
the tree parts of the ${\rm N}^n{\rm LO}$ Lipatov vertex. $n$-parton clusters
were given also a field-theoretical basis in terms of an effective action
describing the interaction between physical partons grouped into
gauge-invariant clusters and the gluons
exchanged in the cross channel~\cite{lipaction}.
In addition to computing the three-parton forward clusters and the 
three-gluon central cluster mentioned above, 
we compute the four-gluon forward cluster,
i.e. the purely gluonic tree part of the NNNLO impact factor.
By taking then the quadruple collinear limit, we obtain the polarized 
triple-splitting functions. They could be used in a gauge-invariant
evaluation of the Altarelli-Parisi evolution at three loops~\cite{uwer}.

The outline of the paper is: 
in \sec{sec:amps} we review the standard color decompositions of the $n$-parton
tree amplitudes, and
we present a color decomposition of the gluon
amplitudes in terms of the linearly independent subamplitudes only. 
In \sec{sec:loif} we review 
the elastic scattering of two partons in the high-energy limit,
which allows for the extraction of the LO impact 
factors. In \sec{sec:nloif}
we review the amplitudes for the production of three partons, with a 
gauge-invariant two-parton forward cluster;
from these, we can extract the tree parts of the 
NLO impact factors; by taking the collinear limit, we obtain
the LO splitting functions. In \sec{sec:nnloif} 
we compute the amplitudes for the production of four partons, with a 
three-parton forward cluster; then we
extract the tree parts of the NNLO impact factors, and by taking the 
triple collinear limit we obtain the polarized and unpolarized
double-splitting functions. In \sec{sec:nnnloif} we compute the amplitude 
for the production of five gluons, with a four-gluon forward cluster.
We extract the tree part of the gluonic NNNLO impact factor, and by taking the 
quadruple collinear limit we obtain the polarized 
triple-splitting functions. In addition, by taking the limit in which
three gluons are emitted in the central-rapidity region, we obtain the
gauge-invariant three-gluon central cluster, i.e. the tree part of the 
NNLO Lipatov vertex.
In \sec{sec:conc} we draw our conclusions.

\section{Tree Amplitudes}
\label{sec:amps}

In this section we review the color decomposition of purely gluonic
and quark-gluon tree amplitudes. For the purely gluonic tree amplitudes,
we introduce a color decomposition in terms of the
linearly independent subamplitudes, \eqn{GluonDecompNew}.

\subsection{Gluon amplitudes}
\label{sec:glueamps}

For an amplitude with $n$ gluons
the usual color decomposition at tree level 
reads~\cite{mpReview,Color,bk,BeGi87,MPX},
\begin{eqnarray}
 i{\cal A}(g_1,\ldots,g_n)&=&  i g^{n-2} \sum_{\sigma\in S_n/{\Bbb Z}_n}
   \tr (\lambda^{\sigma_1}\cdots \lambda^{\sigma_n})
    A(g_{\sigma_1},\ldots,g_{\sigma_n})
\label{GluonDecomp}
\end{eqnarray}
where $S_n/{\Bbb Z}_n$ are the non-cyclic permutations of $n$ elements.
The dependence on the particle helicities and momenta in the subamplitude,
and on the gluon colors in the trace, is implicit
in labelling each leg with the index $i$.
Helicities and momenta are defined as if all particles were outgoing.

The gauge invariant subamplitudes $A$
satisfy the relations~\cite{BeGi87,MPX}, proven for arbitrary
$n$ in Ref.~\cite{BeGi88},
\begin{eqnarray}
&& A(1,2,\ldots,n-1,n) =A(n,1,2,\ldots,n-1) 
   \hspace{3.5cm}  {\rm cyclicity} \nonumber \\
&& A(1,2,\ldots,n) = (-1)^n A(n,\ldots,2,1)  
   \hspace{4.8cm} {\rm reflection}  \label{relations}\\
&& A(1,2,3,\ldots,n) + A(2,1,\ldots,n)+\ldots + A(2,3,\ldots,1,n) = 0
   \hspace{.5cm} {\rm dual \;Ward \;identity} \nonumber 
\end{eqnarray}
The above relations are sufficient to show that, for $n\le 6$  the number
of independent subamplitudes can be reduced from $(n-1)!$ to
$(n-2)!$. For $n\ge 7$ it  is still possible  to
introduce a basis of $(n-2)!$ elements by using Kleiss-Kuijf's
relation~\cite{KlKu89}
\begin{eqnarray}
A(1,x_1,\ldots,x_p,2,y_1,\ldots,y_q)=
(-1)^p \sum_{\sigma\in OP\{\alpha\}\{\beta\}}
 A(1,2,\{\alpha\}\{\beta\})
%M(x_p,x_{p-1},\ldots,x_1,y_1,\ldots,y_q),
\label{merging}
\end{eqnarray}
where $\alpha_i \in \{\alpha\} \equiv \{x_p,x_{p-1},\ldots,x_1\}\,,
\beta_i \in \{\beta\} \equiv \{y_1,\ldots,y_q\}$ and $OP\{\alpha\}\{\beta\}$
is the set of permutations of the $(n-2)$ objects
$\{x_1,\ldots,x_p,y_1,\ldots,y_q\}$ that preserve the ordering
of the $\alpha_i$ within $\{\alpha\}$ and of the $\beta_i$ within
$\{\beta\}$, while allowing for all possible relative orderings of the
$\alpha_i$ with respect to the $\beta_i$.
%$M(x_p,x_{p-1},\ldots,x_1,y_1,\ldots,y_q)$   is a merging of
%$(x_p,x_{p-1},\ldots,x_1)$ and 
%$(y1,\ldots,y_q)$ which leaves the ordering of the two lists unaffected. 
%The summation
%goes over all $(p+q)!/p!q!$ possible mergings. 
The above relation
has been checked up to $n=8$ in Ref.~\cite{BeGi88}, and proven
for arbitrary $n$ in Ref.~\cite{bddk}. Accordingly, the expression for the 
summed amplitude squared can be written as
\begin{eqnarray}
\sum_{a_1, \dots , a_n} |{\cal A}(1,\dots,n)|^2 
&=& \sum_{i,j=1}^{(n-1)!} {c}_{ij} A_i A^*_j \label{square1}\\
&=&{\cal C}_n(N_c)  \sum_{\sigma \in S_{n-1}} 
\left[ |A(1,\sigma_2,\ldots, \sigma_n)|^2 + {\cal O}\left(
\frac{1}{N_c^2}\right) \right]\label{square2} \\
&=& \sum_{i,j=1}^{(n-2)!} \tilde{c}_{ij} A_i A_j^* \,
,\label{square3}
\end{eqnarray}
where ${c}_{ij}$ in \eqn{square1} is
\begin{equation}
{c}_{ij} = (g^2)^{n-2}\,
\sum_{colors} \tr(P_i(\lambda^{d_1},\ldots,\lambda^{d_n}))
[\tr(P_j(\lambda^{d_1},\ldots,\lambda^{d_n}))]^* \, ,
\end{equation}
with $P_i$ the $i^{th}$ permutation in $S_n/{\Bbb Z}_n$. 
In \eqn{square2}, the coefficient ${\cal C}_n(N_c)$ is
\begin{eqnarray}
{\cal C}_n(N_c) = \frac{ (g^2 N_c) }{2^n}^{n-2} (N_c^2-1) \, \label{calc}.
\end{eqnarray}
The first term in \eqn{square2}
constitutes the Leading Color Approximation (LCA).
Up to $n=5$, the $1/N_c^2$ corrections in \eqn{square2} vanish
and LCA is exact.
The reduced color matrix $\tilde{c}_{ij}$ 
in \eqn{square3}, has been obtained from ${c}_{ij}$ applying
the linear transformations of~\eqn{merging},
thus the labels $i,j$ in \eqn{square3} run only on the permutations
of the linearly independent subamplitudes.

Motivated by the reorganization of the color in the
high-energy limit~\cite{ptlipnl,thuile,lego,ptlip}, 
and using \eqns{relations}{merging} we rewrite \eqn{GluonDecomp} as
\begin{eqnarray}
 i{\cal A}(g_1,\ldots,g_n)&=&    
i \frac{(ig)^{n-2}}{2} \sum_{\sigma\in S_{n-2}}
      f^{a_1 a_2 x_1} f^{x_1 a_3 x_2} \cdots f^{x_{n-3} a_{n-1} a_n}
    A(g_{1},g_{\sigma_2},\ldots,g_{\sigma_{n-1}},g_{n})\, ,\nonumber\\
&=& i \frac{g^{n-2}}{2} \sum_{\sigma\in S_{n-2}}
      (F^{a_2} \cdots F^{a_{n-1}})_{a_1 a_n}
      A(g_{1},g_{\sigma_2},\ldots,g_{\sigma_{n-1}},g_{n})\, ,
\label{GluonDecompNew}
\end{eqnarray}
where $(F^a)_{bc}\equiv i f^{bac}$. We have checked 
\eqn{GluonDecompNew} up to $n=7$. \eqn{GluonDecompNew} enjoys
several remarkable properties.
Firstly, it shows explicitly which is the color
decomposition that allows us to write the full amplitude $i {\cal A}$ 
in terms of the $(n-2)!$ linearly independent subamplitudes only. 
In the following we shall refer to it as to a color ladder. Hence the color
matrix obtained squaring \eqn{GluonDecompNew} yields directly the 
$\tilde{c}_{ij}$ matrix in \eqn{square3}. 
We have checked it against the 
explicit results of Ref.~\cite{BeGiKu90}, up to $n=5$. 
Moreover, it is quite suggestive to note the formal 
correspondence with the amplitudes with a quark-antiquark pair
and $(n-2)$ gluons, \eqn{TwoQuarkGluonDecomp},
where the only difference between the two\footnote{The factor $1/2$ in front
of \eqn{GluonDecompNew} is due to our choice for the  normalization of the
fundamental representation matrices, 
i.e. $\tr(\lambda^a \lambda^b) = \delta^{ab}/2$.}
  is the appropriate representation for 
the color matrices, namely the adjoint for the $n$-gluon  amplitude and 
the fundamental for the one with the ${\bar q} q$ pair.
Finally, the most relevant applications of \eqn{GluonDecompNew} for this work
are  to the study  of the multi-gluon amplitudes in the high-energy limit.
As discussed in the following, the color ladder   
naturally arises~\cite{lego,ptlip} in the configurations  
where the gluons are strongly ordered in rapidity,
i.e. in the multi-Regge kinematics. Indeed in the strong-rapidity 
ordering only the subamplitude with the corresponding order in the 
color coefficient contributes to \eqn{GluonDecompNew}. 
At NLO, where the strong ordering is relaxed for two
adjacent gluons, the leading subamplitudes are the two which differ
just by the exchange of the gluon labels in the color ladders~\cite{thuile}. 
As we shall see this result generalizes at NNLO and beyond.
Nonetheless, in the following we have chosen 
to derive our results starting from \eqn{GluonDecomp} instead of
using directly \eqn{GluonDecompNew}. 
The former, though more laborious, shows explicitly
how the color traces must be recombined to obtain the color ladder
and, more importantly, allows us to find the relations necessary
to prove the factorization in the multi-collinear limits. 

For the {\sl maximally helicity-violating} configurations, $(-,-,+,...,+)$,
in \eqn{GluonDecomp} or \eqn{GluonDecompNew},
there is only one independent color/helicity subamplitude, 
the Parke-Taylor (PT) subamplitude
\begin{equation}
A(g_1, ... ,g_n) = 2^{n/2} {\langle i\, j \rangle^4 \over
\langle 1\, 2 \rangle \cdots \langle (n-1)\, n\rangle \langle n\,
1\rangle}\, ,\label{ptgluon}
\end{equation}
where the $i^{th}$ and the $j^{th}$ gluons have negative helicity.
All other color/helicity amplitudes can be obtained by relabelling and
by use of reflection symmetry, \eqn{relations}, and parity inversion.
Parity inversion flips the helicities of all particles,
and it is accomplished by the substitution
$\spa{i}.j \leftrightarrow \spb{j}.i$. Subamplitudes of
non-PT type, i.e. with three or more gluons of $-$ helicity have a more
complicated structure.

\subsection{Quark-gluon amplitudes}
\label{sec:quarkamps}

For an amplitude with two quarks and $(n-2)$ gluons
the color decomposition at tree-level is~\cite{mpReview,Color,bk,BeGi87,MPX},
\begin{equation}
 i{\cal A}(\bar{q},q; g_1,\ldots,g_{(n-2)})
  =   ig^{n-2} \sum_{\sigma\in S_{n-2}}
   (\lambda^{\sigma_1}\ldots \lambda^{\sigma_{n-2}})_{j}^{~\ib}\
    A(\bar{q},q; g_{\sigma_1},\ldots,g_{\sigma_{n-2}})\, ,
\label{TwoQuarkGluonDecomp}
\end{equation}
where $S_{n-2}$ is the permutation group on $n-2$ elements.

For the {\sl maximally helicity-violating} configurations,$(-,-,+, \ldots ,+)$,
there is one independent color/helicity subamplitude, the Parke-Taylor (PT) 
subamplitude
\begin {equation}
A({\bar q}^+,q^-; g_1, ... ,g_{(n-2)}) = 2^{(n-2)/2}
{\langle {\bar q} i \rangle \langle q i \rangle^3 \over
\langle {\bar q} q\rangle \langle q 1\rangle \cdots\langle (n-2)\,
{\bar q}\rangle}\, ,\label{pt2quark}
\end{equation}
where gluon $g_i$ has negative helicity. Helicity is conserved along
the massless-fermion line.
All other color/helicity amplitudes can be obtained by relabelling and
by use of parity inversion, reflection symmetry and charge 
conjugation. In performing parity inversion, there is
a factor of $-1$ for each pair of quarks participating in the amplitude.
Reflection symmetry is like in \eqn{relations}, for gluons and/or quarks
alike. Charge conjugation swaps quarks and antiquarks without
inverting helicities. In particular, using reflection symmetry and charge 
conjugation on Eq.~(\ref{pt2quark}) we obtain
\begin {equation}
A({\bar q}^-,q^+; g_1, ... ,g_{(n-2)}) = 2^{(n-2)/2}
{\langle {\bar q} i \rangle^3 \langle q i \rangle \over
\langle {\bar q} q\rangle \langle q 1\rangle \cdots\langle (n-2)\,
{\bar q}\rangle}\, ,\label{pt2quark2}
\end{equation}
where gluon $g_i$ has negative helicity.

For an amplitude with four quarks and $(n-4)$ gluons
the color decomposition at tree-level is~\cite{mpReview}
\begin{eqnarray}
\lefteqn{ i\A{}(\bar{q_1},q_1;\bar{q_2},q_2;g_1,\ldots,g_{(n-4)})
= ig^{n-2} \sum_{k=0}^{n-4} \sum_{\sigma\in S_{k}}\,
\sum_{\rho\in S_{l}} }
\label{FourQuarkGluonDecomp}\\
&\times& \Biggl[ (\lambda^{\sigma_1} ... \lambda^{\sigma_k})_{j_2}^{~\ib_1}\
  (\lambda^{\rho_1} ... \lambda^{\rho_l})_{j_1}^{~\ib_2}\
    A(\bar{q_1},q_1;\bar{q_2},q_2;g_{\sigma_1},...,
g_{\sigma_k};g_{\rho_1},...,g_{\rho_l}) \nonumber\\ &-&
{1\over N_c}\, (\lambda^{\sigma_1} ... \lambda^{\sigma_k})_{j_1}^{~\ib_1}\
  (\lambda^{\rho_1} ... \lambda^{\rho_l})_{j_2}^{~\ib_2}\
    B(\bar{q_1},q_1;\bar{q_2},q_2;g_{\sigma_1},...,
g_{\sigma_k};g_{\rho_1},...,g_{\rho_l}) \Biggr]\, ,\nonumber
\end{eqnarray}
with $k+l=n-4$, and where we suppose that the two quark pairs have
distinct flavor. The sums are over the partitions of $(n-4)$
gluons between the two quark lines, and over the permutations of the
gluons within each partition. For $k=0$ or $l=0$, the color strings
reduce to Kronecker delta's. For identical quarks, we must subtract 
from \eqn{FourQuarkGluonDecomp}
the same term with the exchange of the quarks $(q_1\leftrightarrow q_2)$.

\noindent
For the maximally helicity-violating configurations, $(-,-,+,...,+)$,
with like-helicity for all of the gluons, the $A$ and
$B$ subamplitudes factorize into distinct contributions
for the two quark antennae~\cite{mpReview,Color,BeGi87,MPX}. However, 
as we shall
see in Sect.~\ref{sec:nnloqqq}, we need the helicity configurations
with two gluons of opposite helicity. For these the above
mentioned factorization does not occur.

%\begin{eqnarray}
%\lefteqn{ A_n^\tree(\bar{q_1},q_1;\bar{q_2},q_2;g_{\sigma_1},...,
%g_{\sigma_k};g_{\rho_1},...,g_{\rho_l}) = }
%\label{FourQuarkMHVa}\\ & & 2^{(n-4)/2}
%{f(\nu_{q_1},\nu_{q_2}) \over \spa{\bar q_1}.{q_1}
%\spa{\bar q_2}.{q_2} }\,
%{ \spa{q_1}.{\bar q_2} \over \spa{q_1}.{g_{\sigma_1}} \cdots
%\spa{g_{\sigma_k}}.{\bar q_2} }\, { \spa{q_2}.{\bar q_1} \over
%\spa{q_2}.{g_{\rho_1}} \cdots \spa{g_{\rho_l}}.{\bar q_1} } \nonumber\\
%\lefteqn{ B_n^\tree(\bar{q_1},q_1;\bar{q_2},q_2;g_{\sigma_1},...,
%g_{\sigma_k};g_{\rho_1},...,g_{\rho_l}) = } \label{FourQuarkMHVb}
%\\ & & 2^{(n-4)/2}
%{f(\nu_{q_1},\nu_{q_2}) \over \spa{\bar q_1}.{q_1}
%\spa{\bar q_2}.{q_2} }\,
%{ \spa{q_1}.{\bar q_1} \over \spa{q_1}.{g_{\sigma_1}} \cdots
%\spa{g_{\sigma_k}}.{\bar q_1} }\, { \spa{q_2}.{\bar q_2} \over
%\spa{q_2}.{g_{\rho_1}} \cdots \spa{g_{\rho_l}}.{\bar q_2} }\, ,\nonumber
%\end{eqnarray}
%
%with
%
%\begin{eqnarray}
%f(+,+) = \spa{\bar q_1}.{\bar q_2}^2 \quad & \quad
%f(+,-) = \spa{\bar q_1}.{q_2}^2 \nonumber\\
%f(-,+) = \spa{q_1}.{\bar q_2}^2 \quad & \quad
%f(-,-) = \spa{q_1}.{q_2}^2\, .\nonumber
%\end{eqnarray}
%
%This factorization does not occur for other helicity configurations.

\section{The Leading Impact Factors}
\label{sec:loif}

\begin{figure}[t]
\begin{center}
\vspace*{-1cm}
\hspace*{0cm}
\epsfxsize=14cm \epsfbox{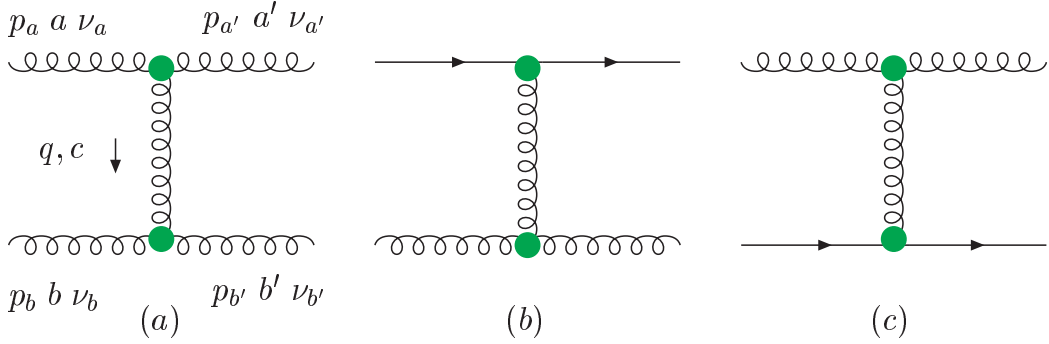}
\vspace*{0.3cm}
\caption{$(a)$ Amplitude for $g\, g \to g\, g$ scattering and $(b),\,(c)$ for
$q\, g \to q\, g$ scattering. We label the external lines with momentum, 
color and helicity, and the internal lines with momentum and color.}
\label{fig:LO}
\end{center}
\end{figure}

We consider the elastic scattering of two partons of momenta $p_a$ and $p_b$
into two partons of momenta $p_{a'}$ and $p_{b'}$, in the high-energy limit,
$s\gg |t|$. 
Firstly, we consider the amplitude for gluon-gluon scattering (\fig{fig:LO}a).
Using Eqs.~(\ref{GluonDecomp}), (\ref{relations}), or \eqn{GluonDecompNew}, and
\eqn{ptgluon}, and \app{sec:appb}, we obtain~\cite{ptlip} 
\begin{equation}
\A{g\, g \rightarrow g\, g}
(p_a^{\nu_a},p_{a'}^{\nu_{a'}} \sep  p_{b'}^{\nu_{b'}},p_b^{\nu_b})
= 2 s
\left[i g\, f^{aa'c}\, C^{g;g}(p_a^{\nu_a};p_{a'}^{\nu_{a'}}) \right]
{1\over t} \left[i g\, f^{bb'c}\, C^{g;g}(p_b^{\nu_b};p_{b'}^{\nu_{b'}}) 
\right]\, ,\label{elas}
\end{equation} 
with $q = p_{b'} + p_b$ and $t \simeq -|q_\perp|^2$.
The LO impact factors $g^*\, g \rightarrow g$, with $g^*$ an off-shell gluon,
are
\begin{equation}
C^{g;g}(p_a^-;p_{a'}^+) = 1 \qquad C^{g;g}(p_b^-;p_{b'}^+) =
{p_{b'\perp}^* \over p_{b'\perp}}\, .\label{centrc}
\end{equation} 
They conserve helicity along the on-shell gluon line and transform
under parity into their complex conjugates,
\begin{equation}
[C^{g;g}(\{k^{\nu}\})]^* = C^{g;g}(\{k^{-\nu}\})\, . 
\end{equation} 
In \eqn{elas} four 
helicity configurations are leading, two for each impact 
factor~\footnote{All throughout this paper, we shall always write only half 
of the helicity configurations contributing to an impact factor, the other half
being obtained by parity.}. The helicity-flip
impact factor $C^{g;g}(p^+;p'^+)$ is subleading in the high-energy limit.

From Eqs.~(\ref{TwoQuarkGluonDecomp})-(\ref{pt2quark}), we obtain
the quark-gluon $q\, g \rightarrow q\, g$
scattering amplitude in the high-energy limit~\cite{thuile},
\begin{eqnarray}
&& \hspace{-.5cm} \A{q\, g \rightarrow q\, g}
(p_a^{\nu_a},p_{a'}^{\nu_{a'}} \sep p_{b'}^{\nu_{b'}},p_b^{\nu_b}) 
= 2 s \left[g\, \lambda^c_{a' \bar a}\,
C^{\bar q;q}(p_a^{-\nu_{a'}};p_{a'}^{\nu_{a'}}) \right] {1\over t} 
\left[i g\, f^{bb'c}\, C^{g;g}(p_b^{\nu_b};p_{b'}^{\nu_{b'}}) \right]
\label{elasqa} \\
&& \hspace{-.5cm} \A{g\, q \rightarrow g\, q}
(p_a^{\nu_a},p_{a'}^{\nu_{a'}} \sep p_{b'}^{\nu_{b'}},p_b^{\nu_b}) 
= 2 s \left[i g\, f^{aa'c}\, 
C^{g;g}(p_a^{\nu_a};p_{a'}^{\nu_{a'}}) \right] {1\over t}
\left[g\, \lambda^c_{b' \bar b}\,
C^{\bar q ;q}(p_b^{-\nu_{b'}};p_{b'}^{\nu_{b'}}) \right] \label{elasqb}
\end{eqnarray}
where we have labelled the incoming quarks
as outgoing antiquarks with negative momentum, e.g.
the antiquark is $p_a$ in \eqn{elasqa} (\fig{fig:LO}b),
and $p_b$ in \eqn{elasqb} (\fig{fig:LO}c). The LO impact factors
$g^*\, q \rightarrow q$ are,
\begin{equation}
C^{\bar q;q}(p_a^-;p_{a'}^+) = -i\, ;\qquad C^{\bar q;q}(p_b^-;p_{b'}^+) =
i \left({p_{b'\perp}^* \over p_{b'\perp}}\right)^{1/2}\, .\label{cbqqm}
\end{equation}
Under parity, the functions (\ref{cbqqm}) transform as
\begin{equation}
[C^{\bar q;q}(\{k^{\nu}\})]^* = S\, C^{\bar q;q}(\{k^{-\nu}\})
\qquad  {\rm with} \quad S = -{\rm sign} (\bar q^0 q^0)\, ,\label{qparity}
\end{equation}
and in
general an impact factor acquires a coefficient $S$ for each pair of quarks
(see Sect.~\ref{sec:amps}).
Analogously, the antiquark-gluon $\bar q\, g \rightarrow \bar q\, g$
amplitude is
\begin{eqnarray}
&& \hspace{-.5cm} \A{\bar q\,g\rightarrow\bar q\,g}
(p_a^{\nu_a},p_{a'}^{\nu_{a'}} \sep p_{b'}^{\nu_{b'}},p_b^{\nu_b}) 
= 2 s \left[g\,\lambda^c_{a\bar a'}
\, C^{q;\bar q}(p_a^{-\nu_{a'}};p_{a'}^{\nu_{a'}}) \right] {1\over t} 
\left[i g\, f^{bb'c}\, C^{g;g}(p_b^{\nu_b};p_{b'}^{\nu_{b'}}) \right]
\label{elasaqa}\\ 
&& \hspace{-.5cm} \A{g\,\bar q\rightarrow g\,\bar q}
(p_a^{\nu_a},p_{a'}^{\nu_{a'}} \sep p_{b'}^{\nu_{b'}},p_b^{\nu_b})  
= 2 s \left[i g\, f^{aa'c}\, 
C^{g;g}(p_a^{\nu_a};p_{a'}^{\nu_{a'}}) \right] {1\over t}
\left[g\, \lambda^c_{b \bar b'}\,
C^{q;\bar q}(p_b^{-\nu_{b'}};p_{b'}^{\nu_{b'}}) \right] \label{elasaqb}
\end{eqnarray}
where the antiquark is $p_{a'}$ in \eqn{elasaqa} 
and $p_{b'}$ in \eqn{elasaqb}, and the 
LO impact factors $g^*\, \bar q \rightarrow \bar q$ are,
\begin{equation}
C^{q;\bar q}(p_a^-;p_{a'}^+) = i\, ;\qquad C^{q;\bar q}(p_b^-,p_{b'}^+) =
-i \left({p_{b'\perp}^* \over p_{b'\perp}}\right)^{1/2}\, .\label{cbqqp}
\end{equation}
In the amplitudes (\ref{elas}), (\ref{elasqa}), (\ref{elasqb}),
(\ref{elasaqa}), (\ref{elasaqb}),
the leading contributions from all the 
Feynman diagrams have been included. However, the amplitudes have the
effective form of a gluon exchange in the $t$ channel (\fig{fig:LO}),
and differ only for the relative color strength in the production vertices
\cite{CM}.
This allows us to replace an incoming gluon with a quark, for instance
on the upper line, via the simple substitution
\begin{equation}
i g\, f^{aa'c}\, C^{g;g}(p_a^{\nu_a};p_{a'}^{\nu_{a'}}) \leftrightarrow g\, 
\lambda^c_{a' \bar a}\, C^{\bar q;q}(p_a^{-\nu_{a'}};p_{a'}^{\nu_{a'}})\, 
,\label{qlrag}
\end{equation}
and similar ones for an antiquark and/or for the lower line.
For example, the quark-quark $q\,q\to q\,q$ scattering amplitude in 
the high-energy limit is
\begin{equation}
\A{q\,q\to q\,q}
(p_a^{\nu_a},p_{a'}^{\nu_{a'}} \sep p_{b'}^{\nu_{b'}},p_b^{\nu_b}) 
= 2 s \left[g\, 
\lambda^c_{a' \bar a}\,
C^{\bar q;q}(p_a^{-\nu_{a'}};p_{a'}^{\nu_{a'}}) \right] {1\over t} 
\left[g\, \lambda^c_{b' \bar b}\,
C^{\bar q;q}(p_b^{-\nu_{b'}};p_{b'}^{\nu_{b'}}) \right]\,
.\label{elasq}
\end{equation}

\section{The Next-to-leading Impact Factors}
\label{sec:nloif}

Let three partons be produced with momenta $k_1$, $k_2$ and $p_{b'}$ 
in the scattering between two partons of momenta $p_a$ and $p_b$, and
to be specific, we shall take partons $k_1$ and $k_2$ in the forward-rapidity
region of parton $p_a$, the analysis for $k_1$ and $k_2$ 
in the forward-rapidity region of $p_b$ being similar.
Parametrizing the momenta as in \eqn{in}, we have,
\begin{equation}
y_1 \simeq y_2 \gg y_{b'}\,;\qquad |k_{1\perp}|\simeq|k_{2\perp}|
\simeq|p_{b'\perp}|\, .\label{nloreg}
\end{equation}
\begin{figure}[t]
\begin{center}
\vspace*{0cm}
\hspace*{0cm}
\epsfxsize=14cm \epsfbox{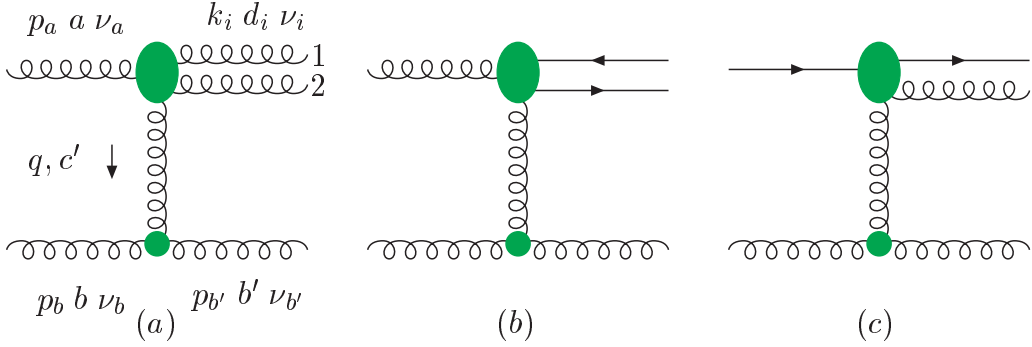}
\caption{Amplitudes for the production of three partons, with partons
$k_1$ and $k_2$ in the forward-rapidity region of parton $p_a$.}
\label{fig:NLO}
\end{center}
\end{figure}

\subsection{The NLO impact factor $g\, g^* \rightarrow g\, g$}
\label{sec:nloggg}

We consider the amplitude for the scattering $g\, g\, \to g\,
g\, g$ (\fig{fig:NLO}a). Only PT subamplitudes contribute,
thus using Eqs.~(\ref{GluonDecomp}), (\ref{relations}) and
(\ref{ptgluon}), and \app{sec:appc}, we obtain~\cite{fl,ptlipnl}
\begin{eqnarray}
& & \A{g\,g \to 3g}(p_a^{\nu_a}, k_1^{\nu_1}, k_2^{\nu_2} \sep 
p_{b'}^{\nu_{b'}}, p_b^{\nu_b}) \nonumber\\ & & = 
4\, g^3\, {s\over |q_{\perp}|^2}\, 
C^{g;g}(p_b^{\nu_b};p_{b'}^{\nu_{b'}})\, \sum_{\sigma\in S_2} \left[
A^{g;gg}(p_a^{\nu_a}; k_{\sigma_1}^{\nu_{\sigma_1}}, 
k_{\sigma_2}^{\nu_{\sigma_2}}) \right. \label{trepos}\\ & & \times
{\rm tr} \left( \lambda^a \lambda^{d_{\sigma_1}} \lambda^{d_{\sigma_2}} 
\lambda^{b'} \lambda^b - \lambda^a \lambda^{d_{\sigma_1}} 
\lambda^{d_{\sigma_2}} \lambda^b 
\lambda^{b'} + \lambda^a \lambda^{b'} \lambda^b \lambda^{d_{\sigma_2}} 
\lambda^{d_{\sigma_1}} - \lambda^a \lambda^b \lambda^{b'} 
\lambda^{d_{\sigma_2}} 
\lambda^{d_{\sigma_1}} \right) \nonumber\\ & &
\left. + B^{g;gg}(p_a^{\nu_a}; k_{\sigma_1}^{\nu_{\sigma_1}}, 
k_{\sigma_2}^{\nu_{\sigma_2}})\, {\rm tr} \left(
\lambda^a \lambda^{d_{\sigma_1}} \lambda^{b'} \lambda^b \lambda^{d_{\sigma_2}}
- \lambda^a \lambda^{d_{\sigma_2}} \lambda^b \lambda^{b'} 
\lambda^{d_{\sigma_1}} \right) \right] \, ,\nonumber
\end{eqnarray}
with the sum over the permutations of the two gluons 1 and 2, the 
LO impact factor, $C^{g;g}(p_b^{\nu_b};p_{b'}^{\nu_{b'}})$, as in 
\eqn{centrc}, and 
\begin{eqnarray}
A^{g;gg}(p_a^{\nu_a}; k_1^{\nu_1}, k_2^{\nu_2}) &=&
C^{g;gg}(p_a^{\nu_a}; k_1^{\nu_1}, k_2^{\nu_2})
A^{\bar\nu}(k_1, k_2) \nonumber\\
B^{g;gg}(p_a^{\nu_a}; k_1^{\nu_1}, k_2^{\nu_2}) &=&
C^{g;gg}(p_a^{\nu_a}; k_1^{\nu_1}, k_2^{\nu_2})
B^{\bar\nu}(k_1, k_2) \label{treposb}
\end{eqnarray}
with $\bar\nu={\rm sign}(\nu_a+\nu_1+\nu_2)$, and
\begin{eqnarray}
C^{g;gg}(p_a^-;k_1^+,k_2^+) &=& 1 \label{trecoeff}\\
C^{g;gg}(p_a^+;k_1^-,k_2^+) &=& x_1^2 \nonumber\\
C^{g;gg}(p_a^+;k_1^+,k_2^-) &=& x_2^2\, .\nonumber
\end{eqnarray}
%
\iffalse
\begin{eqnarray}
C^{g;gg}(p_a^{\nu_a}; k_1^{\nu_1}, k_2^{\nu_2}) &=& 
\left\{
\begin{array}{l}
1{~}\qquad \nu_a=-1 \\
x_i^2 \qquad \nu_i=-1 \qquad i=1,2\\
\end{array}\right.
\quad {\rm with}\; \bar\nu=1, \label{treposb}
\end{eqnarray}
\fi
The momentum fractions are defined as
\begin{equation}
x_i= \frac{k_i^+}{k_1^+ + k_2^+}   
\qquad i=1,2 \quad (x_1+x_2=1)\, \label{mtmfrac},
\end{equation}
and the function $A^+$ as follows:
\begin{equation}
A^+(k_1,k_2) = - \sqrt{2}\, {q_\perp\over k_{1\perp}}
\sqrt{x_1\over x_2}\, {1\over \langle 1 2\rangle}\, ,\label{avert}
\end{equation}
with $\langle 1 2\rangle$ a shorthand for 
$\langle k_1 k_2\rangle$. Using the dual Ward identity~\cite{mpReview}, 
or U(1) decoupling equations~\cite{bk,bg},
the function $B^{\bar\nu}$ in \eqn{treposb}, and thus the function 
$B^{g;gg}$, can be written as
\begin{equation}
B^{\bar\nu}(k_1,k_2) = -\left[ A^{\bar\nu}(k_1,k_2) + 
A^{\bar\nu}(k_2,k_1)\right] \, .\label{ab}
\end{equation}
The function
$C^{g;gg}(p_a^+;k_1^+,k_2^+)$ is subleading to the required accuracy.
The function $A^{\bar\nu}$ has a collinear divergence as $2k_1\cdot 
k_2\to 0$, but the divergence cancels out in the function $B^{\bar\nu}$
where gluons 1 and 2 are not adjacent in color ordering \cite{ptlipnl}.

Using \eqn{ab}, and fixing $t \simeq
- |q_{\perp}|^2$, the amplitude (\ref{trepos}) may be rewritten as,
\begin{eqnarray}
& & \A{g\,g\to 3g}(p_a^{\nu_a}, k_1^{\nu_1}, k_2^{\nu_2} \sep  
p_{b'}^{\nu_{b'}}, p_b^{\nu_b}) \label{nllfg}\\ & & = 2 s \left\{ 
(ig)^2\, \sum_{\sigma\in S_2} f^{a d_{\sigma_1} c } f^{c d_{\sigma_2} c'} 
A^{g;gg}(p_a^{\nu_a}; k_{\sigma_1}^{\nu_{\sigma_1}}, 
k_{\sigma_2}^{\nu_{\sigma_2}}) \right\}\, {1\over t}\, \left[ig\, f^{bb'c'} 
C^{g;g}(p_b^{\nu_b};p_{b'}^{\nu_{b'}}) \right]\, ,\nonumber
\end{eqnarray}
where the NLO impact factor for 
$g^*\, g \rightarrow g\, g$ is enclosed in curly brackets, and includes
six helicity configurations. 

In the multi-Regge limit $y_1\gg y_2$, 
\begin{eqnarray}
\lim_{y_1\gg y_2} 
A^{g;gg}(p_a^{\nu_a}; k_1^{\nu_1}, k_2^{\nu_2})
= C^{g;g}(p_a^{\nu_a};k_1^{\nu_1})
{1\over t_1}\, C^g(q_1,k_2^{\nu_2},q)\, ,\label{camr}
\end{eqnarray}
with $q_1 = -(p_a + k_1)$, and $t_1 \simeq - |q_{1\perp}|^2$,
and with LO Lipatov vertex, $g^*\, g^* \rightarrow g$ \cite{ptlip,lipat},
\begin{equation}
C^g(q_1,k^+,q_2) = \sqrt{2}\, {q^*_{1\perp} q_{2\perp}\over k_\perp}\, 
.\label{lip}
\end{equation}
Accordingly, the amplitude (\ref{nllfg}) is reduced to an amplitude in
multi-Regge kinematics~\cite{fkl,ptlip}, with
the effective form of a gluon-ladder exchange in the $t$ channel,
\begin{eqnarray}
& & \qquad \A{g g \to 3g}(p_a^{\nu_a}, k_1^{\nu_1} \sep  k_2^{\nu_2} \sep
p_{b'}^{\nu_{b'}}, p_b^{\nu_b}) = \label{three}\\ 
& & 2 s \left[i g\, f^{ad_1c}\, C^{g;g}
(p_a^{\nu_a};k_1^{\nu_1}) \right]\, {1\over t_1} 
\left[i g\, f^{cd_2c'}\, C^g(q_1,k_2^{\nu_2},q_2)\right]\, {1\over t_2}\, 
\left[i g\, f^{bb'c'}\, C^{g;g}(p_b^{\nu_b};p_{b'}^{\nu_{b'}}) \right]\,
,\nonumber
\end{eqnarray}
with $q_2 = p_b+p_{b'}$ and $t_2 \simeq - |q_{2\perp}|^2$.

\subsection{The NLO impact factor $g\, g^* \rightarrow \bar q q$}
\label{sec:nlogqbarq}

The amplitude $g\, g\to \bar{q}\,q\, g$ for the production of a 
$q\bar q$ pair in the forward-rapidity region of gluon $a$ 
(\fig{fig:NLO}c) is obtained by taking the amplitudes 
(\ref{TwoQuarkGluonDecomp})-(\ref{pt2quark2}) 
in the kinematics (\ref{nloreg}) \cite{ptlipqq},
\begin{eqnarray}
& & \A{g\, g\to \bar{q}\,q\, g}(p_a^{\nu_a}, k_1^{\nu_1}, k_2^{-\nu_1} \sep 
p_{b'}^{\nu_{b'}}, p_b^{\nu_b}) \label{forwqq}\\ & & = 2 s \left\{g^2\, 
\left[\left(\lambda^{c'} \lambda^a\right)_{d_2\bar{d_1}} 
A^{g;\bar{q}q}(p_a^{\nu_a}; k_1^{\nu_1}, k_2^{-\nu_1})
+ \left(\lambda^a \lambda^{c'}\right)_{d_2\bar{d_1}} 
A^{g;\bar{q}q}(p_a^{\nu_a}; k_2^{-\nu_1}, k_1^{\nu_1})
\right] \right\} 
\nonumber\\ & & \times {1\over t}\, \left[ig\, f^{bb'c'} 
C^{g;g}(p_b^{\nu_b};p_{b'}^{\nu_{b'}}) \right]\, ,\nonumber
\end{eqnarray}
with $k_1$ the antiquark, the NLO impact factor $g^*\, g 
\rightarrow \bar q q$ in curly brackets, and with
\begin{eqnarray}
A^{g;\bar{q}q}(p_a^{\nu_a}; k_1^{\nu_1}, k_2^{-\nu_1}) &=&
C^{g;\bar{q}q}(p_a^{\nu_a}; k_1^{\nu_1}, k_2^{-\nu_1})
A^{\bar\nu}(k_1, k_2) \nonumber\\
C^{g\;\bar{q}q}(p_a^+;k_1^+,k_2^-) &=& \sqrt{x_1\,x_2^3}
\label{cqqa}\\
C^{g;\bar{q}q}(p_a^+;k_1^-,k_2^+) &=& \sqrt{x_1^3\,x_2}
\, ,\nonumber
\end{eqnarray}
with momentum fractions as in \eqn{mtmfrac},
$A^{\bar\nu}$ in \eqn{avert} and $\bar\nu=\nu_a$. 
The NLO impact factor $g^*\, g \rightarrow \bar q q$ 
allows for four helicity configurations.

In the multi-Regge limit $k_1^+\gg k_2^+$, the NLO impact factor $g^*\, g 
\rightarrow \bar q q$ vanishes, since
quark production along the multi-Regge ladder is suppressed.

\subsection{The NLO impact factor $q\, g^* \rightarrow q\, g$}
\label{sec:nloqqg}

The amplitude $q\,g\to q\,g\,g$ for the production of a $q\, g$ pair 
in the forward-rapidity region of quark $a$ (\fig{fig:NLO}b) 
is obtained by taking the amplitudes 
(\ref{TwoQuarkGluonDecomp})-(\ref{pt2quark2}) in the kinematics (\ref{nloreg})
\cite{thuile} 
\begin{eqnarray}
& & \A{q\,g\to q\,g\,g}(p_a^{-\nu_1}, k_1^{\nu_1}, k_2^{\nu_2} \sep 
p_{b'}^{\nu_{b'}}, p_b^{\nu_b}) \label{forwqg}\\ & & = 2 s \left\{ g^2\, 
\left[\left(\lambda^{d_2}
\lambda^{c'}\right)_{d_1\bar{a}} 
A^{\bar{q};qg}(p_a^{-\nu_1};k_1^{\nu_1},k_2^{\nu_2}) 
+ \left(\lambda^{c'}\lambda^{d_2}\right)_{d_1\bar{a}} 
B^{\bar{q};qg}(p_a^{-\nu_1};k_1^{\nu_1},k_2^{\nu_2}) \right] \right\} 
\nonumber\\ & & \times {1\over t}\, \left[ig\, f^{bb'c'} 
C^{g;g}(p_b^{\nu_b};p_{b'}^{\nu_{b'}}) \right]\, ,\nonumber
\end{eqnarray}
with $k_1$ the final-state quark, and the NLO impact factor
$q\, g^* \rightarrow q\, g$ in curly brackets. As above, the NLO impact factor
includes four helicity configurations,
\begin{eqnarray}
A^{\bar{q};qg}(p_a^{-\nu_1};k_1^{\nu_1},k_2^{\nu_2}) &=&
C^{\bar{q};qg}(p_a^{-\nu_1};k_1^{\nu_1},k_2^{\nu_2})
A^{\bar\nu}(k_1, k_2) \nonumber\\
B^{\bar{q};qg}(p_a^{-\nu_1};k_1^{\nu_1},k_2^{\nu_2}) &=&
C^{\bar{q};qg}(p_a^{-\nu_1};k_1^{\nu_1},k_2^{\nu_2})
B^{\bar\nu}(k_1, k_2) \nonumber\\
C^{\bar{q};qg}(p_a^-;k_1^+,k_2^+) &=& -i \sqrt{x_1} \label{cqg}\\
C^{\bar{q};qg}(p_a^+;k_1^-,k_2^+) &=& i \sqrt{x_1^3}\, ,\nonumber
\end{eqnarray}
with $A^{\bar\nu}$ in \eqn{avert}, and $B^{\bar\nu}$ given by \eqn{ab}, 
with $\bar\nu =\nu_2$\footnote{In this context, \eqn{ab} is only a 
bookkeeping,
since the U(1) decoupling equation is valid only for the gluino-gluon
subamplitudes corresponding to the quark-gluon subamplitudes used in
\eqn{forwqg}.}. As in \sec{sec:nloggg}, the function $B^{\bar{q};qg}$
vanishes in the collinear limit.

In the multi-Regge limit $k_1^+\gg k_2^+$ the amplitude (\ref{forwqg})
reduces to \eqn{three}, with the substitution 
(\ref{qlrag}) for the upper line, and the LO impact factor
$C^{\bar{q}; q}$ in \eqn{cbqqm}.

The treatment of the amplitude $\bar q\,g\to \bar q\,g\,g$ for the 
production of a $\bar q\, g$ pair in the forward-rapidity region of 
antiquark $a$ is identical to the former,
thus the NLO impact factor $\bar q\, g^* \rightarrow \bar q\, g$
is the same as in \eqn{forwqg} up to inverting the color flow on the quark
line~\cite{thuile}. The corresponding functions $A$ and $B$ are the same 
as in \eqn{cqg}.

\subsection{NLO impact factors in the collinear limit}
\label{sec:nlocoll}

The collinear factorization for a 
generic amplitude occurs both on the subamplitude and on
the full amplitude \cite{mpReview}, since in Eqs.~(\ref{GluonDecomp}),
(\ref{TwoQuarkGluonDecomp}) and (\ref{FourQuarkGluonDecomp}) color
orderings where the collinear partons are not adjacent do not have 
a collinear divergence. Hence in the collinear limit for partons
$i$ and $j$, with $k_i = z P$ and $k_j = (1-z) P$,
a generic amplitude (\ref{GluonDecomp}) can be written as
\begin{equation}
\lim_{k_i || k_j} \A{... d_i d_j ...}(..., k_i^{\nu_i}, k_j^{\nu_j}, ...)
= \sum_\nu \A{... c ...}(..., P^{\nu}, ...) {\rm Split}_{-\nu}^{f \to f_i f_j}
(k_i^{\nu_i}, k_j^{\nu_j})\, ,\label{gencoll}
\end{equation}
with $f$ denoting the parton species.
Accordingly, for $k_1 = z P$ and $k_2 = (1-z) P$, we can write the amplitudes 
(\ref{nllfg}), (\ref{forwqq}) and (\ref{forwqg}) as
\begin{eqnarray}
& &\lim_{k_1 || k_2} \A{f g \to f_1 f_2 g}(p_a^{\nu_a}, k_1^{\nu_1}, 
k_2^{\nu_2} | p_{b'}^{-\nu_b}, p_b^{\nu_b}) \nonumber\\ & & = 
\A{f g \to f g}(p_a^{\nu_a}, P^{-\nu_a} \sep p_{b'}^{-\nu_b}, 
p_b^{\nu_b}) \cdot 
{\rm Split}_{\nu_a}^{f \to f_1 f_2}(k_1^{\nu_1}, k_2^{\nu_2})\, ,\label{coll}
\end{eqnarray}
with $\A{f g \to f g}$ as in 
\eqn{elas}, (\ref{elasqa}) and (\ref{elasaqa}), respectively,
and where we have used helicity conservation in the $s$ channel
(\sec{sec:loif}). 
For the collinear factors, ${\rm Split}_{-\nu}^{f \to f_1 f_2}$, we obtain
\begin{eqnarray}
{\rm Split}_{-\nu}^{g \to g g}(k_1^{\nu_1}, k_2^{\nu_2}) &=&  i g\,
f^{cd_1d_2}\, {\rm split}_{-\nu}^{g \to g g}(k_1^{\nu_1}, k_2^{\nu_2})
\nonumber\\
{\rm Split}_{-\nu}^{g \to \bar{q}q}(k_1^{\nu_1}, k_2^{\nu_2}) &=&  g\,
(\lambda^c)_{d_2\bar{d_1}}\,
{\rm split}_{-\nu}^{g \to \bar{q}q}(k_1^{\nu_1}, k_2^{\nu_2})\label{colfac}\\
{\rm Split}_{-\nu}^{q \to q\, g}(k_1^{\nu_1}, k_2^{\nu_2}) &=& g\,
(\lambda^{d_2})_{d_1\bar{c}}\,
{\rm split}_{-\nu}^{q \to q\, g}(k_1^{\nu_1}, k_2^{\nu_2})\nonumber\\
{\rm Split}_{-\nu}^{\bar q \to \bar q\, g}(k_1^{\nu_1}, k_2^{\nu_2}) &=& g\,
(\lambda^{d_2})_{c\bar d_1}\,
{\rm split}_{-\nu}^{\bar q \to \bar q\, g}(k_1^{\nu_1}, k_2^{\nu_2})\, 
.\nonumber
\end{eqnarray}
with splitting factors \cite{mpReview,bddk},
\begin{eqnarray}
{\rm split}_-^{g \to g g}(k_1^+, k_2^+) &=& \sqrt{2}\, {1\over \sqrt{z(1-z)}
\langle 1 2\rangle}\, \nonumber\\
{\rm split}_+^{g \to g g}(k_1^-, k_2^+) &=& \sqrt{2}\,{z^2\over \sqrt{z(1-z)}
\langle 1 2\rangle}\, \nonumber\\
{\rm split}_+^{g \to g g}(k_1^+, k_2^-) &=& \sqrt{2}\, 
{(1-z)^2\over \sqrt{z(1-z)}\langle 1 2\rangle}\, \nonumber\\
{\rm split}_+^{g \to \bar{q}q}(k_1^+, k_2^-) &=& \sqrt{2}\, {1-z
\over \langle 1 2\rangle} \nonumber\\
{\rm split}_+^{g \to \bar{q}q}(k_1^-, k_2^+) &=& \sqrt{2}\, {z
\over \langle 1 2\rangle}\label{split}\\
{\rm split}_-^{q \to q\, g}(k_1^+, k_2^+) &=& 
{\rm split}_-^{\bar q \to \bar q\, g}(k_1^+, k_2^+) =
\sqrt{2}\, {1\over \sqrt{1-z}\langle 1 2\rangle} \nonumber\\
{\rm split}_+^{q \to q\, g}(k_1^-, k_2^+) &=& 
{\rm split}_+^{\bar q \to \bar q\, g}(k_1^-, k_2^+) =
\sqrt{2}\, {z\over \sqrt{1-z}\langle 1 2\rangle}\nonumber
\end{eqnarray}
and ${\rm split}_{\nu}^{f \to f_1 f_2}
(k_1^{-\nu_1}, k_2^{-\nu_2})$ obtained from ${\rm split}_{-\nu}
^{f \to f_1 f_2}(k_1^{\nu_1}, k_2^{\nu_2})$ by exchanging $\langle k_1
k_2\rangle$ with $[k_2 k_1]$, and multiplying by the coefficient $S$,
\eqn{qparity}, if the splitting factor includes a quark pair. 

Summing over the two helicity states of partons 1 and 2, we obtain a
two-dimensional matrix, whose entries are the Altarelli-Parisi splitting 
functions at fixed color and helicity of the parent parton~\cite{ap}
\begin{equation}  
\sum_{\nu_1\nu_2}\, 
{\rm Split}_{\lambda}^{f \to f_1 f_2} (k_1^{\nu_1}, k_2^{\nu_2})
[{\rm Split}_{\rho}^{f \to f_1 f_2} (k_1^{\nu_1}, k_2^{\nu_2})]^*
= \delta^{cc'} {2g^2\over s_{12}} e^{i(\phi_\lambda-\phi_\rho)} 
P^{f\to f_1f_2}_{\lambda\rho}\, ,\label{apmatrix}
\end{equation}
with $e^{i(\phi_\lambda-\phi_\rho)}$ a phase, where
$e^{i(\phi_+ -\phi_-)}=[2 1]/\langle 1 2\rangle$, and where by
definition $P^{f\to f_1f_2}_{++} = P^{f\to f_1f_2}_{--}$, and
$P^{f\to f_1f_2}_{+-} = P^{f\to f_1f_2}_{-+}$, and
\begin{eqnarray}
P^{g \to g g}_{++} &=& 2C_A\, \left[
{z\over 1-z} + {1-z\over z} + z(1-z)\right] \nonumber\\
P^{g \to g g}_{+-} &=& 2C_A\, z(1-z) \nonumber\\
P^{g \to \bar{q}q}_{++} &=& {1\over 2} 
\left[ z^2 + (1-z)^2 \right] \label{appol}\\
P^{g \to \bar{q}q}_{+-} &=&  z(1-z) \nonumber\\
P^{q \to q\, g}_{++} &=& P^{\bar q \to \bar q\, g}_{++} = 
C_F\, {1+z^2\over 1-z}\, ,\nonumber\\
P^{q \to q\, g}_{+-} &=& P^{\bar q \to \bar q\, g}_{+-} = 0
\, .\nonumber
\end{eqnarray}
For $P^{q \to q\, g}$ helicity conservation on the quark line
sets the off-diagonal elements equal to zero. 
$P^{q \to g\, q}$ is obtained from $P^{q \to q\, g}$ by exchanging
$(z\leftrightarrow 1-z)$. Since we sum over
two helicity states of the external partons, \eqn{appol} is valid in
the dimensional reduction (DR) scheme~\cite{fdh,dimred}. 
\eqn{appol} agrees with the corresponding spin-correlated
splitting functions of Ref.~\cite{csz} in the DR scheme,
after contracting the ones of type $P^{g\to f_1f_2}$ 
with a parent-gluon polarization as in \app{sec:appe}.
The connection of \eqn{appol} with other regularization schemes (RS) 
is also given in Ref.~\cite{csz}.

Averaging over the trace of $P^{f\to f_1f_2}$ in
\eqn{apmatrix}, i.e. over
color and helicity of the parent parton on the left hand side
of \eqn{apmatrix}, we obtain
the unpolarized Altarelli-Parisi splitting functions~\footnote{Note that
in the DR scheme the unpolarized splitting functions do not coincide
with the azimuthally-averaged ones. The latter are given in any RS in
Ref.~\cite{csz}.}
\begin{equation}
{1\over 2{\cal C}} \sum_{\nu\nu_1\nu_2}\, 
|{\rm Split}_{-\nu}^{f \to f_1 f_2} (k_1^{\nu_1}, k_2^{\nu_2})|^2 =
\frac{2g^2}{s_{12}} \langle P^{f\to f_1f_2} \rangle\, ,\label{unpolap}
\end{equation}
with ${\cal C}= N_c^2-1$ for a parent gluon and ${\cal C}= N_c$
for a parent quark, and where the averaged trace of $P^{f\to f_1f_2}$ 
is $\langle P^{f\to f_1f_2} \rangle
= \tr P^{f\to f_1f_2}/2 = P^{f\to f_1f_2}_{++}$.

\section{The Next-to-next-to-leading Impact Factors}
\label{sec:nnloif}

In order to derive the next-to-next-to-leading (NNLO) impact factors,
we repeat the analysis of Sect.~\ref{sec:nloif} with one more final-state 
parton.
Let four partons be produced with momenta $k_1$, $k_2$, $k_3$ and $p_{b'}$ 
in the scattering between two partons of momenta $p_a$ and $p_b$,
with a cluster of three partons, $k_1$, $k_2$ and $k_3$, 
in the forward-rapidity region of parton $p_a$,
\begin{equation}
y_1 \simeq y_2 \simeq y_3 \gg y_{b'}\,;\qquad |k_{1\perp}|\simeq|k_{2\perp}|
\simeq|k_{3\perp}| \simeq|p_{b'\perp}|\, .\label{nnloreg}
\end{equation}

\subsection{The NNLO impact factor $g\, g^* \rightarrow g\, g\, g$}
\label{sec:nnlogggg}

\begin{figure}[t]
\begin{center}
\vspace*{-0.2cm}
\hspace*{0cm}
\epsfxsize=14cm \epsfbox{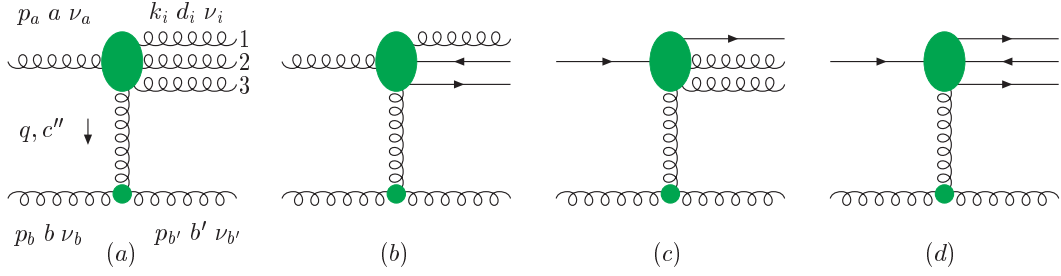}
\vspace*{-0.2cm}
\caption{Amplitudes for the production of four partons, with partons
$k_1$, $k_2$ and $k_3$ in the forward-rapidity region of parton $p_a$.}
\label{fig:NNLO}
\end{center}
\end{figure}

We begin with the amplitude for the scattering $g\, g\, \to g\,
g\, g\, g$ (\fig{fig:NNLO}a) in the kinematics (\ref{nnloreg}).
Using Eqs.~(\ref{GluonDecomp}), (\ref{relations}) and
(\ref{ptgluon}), and the subamplitudes of non-PT type,
with three gluons of $+$ helicity and three gluons of $-$ 
helicity,~\cite{mpReview}, and \app{sec:appd},
we obtain
\begin{eqnarray}
& & \A{g\,g \to 4g}(p_a^{\nu_a}, k_1^{\nu_1}, k_2^{\nu_2}, 
k_3^{\nu_3} \sep p_{b'}^{\nu_{b'}}, p_b^{\nu_b}) \label{threeggg}\\ & & = 
4\, g^4\, {s\over |q_{\perp}|^2}\, 
C^{g;g}(p_b^{\nu_b};p_{b'}^{\nu_{b'}})\, \sum_{\sigma\in S_3} \left[
A^{g;3g}(p_a^{\nu_a}; k_{\sigma_1}^{\nu_{\sigma_1}}, 
k_{\sigma_2}^{\nu_{\sigma_2}}, k_{\sigma_3}^{\nu_{\sigma_3}})\,
\tr \left( \lambda^a \lambda^{d_{\sigma_1}} \lambda^{d_{\sigma_2}} 
\lambda^{d_{\sigma_3}} \lambda^{b'} \lambda^b 
\right. \right. \nonumber \\ & & \left. \qquad
- \lambda^a \lambda^{d_{\sigma_1}} \lambda^{d_{\sigma_2}} 
\lambda^{d_{\sigma_3}} \lambda^b \lambda^{b'}
+ \lambda^b \lambda^{b'} \lambda^{d_{\sigma_3}} \lambda^{d_{\sigma_2}} 
\lambda^{d_{\sigma_1}} \lambda^a
- \lambda^{b'} \lambda^b \lambda^{d_{\sigma_3}} \lambda^{d_{\sigma_2}} 
\lambda^{d_{\sigma_1}} \lambda^a \right) \nonumber\\ & & 
+ B^{g;3g}(p_a^{\nu_a}; k_{\sigma_1}^{\nu_{\sigma_1}}, 
k_{\sigma_2}^{\nu_{\sigma_2}}, k_{\sigma_3}^{\nu_{\sigma_3}})\,
\tr \left( \lambda^a 
\lambda^{d_{\sigma_1}} \lambda^{d_{\sigma_2}} \lambda^{b'} \lambda^b 
\lambda^{d_{\sigma_3}} \right. \nonumber\\ & & \left. \left. \qquad
- \lambda^a \lambda^{d_{\sigma_1}} \lambda^{d_{\sigma_2}} \lambda^b 
\lambda^{b'} \lambda^{d_{\sigma_3}}
+ \lambda^b \lambda^{b'} \lambda^{d_{\sigma_2}} 
\lambda^{d_{\sigma_1}} \lambda^a \lambda^{d_{\sigma_3}}
- \lambda^{b'} \lambda^b \lambda^{d_{\sigma_2}} 
\lambda^{d_{\sigma_1}} \lambda^a \lambda^{d_{\sigma_3}}
\right) \right]\, ,\nonumber
\end{eqnarray}
with the sum over the permutations of the three gluons 1, 2 and 3, and the
LO impact factor, $C^{g; g}(p_b^{\nu_b};p_{b'}^{\nu_{b'}})$, as in 
\eqn{centrc}. From the PT subamplitudes (\ref{ptgluon}) we obtain the
function of $(-+++)$ helicities
\begin{eqnarray}
A^{g;3g}(p_a^{\nu_a}; k_1^{\nu_1}, k_2^{\nu_2}, k_3^{\nu_3}) &=&
C^{g;3g}(p_a^{\nu_a}; k_1^{\nu_1}, k_2^{\nu_2}, k_3^{\nu_3})
A^{\bar\nu}(k_1, k_2, k_3)\, ,\label{pt3g}
\end{eqnarray}
where $\bar\nu={\rm sign}(\nu_a+\nu_1+\nu_2+\nu_3)$ and
\begin{eqnarray}
A^+(k_1,k_2,k_3) &=& - 2\, {q_\perp\over k_{1\perp}}
\sqrt{x_1\over x_3}\, {1\over \langle 1 2\rangle
\langle 2 3\rangle}\, ,\label{apiu}
\end{eqnarray}
and
\begin{equation}
x_i= \frac{k_i^+}{k_1^+ + k_2^+ + k_3^+}   
\qquad i=1,2,3 \quad (x_1+x_2+x_3=1)\, .\label{trefrac}
\end{equation}
The functions $C^{g;3g}$ are a straightforward generalization of 
the functions $C^{g;gg}$  defined in \eqn{trecoeff} and read,
\begin{eqnarray}
C^{g;3g}(p_a^{\nu_a}; k_1^{\nu_1}, k_2^{\nu_2},k_3^{\nu_3}) &=& 
\left\{
\begin{array}{l}
1{~}\qquad \nu_a=- \\
x_i^2 \qquad \nu_i=- \qquad i=1,2,3\\
\end{array}\right.
\quad {\rm with}\; \bar\nu=+, \label{fourpt}
\end{eqnarray}
%C^{g;3g}(p_a^-;k_1^+,k_2^+,k_3^+) &=& 1 \nonumber\\
%C^{g;3g}(p_a^+;k_1^-,k_2^+,k_3^+) &=& x_1^2 \label{fourpt}\\
%C^{g;3g}(p_a^+;k_1^+,k_2^-,k_3^+) &=& x_2^2 \nonumber\\
%C^{g;3g}(p_a^+;k_1^+,k_2^+,k_3^-) &=& x_3^2 \nonumber\\
%
From the non-PT subamplitudes~\cite{mpReview} we obtain the
function of $(--++)$ helicities
\begin{eqnarray}
&&A^{g;3g}(p_a^-,k_1^+,k_2^+,k_3^-) = \nonumber 
\frac{2}{s_{12} |k_{1\perp}|^2 } \times \\
&&  \left[- 
\frac{s_{a12} s_{123} } {s_{23}}
\left(\beta(k_1,k_2,k_3) x_1 + 
\frac{\gamma(k_1,k_2,k_3) (x_1 x_2 + \beta(k_1,k_2,k_3) (x_2 + x_3))}
{x_2\, x_3}\right)\right.\nonumber\\
&&\left.-\frac{\beta(k_1,k_2,k_3)^2 s_{a12} }{x_2}  
+\frac{\gamma(k_1,k_2,k_3)^2 
 s_{123} |k_{1\perp}|^2}{s_{23} x_1 x_2 x_3}
+\frac{s_{12} x_1^2 x_2 |q_{\perp}|^2 }{s_{23} (x_2 + x_3)}\right]
\nonumber\\[10pt]
&&A^{g;3g}(p_a^-,k_1^+,k_2^-,k_3^+) =  
\nonumber\frac{2}{s_{12} |k_{1\perp}|^2} \times \\
&&
\left[
- \frac{s_{a12} s_{123} } {s_{23}}
\left(-\alpha(k_1,k_3,k_2) x_1 + 
\frac{\gamma(k_1,k_3,k_2) (x_1 x_2 - \alpha(k_1,k_3,k_2) (x_2 + x_3))}
{x_2\, x_3}\right)\right.\label{fournonpt}\\&&\left.
-\frac{\alpha(k_1,k_3,k_2)^2 s_{a12} }{x_2} 
+ \frac{\gamma(k_1,k_3,k_2)^2 s_{123} 
|k_{1\perp}|^2}{s_{23} x_1 x_2 x_3}
+ \frac{s_{12} x_1^2 x_2 |q_{\perp}|^2}
{s_{23} (x_2 + x_3)}\right] \nonumber \\[10pt]
&&A^{g;3g}(p_a^-,k_1^-,k_2^+,k_3^+) =\nonumber \\
&&\frac{2}{s_{12} }\left[
\frac{\gamma(k_2,k_3,k_1)^2 s_{123} }{ s_{23} x_1 x_2 x_3} - 
\frac{\alpha(k_2,k_3,k_1)^2 s_{a12} }{ x_2 |k_{3\perp}|^2  } + 
\frac{\alpha(k_2,k_3,k_1) \gamma(k_2,k_3,k_1) s_{a12} s_{123} (x_2 + x_3) }
{ s_{23} x_2 x_3 |k_{3\perp}|^2}\right]\, ,\nonumber\\[10pt]\nonumber
\end{eqnarray}
with $s_{ijk}=(p_i+p_j+p_k)^2$ the three-particle invariant, and
\begin{eqnarray}
&&\alpha(k_1,k_2,k_3) \equiv \frac{\sqrt{x_1}\, 
k_{3\perp} (\sqrt{x_1} \,q_{\perp}^* + \sqrt{x_2}\, [1\,2])}{s_{a13} } 
\nonumber\\
&&\beta(k_1,k_2,k_3) \equiv \frac{(k_{1\perp}+k_{2\perp}) [1\,2] 
\sqrt{x_1 \,x_2}}{s_{a12}} \label{alfabetagamma}\\
&&\gamma(k_1,k_2,k_3) \equiv \frac{\sqrt{x_1\,x_2\, x_3}\, [1\,2] 
(\sqrt{x_1}\, \langle 1 \,3\rangle+\sqrt{x_2}\, \langle 2 \,3 \rangle )}
{s_{123} }\, .\nonumber
\end{eqnarray}

Using the U(1) decoupling equations~\cite{bg,bk}, the function $B$ in 
\eqn{threeggg} can be written as
\begin{eqnarray}
& & B^{g;3g}(p_a^{\nu_a}; k_1^{\nu_1}, k_2^{\nu_2}, k_3^{\nu_3}) = 
\label{baaa}\\
& & - \left[A^{g;3g}(p_a^{\nu_a}; k_1^{\nu_1}, k_2^{\nu_2}, k_3^{\nu_3}) +
A^{g;3g}(p_a^{\nu_a}; k_1^{\nu_1}, k_3^{\nu_3}, k_2^{\nu_2}) +
A^{g;3g}(p_a^{\nu_a}; k_3^{\nu_3}, k_1^{\nu_1}, k_2^{\nu_2})\right] \, 
.\nonumber
\end{eqnarray}
In the triple collinear limit, $k_1 || k_2 || k_3$, Sect.~\ref{sec:nnlocoll},
the function $A$ has a double collinear divergence,
while the function $B$, whose gluon 3 is not
color adjacent to gluons 1 and 2, has only a single collinear divergence.

Using \eqn{baaa}, we can rewrite \eqn{threeggg} as
\begin{eqnarray}
& & \A{g\,g \to 4g}(p_a^{\nu_a}, k_1^{\nu_1}, k_2^{\nu_2}, 
k_3^{\nu_3} \sep p_{b'}^{\nu_{b'}}, p_b^{\nu_b}) = \label{nnloggg}\nonumber\\ 
& & 2\, s\, \left\{ (ig)^3\, \sum_{\sigma\in S_3}
f^{a d_{\sigma_1} c } f^{c d_{\sigma_2} c'} f^{c' d_{\sigma_3} c''}  
A^{g;3g}(p_a^{\nu_a}; k_{\sigma_1}^{\nu_{\sigma_1}}, 
k_{\sigma_2}^{\nu_{\sigma_2}}, k_{\sigma_3}^{\nu_{\sigma_3}}) \right\}
%\nonumber \\& & \times
{1\over t}\, \left[ig\, f^{b b' c''} 
C^{g ;g}(p_b^{\nu_b};p_{b'}^{\nu_{b'}})
\right]\, ,\nonumber\\[-5pt]
\end{eqnarray}
where the NNLO impact factor $g^*\, g \rightarrow g\, g\, g$
is enclosed in curly brackets, and includes
14 helicity configurations.

\subsection{The NNLO impact factor $g\, g^* \rightarrow g \,\bar q \, q$}
\label{sec:nnloqqgg}

We consider the amplitude for the scattering $g\, g\, \to g\, \bar q\, q\,
\, g$ (\fig{fig:NNLO}b), in the kinematics (\ref{nnloreg}).
Using Eqs.~(\ref{TwoQuarkGluonDecomp})-(\ref{pt2quark2}) and the 
subamplitudes of non-PT type,
with two gluons of $+$ helicity and two gluons of $-$ 
helicity~\cite{mpReview}, we obtain
\begin{eqnarray}
&&\A{g g \to g\,\bar{q} q\, g}
(p_a^{\nu_a}, k_1^{\nu_1}, k_2^{\nu_2}, 
k_3^{-\nu_2} \sep  p_{b'}^{\nu_{b'}}, p_b^{\nu_b}) \nonumber\\
&& = 2 s\, I^{g;g\bar{q}q}(p_a^{\nu_a};k_1^{\nu_1},k_2^{\nu_2},k_3^{-\nu_2})
{1\over t}\, \left[ig\, f^{b b' c'} 
C^{g;g}(p_b^{\nu_b};p_{b'}^{\nu_{b'}})
\right]\, ,\label{nnlogqq}
\end{eqnarray}
with $k_3$ the quark, 
and with NNLO impact factor $g\, g^*  \rightarrow g \, \bar{q}\, q $,
\begin{eqnarray}
&& I^{g;g\bar{q}q}(p_a^{\nu_a};k_1^{\nu_1},k_2^{\nu_2},k_3^{-\nu_2}) =
\label{ggqqif}\\ && g^3 \left[
\left(\lambda^{c'} \lambda^{a} \lambda^{d_1}\right)_{d_3 \bar{d}_2}
A_1^{g;g\bar{q}q}(p_a^{\nu_a};k_1^{\nu_1},k_2^{\nu_2},k_3^{-\nu_2}) +
\left(\lambda^{a} \lambda^{c'} \lambda^{d_1}\right)_{d_3 \bar{d}_2}   
A_2^{g;g\bar{q}q}(p_a^{\nu_a};k_1^{\nu_1},k_2^{\nu_2},k_3^{-\nu_2}) 
\right.\nonumber\\
&&+\left(\lambda^{d_1} \lambda^{c'} \lambda^{a}\right)_{d_3 \bar{d}_2}   
A_3^{g;g\bar{q} q}(p_a^{\nu_a};k_1^{\nu_1},k_2^{\nu_2},k_3^{-\nu_2}) +
\left(\lambda^{d_1} \lambda^{a} \lambda^{c'}\right)_{d_3 \bar{d}_2}   
A_4^{g;g\bar{q} q}(p_a^{\nu_a};k_1^{\nu_1},k_2^{\nu_2},k_3^{-\nu_2}) 
\nonumber\\
&&+\left(\lambda^{a} \lambda^{d_1} \lambda^{c'}\right)_{d_3 \bar{d}_2}   
B_1^{g;g\bar{q} q}(p_a^{\nu_a};k_1^{\nu_1},k_2^{\nu_2},k_3^{-\nu_2}) +
\left(\lambda^{c'} \lambda^{d_1} \lambda^{a}\right)_{d_3 \bar{d}_2}   
B_2^{g;g\bar{q} q}(p_a^{\nu_a};k_1^{\nu_1},k_2^{\nu_2},k_3^{-\nu_2})
\left. \frac{}{} \right]\nonumber 
\end{eqnarray}
The NNLO impact factor allows for
eight helicity configurations. From the PT subamplitudes 
(\ref{pt2quark})-(\ref{pt2quark2}) we obtain
\begin{eqnarray}
&&A_1^{g; g\bar{q} q}(p_a^{+};k_1^{+},k_2^{-},k_3^{+})  = 
- 2 \, {q_\perp \over k_{1\perp}} \sqrt{x_1 x_2^3 }
{1\over \langle 1\,2\rangle\; \langle 2\, 3\rangle}\,\nonumber\\
&&A_2^{g; g\bar{q} q}(p_a^{+};k_1^{+},k_2^{-},k_3^{+})  = 
  2 \, {q_\perp \over k_{3\perp}} \sqrt{\frac{x_2^3  x_3^2}{x_1}}
{1\over \langle 1\,2\rangle\; \langle 2\, 3\rangle}\,\nonumber\\
&&A_3^{g; g\bar{q} q}(p_a^{+};k_1^{+},k_2^{-},k_3^{+})  = 
- 2 \, {q_\perp \over k_{2\perp}} x_2^2 \sqrt{\frac{x_3}{x_1}}
{1\over \langle 1\,3\rangle\; \langle 3\, 2\rangle}\,\nonumber\\
&&A_4^{g; g\bar{q} q}(p_a^{+};k_1^{+},k_2^{-},k_3^{+})  = 
  2 \, {q_\perp \over k_{1\perp}} \sqrt{x_1 x_2^2  x_3}
{1\over \langle 1\,3\rangle\; \langle 3\, 2\rangle}\,\label{qqgcoeff}\\
&&B_1^{g; g\bar{q} q}(p_a^{+};k_1^{+},k_2^{-},k_3^{+})  =
- 2 \, {q_\perp\over k_{3\perp}} x_2 x_3
{1\over k_{1\perp} \langle 2\, 3\rangle}\,\nonumber\\
&&B_2^{g; g\bar{q} q}(p_a^{+};k_1^{+},k_2^{-},k_3^{+})  = 
- 2 \, {q_\perp\over k_{2\perp}} x_2^2
{1\over k_{1\perp} \langle 2\, 3\rangle}\,\nonumber
\end{eqnarray}
with momentum fractions as in \eqn{trefrac}.
The impact factors from the non-PT subamplitudes~\cite{mpReview}
are, 
\begin{eqnarray}
\lefteqn{A_1^{g;g\bar{q} q}(p_a^-;k_1^+,k_2^-,k_3^+ ) = } \nonumber \\
&& 2\left\{ \frac{ \gamma  (k_1,k_3,k_2)^2 s_{1 2 3}   }{
 \spb1.3  \spa1.2 s_{2 3} x_1 x_2 x_3} 
 + \frac{ \sqrt{x_1} \gamma  (k_1,k_3,k_2) s_{1 2 3}
     \alpha  (k_1,k_3,k_2) s_{3 b b'}   }{
  s_{1 2} \spa2.3 k_{1 \perp}^*   x_3 \sqrt{x_3} \spb1.3 k_{2 \perp}} \right. \nonumber \\  
&& -  \frac{  (k_{3 \perp} + q_\perp) 
     \alpha   (k_1,k_3,k_2)^2 s_{3 b b'}  }{
    \sqrt{x_2 x_3}
     s_{1 2} |k_{1 \perp}|^2  k_{2 \perp} } 
 -  \frac{ |q_\perp|^2   x_1^2 x_2^{3/2}   }{
   \sqrt{x_3} s_{2 3} |k_{1 \perp}|^2   
    (x_2 +  x_3)} \nonumber \\
&&  - \frac{  x_1 s_{3 b b'}}{s_{1 2} s_{2 3} |k_{1 \perp}|^2  x_3} 
\left[ \frac{}{}s_{1 2 3}  \sqrt{ x_2 x_3} \alpha   (k_1,k_3,k_2) 
     \right.   
     \nonumber \\      
&& \left.   \left. + 
 \frac{\gamma   (k_1,k_3,k_2) s_{1 2 3} }{\sqrt{x_2 x_3}}
        \left[ -  x_2 + 
          {\alpha   (k_1,k_3,k_2) 
           (x_2 + x_3)\over x_1} \, \right]   \right] \right\} 
\label{qqgnpt1}
\end{eqnarray}
\begin{eqnarray}
\lefteqn{A_2^{g;g\bar{q} q}(p_a^-;k_1^+,k_2^-,k_3^+ ) = } \nonumber \\
&& 2 \left\{ -\frac{ \gamma  (k_1,k_3,k_2)^2 s_{1 2 3}   }{
 \spb1.3  \spa1.2 s_{2 3} x_1 x_2 x_3}  
 - \sqrt{x_2 x_3\over x_1} \frac{ (- \spb2.3 \sqrt{x_1} + 
      \spb1.3  \sqrt{x_2} ) }{ s_{1 2} \spb2.3 } \right. \nonumber \\  
&& \left. +\frac{  k_{2 \perp} (k_{3 \perp}^*  - \spb2.3 \sqrt{x_2 x_3})^2  }{  
   s_{2 3} k_{ 3\perp}^*   s_{1 b b'} \sqrt{x_2 x_3}} 
 +\frac{   \sqrt{x_2 x_3}   }{   s_{2 3}} 
 +\frac{   \sqrt{x_{2} x_3 } \gamma  (k_1,k_3,k_2) s_{1 2 3} }{ 
 x_1  \spa1.2 \spa2.3 \spb1.3 k_{3 \perp}^* } \right\}
\label{qqgnpt2}
\end{eqnarray}
\begin{eqnarray}
\lefteqn{A_3^{g;g\bar{q} q}(p_a^-;k_1^+,k_2^-,k_3^+ ) = } \nonumber \\
&& 2 \left\{ \frac{  k_{ 3\perp} \sqrt{x_2} \alpha  (k_3,k_1,k_2)^2 s_{1 b b'}  }{
     k_{2 \perp}   |k_{2 \perp}|^2  s_{2 3}  \sqrt{x_3}} 
 -\frac{   \gamma   (k_1,k_3,k_2)^2 s_{1 2 3}  \spb1.2 }{  
  x_1 x_2 x_3  \spb1.3 s_{1 3} s_{2 3} } \right. \nonumber \\ 
&& +\frac{\spa1.2  x_1} {   \spa1.3   s_{2 3}}  
  + \frac{ \alpha   (k_3,k_1,k_2) s_{1 b b'}   \sqrt{x_2} 
 (\spa1.3 - \spa1.2 \sqrt{ x_2 x_3}) }{ \spa1.3 |k_{2 \perp}|^2  s_{2 3}\sqrt{x_3}}  \nonumber \\  
&& \left. +\frac{   \spb1.3 x_2 x_3 \sqrt{x_1} +  
        (k_{2 \perp}^*  x_1 - q_\perp^*  x_2 x_3)\sqrt{x_3}  }{   
  \sqrt{x_1} \spa1.3 k_{2 \perp}^*  \spb2.3} \right\}
\label{qqgnpt3}
\end{eqnarray}
\begin{eqnarray}
\lefteqn{A_4^{g;g\bar{q} q}(p_a^-;k_1^+,k_2^-,k_3^+ ) = }  \nonumber \\
&& 2 \left\{ \frac{ (k_{ 2\perp} + q_{\perp})^2 \sqrt{x_1}
     \alpha   (k_1,k_2,k_3)   }{  
   \spa1.3 |k_{ 1\perp}|^2  k_{ 3\perp} \sqrt{x_2}} 
 -\frac{  (k_{2 \perp} + q_\perp) q_\perp^*  x_1^2 
     \gamma   (k_1,k_3,k_2) s_{1 2 3}   }{   
   \spa2.3 \spb1.3 |k_{ 1\perp}|^2 s_{1 3} \sqrt{x_1  x_3} x_2 } \right. \nonumber \\
&&+\frac{   \spb1.2 \gamma   (k_1,k_3,k_2)^2 s_{1 2 3}   }{     
   \spb1.3 s_{1 3} s_{2 3} x_1 x_2 x_3}  
 +\frac{  x_1^2    |q_{\perp}|^2     \sqrt{ x_2 x_3}}
        { |k_{ 1\perp}|^2   s_{2 3}  (x_2 + x_3)} \nonumber \\
&& -\frac{  \spb1.2 x_1  s_{1 2 3} \left[-(k_{ 2\perp} + q_\perp)  
          x_2 \sqrt{x_1 x_3} \spb1.3  + s_{2 b b'} 
        \gamma   (k_1,k_3,k_2) \right]    }{ x_2   
  |k_{1 \perp}|^2   s_{1 3} s_{2 3} \spb1.3 }  \nonumber \\
&&\left.  + \frac{x_1 (k_{ 2\perp} + q_\perp) \spb1.2 
        \gamma   (k_1,k_3,k_2) s_{1 2 3} (x_2 + x_3)     }{   
   s_{1 3} |k_{ 1\perp}|^2   s_{2 3} \sqrt{x_1  x_3} x_2 }  
   \right\} \label{qqgnpt4}
\end{eqnarray}
\begin{eqnarray}
\lefteqn{B_1^{g;g\bar{q} q}(p_a^-;k_1^+,k_2^-,k_3^+ ) = }  \nonumber \\
&& 2 \left\{ \frac{ -  \alpha   (k_3,k_1,k_2)^2 s_{1 b b'} 
       }{   k_{3 \perp}^*  s_{2 3} k_{2 \perp} \sqrt{x_2 x_3}} 
 +\frac{   (k_{2 \perp} + q_\perp)^2 \spb1.3 \sqrt{x_1 }
      x_3  \alpha   (k_1,k_2,k_3) }{  
   |k_{1 \perp}|^2 |k_{ 3\perp}|^2  k_{ 3\perp} \sqrt{x_2}} \right. \nonumber \\  
&& -\frac{   \spb1.3 \sqrt{x_1 x_3} 
     \left[ \spa2.3 \sqrt{x_2} |k_{3 \perp}|^2  - 
            (k_{2 \perp} + q_\perp) s_{23} (x_1 + x_2) \sqrt{x_3} \,  
             + k_{1 \perp} k_{2 \perp} \spb2.3 x_3\sqrt{x_2} \right]   }{   
    \sqrt{x_2} |k_{1 \perp}|^2 |k_{ 3\perp}|^2   s_{2 3} }  \nonumber \\
&&\left. - \frac{ \sqrt{x_1}}{ |k_{1 \perp}|^2  s_{2 3} \sqrt{x_2} }
 \left[ - \spb1.3 k_{1 \perp} x_2  + q_{\perp}^*  \sqrt{x_1x_3} 
        \left( {q_\perp x_2 \over x_2 + x_3} + k_{ 2\perp} \right) \right]  
  \right\} \label{qqgnpt5}
\end{eqnarray}
\begin{eqnarray}
\lefteqn{B_2^{g;g\bar{q} q}(p_a^-;k_1^+,k_2^-,k_3^+ ) = } \nonumber \\
&& 2 \left\{ \frac{ - (k_{1 \perp} + k_{2 \perp}) (k_{1 \perp}^*  + \spb1.2 \sqrt{x_1 x_2})^2 
     \sqrt{x_2}   }{   |k_{1 \perp}|^2  k_{ 2\perp}^*  s_{3 b b'} \sqrt{x_3}}   
 +\frac{  k_{3 \perp} \sqrt{x_2} ( k_{3 \perp}^{*  }- 
       \spb2.3 \sqrt{x_2 x_3}\, )^2   }{  
   k_{ 2\perp}^*  s_{2 3} s_{1 b b'} x_3^{3 / 2}} \right.  \nonumber \\  
&& - \frac{  x_2}{   |k_{1 \perp}|^2  |k_{ 2\perp}|^2  s_{2 3} x_3}
\left[ \frac{}{} - k_{3 \perp} \spb2.3 s_{3 b b'} x_1 x_2 
 \right. \nonumber \\  
&& \left.    +(k_{ 1\perp}^*  + \spb1.2 \sqrt{x_1 x_2}) 
( \spa2.3 q_\perp        
        (k_{3 \perp}^*  - \spb2.3 \sqrt{x_2 x_3}) - k_{3 \perp} \spb2.3 
          k_{2 \perp}  (x_2 + x_3)) \frac{}{} \right]     \nonumber \\  
&& - \frac{  \sqrt{x_2} }{  |k_{1 \perp}|^2  |k_{ 2\perp}|^2 s_{2 3} x_3^{3 / 2} 
     } \left[
|k_{3 \perp} |^2  s_{3 b b'} x_1 x_2
 + |q_\perp|^2 |k_{ 2\perp}|^2 { x_1 x_3^2 \over x_2 + x_3}
         \right. \nonumber \\
 &&   \left. \frac{}{} \left.    + 
        (k_{1 \perp}^*  + \spb1.2 \sqrt{x_1 x_2}) (k_{ 3\perp} s_{1 b b'} x_2 x_3 
        +k_{ 2\perp} |k_{3 \perp}|^2  ( x_2 + x_3) )  
      \frac{}{}  \right] \frac{}{} \right\} \, .\label{qqgnpt6}
\end{eqnarray}
The functions $A$ and $B$ for the remaining helicity configurations
are derived using the relations,
\begin{eqnarray}
&&A_i^{g; g\bar{q} q}(p_a^{\nu_a};k_1^{+},k_2^{+},k_3^{-})  = 
  - A_{5-i}^{g; g\bar{q} q}(p_a^{\nu_a};k_1^{+},k_3^{-},k_2^{+})    
\qquad i=1,2,3,4 \label{qqgab}\\
&&B_i^{g; g\bar{q} q}(p_a^{\nu_a};k_1^{+},k_2^{+},k_3^{-})  = -
  B_{3-i}^{g; g\bar{q} q}(p_a^{\nu_a};k_1^{+},k_3^{-},k_2^{+})    
\qquad i=1,2 \nonumber
\end{eqnarray}

\subsection{The NNLO impact factor $q\, g^* \rightarrow q\, g\, g$}
\label{sec:nnloqgg}

We consider the amplitude $q\,g\to q\, g\, g\, g$ for the production of a 
quark and two gluons in the forward-rapidity region of quark $a$ 
(\fig{fig:NNLO}c) in the kinematics (\ref{nnloreg}).
Using Eqs.~(\ref{TwoQuarkGluonDecomp})-(\ref{pt2quark2}) and the 
subamplitudes of non-PT type,
with two gluons of $+$ helicity and two gluons of $-$ 
helicity~\cite{mpReview}, we obtain
\begin{eqnarray}
& & \A{q\,g\to q\,3g}(p_a^{-\nu_1}, k_1^{\nu_1}, k_2^{\nu_2},
k_3^{\nu_3} \sep p_{b'}^{\nu_{b'}}, p_b^{\nu_b}) \nonumber\\ && = 2 s\,
I^{\bar q;qgg}(p_a^{-\nu_1};k_1^{\nu_1},k_2^{\nu_2},k_3^{\nu_3})
{1\over t}\, \left[ig\, f^{b b' c'} 
C^{g; g}(p_b^{\nu_b};p_{b'}^{\nu_{b'}})
\right]\, ,\label{nnloqgg}
\end{eqnarray}
with $k_1$ the final-state quark, and
the NNLO impact factor $q\, g^*  \rightarrow q\,g\,g $,
\begin{eqnarray}
&& I^{\bar q;qgg}(p_a^{-\nu_1};k_1^{\nu_1},k_2^{\nu_2},k_3^{\nu_3})
\label{qqggif}\\ && = g^3 \sum_{\sigma\in S_2} \left[
\left(\lambda^{d_{\sigma_2}} \lambda^{d_{\sigma_3}} 
\lambda^{c'}\right)_{d_1 \bar{a}}   
A^{\bar q;qgg}(p_a^{-\nu_1};k_1^{\nu_1},k_{\sigma_2}^{\nu_{\sigma_2}},
k_{\sigma_3}^{\nu_{\sigma_3}}) \right. \nonumber\\ &&+
\left(\lambda^{c'} \lambda^{d_{\sigma_2}} 
\lambda^{d_{\sigma_3}}\right)_{d_1 \bar{a}}   
B_1^{\bar q;qgg}(p_a^{-\nu_1};k_1^{\nu_1},k_{\sigma_2}^{\nu_{\sigma_2}},
k_{\sigma_3}^{\nu_{\sigma_3}}) \nonumber\\ &&+
\left(\lambda^{d_{\sigma_2}} \lambda^{c'} 
\lambda^{d_{\sigma_3}}\right)_{d_1 \bar{a}}   
B_2^{\bar q ;qgg}(p_a^{-\nu_1};k_1^{\nu_1},k_{\sigma_2}^{\nu_{\sigma_2}},
k_{\sigma_3}^{\nu_{\sigma_3}})
\left. \frac{}{} \right]\, .\nonumber
\end{eqnarray}
The NNLO impact factor allows for
eight helicity configurations. From the PT subamplitudes 
(\ref{pt2quark})-(\ref{pt2quark2}) we obtain
\begin{eqnarray}
&&A^{\bar q ; qgg}(p_a^{-};k_1^{+},k_2^{+},k_3^{+})  = 
2i \, {q_\perp\over k_{1\perp}} \frac{x_1}{\sqrt{x_3}}
{1\over \langle 1\,2\rangle\; \langle 2\, 3\rangle}\,\nonumber\\
&&B_1^{\bar q ; qgg}(p_a^{-};k_1^{+},k_2^{+},k_3^{+})  =
2i \, {q_\perp\over k_{3\perp}} 
 \sqrt{\frac{x_1 x_3}{x_2}}
{1\over k_{1\perp} \langle 2\, 3\rangle}\,\label{qggcoeff}\\
&&B_2^{\bar q ; qgg}(p_a^{-};k_1^{+},k_2^{+},k_3^{+})  =
2i \, {q_\perp\over k_{1\perp}} 
 \frac{x_1}{\sqrt{x_2}}
{1\over k_{3\perp} \langle 1\, 2\rangle}\,\nonumber
\end{eqnarray}
and
\begin{eqnarray}
&&A^{\bar q; qgg}(p_a^{+},k_1^{-},k_2^{+},k_3^{+})  = 
 - x_1  A^{\bar q ; qgg}(p_a^{-};k_1^{+},k_2^{+},k_3^{+}) \nonumber\\
&&B_i^{\bar q ; qgg}(p_a^{+};k_1^{-},k_2^{+},k_3^{+})  = 
 - x_1  B_i^{\bar q ; qgg}(p_a^{-};k_1^{+},k_2^{+},k_3^{+}) \qquad i=1,2 
\label{abrel}
\end{eqnarray}
The impact factors from the non-PT subamplitudes~\cite{mpReview} are,
\begin{eqnarray}
\lefteqn{A^{\bar{q};q g g}(p_a^-;k_1^+,k_2^-,k_3^+ ) = }  \nonumber \\
&& 2i \left\{ \frac{ k_{ 2\perp} (k_{1 \perp}^*  + \spb1.2 \sqrt{x_1 x_2})^2   }{  
    |k_{1 \perp}|^2  \spb1.2 s_{3 b b'} \sqrt{x_2}}   
+\frac{   \gamma  (k_1,k_2,k_3)^* \gamma  (k_1,k_3,k_2)^2 s_{1 2 3}^2
       }{   
 (x_1 x_2 x_3)^{3/2}  s_{1 2} s_{2 3} \spa2.1 \spb1.3} \right. \nonumber \\ 
&& +\frac{   \spb1.3  \spa1.2 x_1 (\sqrt{x_3} \spa2.3  - 
       k_{1 \perp} \sqrt{x_2})+ k_{1 \perp} \sqrt{x_3} 
         \gamma  (k_1,k_3,k_2) s_{1 2 3}      }{  
   s_{1 2} s_{2 3} k_{ 1\perp}  \sqrt{x_1 x_3}} \nonumber \\  
&& +\frac{   \sqrt{x_1}}{s_{2 3}  |k_{ 1\perp}|^2   x_3 }
 \left[ k_{1 \perp}^*  x_3  
        (q_\perp x_2 + k_{2 \perp} (x_2 + x_3)) 
        - |q_\perp|^2  {x_1x_2 x_3 \over x_2+x_3} \right. \nonumber \\ 
&& \left. \frac{}{} \left.    +  x_1 (-q_\perp k_{3 \perp}^*  x_2 + 
          k_{2 \perp} \sqrt{x_3} (k_{ 1\perp}^*  \sqrt{ x_3} - \spb2.3 
\sqrt{x_2}))
       \frac{}{}   \right] \frac{}{} \right\} \label{qggnpt1}
\end{eqnarray}
\begin{eqnarray}
\lefteqn{A^{\bar{q} ;qgg}(p_a^-;k_1^+,k_2^+,k_3^- ) = } \nonumber \\
&&2 i \left\{ 
      \frac{ \beta  (k_1,k_2,k_3)^2 s_{3 b b'}  k_{2 \perp}^*     }{   
   \spb1.2 s_{1 2} |k_{1 \perp}|^2  x_2^{3/2}}   
   +\frac{ \beta  (k_1,k_2,k_3) s_{3 b b'}  k_{2 \perp}^* \sqrt{x_1} (x_3+x_2)    }{   
    s_{1 2} |k_{1 \perp}|^2  \sqrt{x_3 x_2} \spb3.2 } \right. \nonumber \\  
&& -\frac{  k_{2 \perp}^*  x_1^{3  /  2} }{ s_{1 2}  |k_{1 \perp}|^2   \spb3.2}
      \left[ { q_\perp \gamma  (k_1,k_2,k_3)^* s_{1 3 2} \over 
\sqrt{x_1 x_3} \spa1.2 }
      + s_{1 3 2} \sqrt{x_3 x_2}  
       \right. \nonumber \\
&&\left.  -  { \sqrt{x_2} \beta  (k_1,k_2,k_3) s_{3 b b'} q_{\perp}^*  
 \over x_1 \sqrt{x_3} k_{2\perp}^* }  \right]
 +\frac{   \gamma  (k_1,k_3,k_2)^* \gamma  (k_1,k_2,k_3)^2 s_{1 3 2}^2
       }{   
 (x_1 x_3 x_2)^{3/2}  s_{1 2} s_{3 2} \spa3.1 \spb1.2} 
 \nonumber \\  
&&+\frac{   x_1 \left[ -s_{1 2}  |q_\perp|^2 x_2^2 \sqrt{x_1 x_3}
+       \spa1.3 \beta  (k_1,k_2,k_3) s_{3 b b'} k_{ 2\perp}^*  
(x_3 + x_2)^2    
        \right]  }{   
   s_{1 2} s_{3 2}  |k_{ 1\perp}|^2  x_2 \sqrt{x_3} 
     (x_3 + x_2)} \nonumber \\
 &&   \left.  +\frac{  \sqrt{x_1} s_{3 b b'}
     \left[ \spb1.2 \spa1.3 k_{2 \perp}^*   \sqrt{x_1^3x_2} + 
       \beta  (k_1,k_2,k_3)  (k_{2 \perp}^*  s_{1 3 2} \sqrt{x_1x_3} 
           - \spa1.3 k_{1 \perp}^* q_\perp^* x_2)\right]   }{  
  \spb1.2 s_{1 2} s_{3 2}   |k_{ 1\perp}|^2  \sqrt{x_2x_3}} \right\} 
\nonumber \\
  &&  \label{qggnpt2}
\end{eqnarray}
\begin{eqnarray}
\lefteqn{B_1^{\bar{q} ;qgg}(p_a^-;k_1^+,k_2^-,k_3^+ ) = }  \nonumber \\
&& 2 i  \left\{ \frac{  (k_{ 2\perp} + q_{\perp})^2 \spb1.3 x_1 \sqrt{x_3}  }{
   |k_{1 \perp}|^2 k_{3 \perp}   s_{2 b b'}} 
 +\frac{   k_{2 \perp} \spb1.3 x_1 x_3  }{
   \spa2.3  k_{ 1\perp}^*  |k_{ 3\perp}|^2  \sqrt{x_2}} 
 -\frac{    
     \alpha  (k_2,k_1,k_3)^* \alpha  (k_3,k_1,k_2)^2  s_{1 b b'}^2  }{   
   |k_{ 3\perp}|^2  k_{ 3\perp}^* k_{ 2\perp} s_{2 3} \sqrt{x_1} x_2} \right. \nonumber \\  
&&+\frac{   \spb1.3 \sqrt{x_3}  }{
 |k_{1 \perp}|^2 |k_{ 3\perp}|^2   s_{2 3} x_2 }
      \left[ -|k_{3 \perp}|^2 (k_{ 2\perp} + q_{\perp})   x_1 x_2 
 +        |k_{1 \perp}|^2 k_{2 \perp}  (x_1 + x_2) x_3 \right] \nonumber \\ 
&&     +\frac{ x_3}{
 |k_{1 \perp}|^2 |k_{ 3\perp}|^2   s_{2 3} x_2 \sqrt{x_1}}
 \left[ k_{ 2\perp}  k_{1 \perp}^*  |k_{ 3\perp}|^2  x_1 
            + k_{ 2\perp}  |k_{ 3\perp}|^2  \spb1.2 
           x_1^{3  /  2} \sqrt{x_2}  \right. \nonumber \\ 
 &&  \left. \frac{}{} \left.  - q_\perp^*  \left( {q_{\perp} |k_{ 3\perp}|^2 
x_1^2 x_2 \over x_2+x_3} + 
             |k_{ 1\perp}|^2 k_{ 2\perp}   (x_1 + x_2) x_3 \right)  \right]  
\frac{}{} \right\}  \label{qggnpt3}
 \end{eqnarray}

\begin{eqnarray}
\lefteqn{B_1^{\bar{q} ;qgg}(p_a^-;k_1^+,k_2^+,k_3^- ) = } \nonumber \\
&&  2i \left\{  -\frac{    
     \alpha  (k_3,k_1,k_2)^* \alpha  (k_2,k_1,k_3)^2  s_{1 b b'}^2  }{   
   k_{3\perp}  k_{ 2\perp}^* |k_{ 3\perp}|^2 s_{3 2} \sqrt{x_1 } x_2}  
-\frac{   \spa1.3  \sqrt{x_3} 
     \alpha  (k_1,k_2,k_3)^2  s_{2 b b'}  }{   
   |k_{1 \perp}|^2   |k_{ 3\perp}|^2 k_{ 3\perp} } \right. \nonumber \\
 && +\frac{  \sqrt{x_3} }
     {|k_{1 \perp}|^2 |k_{3 \perp}|^2  s_{23} }
        \left[     \frac{}{} \spb1.2 \, 
          \spa1.3 {\sqrt{x_1} x_3 \over \sqrt{x_2}}
             (s_{2 b b'} x_1 - s_{1 b b'} x_2)                          \right. \nonumber \\
&&       +  \sqrt{x_1} (-k_{ 1\perp} q_{\perp} x_3^{3 / 2} 
  + \spa1.3 k_{3 \perp} \sqrt{x_1} (x_3 + x_2)) \left( q_\perp^{* 2}-\spb1.2^2
      - {x_1 - x_2\over \sqrt{x_1 x_2}} q_{\perp}^* \spb1.2  \right)
          \nonumber \\
&&  \left.   \frac{}{} \left.  \frac{}{}  - q_{\perp}^*  x_1 \sqrt{x_3} \, \,
        \left( {|k_{ 3\perp}|^2 q_\perp   \sqrt{x_1} \over x_3+x_2} + 
        \spa1.3 (s_{1 b b'} + s_{2 b b'}) \sqrt{x_3} \right) \right]  \right\}
\label{qggnpt4}  
\end{eqnarray}
\begin{eqnarray}
\lefteqn{B_2^{\bar{q} ;qgg}(p_a^-;k_1^+,k_2^-,k_3^+ ) = } \nonumber \\
&&\frac{2i}{|k_{1\perp}|^2} \left\{ \frac{ (k_{2\perp} + q_{\perp}) x_1 
     (k_{1\perp}^*  + \spb1.2 \sqrt{x_1 x_2})  }{
    k_{3\perp}   \spb1.2 \sqrt{x_2}} 
 - \frac{ k_{2\perp} (k_{1\perp}^*  + \spb1.2 \sqrt{x_1 x_2})^2  }{
     \spb1.2 s_{3 b b'} \sqrt{x_2}}  
     \right.  \nonumber \\
&&\left. -\frac{   (k_{2\perp} + q_{\perp})^2 \spb1.3 x_1 \sqrt{x_3}  }{
   k_{3\perp}   s_{2 b b'}} \right\}  \label{qggnpt5}
\end{eqnarray}
\begin{eqnarray}
\lefteqn{B_2^{\bar{q} ;qgg}(p_a^-;k_1^+,k_2^+,k_3^- ) = } \nonumber \\
&&\frac{2 i}{|k_{ 1\perp}|^2 } \left\{ \frac{  \spa1.3  \sqrt{x_3}
 \alpha  (k_1,k_2,k_3)^2  s_{2 b b'} }{  
     k_{ 3\perp} |k_{ 3\perp}|^2 } 
  +\frac{  \beta  (k_1,k_2,k_3)^2  s_{3 b b'} k_{2 \perp}^*   }{  
  s_{1 2} \spb2.1    x_2^{3/2}} 
   \right.
   \nonumber \\ 
&&  -\frac{  x_1}{   \spa1.2  k_{3 \perp}^*}
 \left[ {\spa1.2  x_3 \alpha  (k_1,k_2,k_3) s_{2 b b'} \over
\sqrt{x_1} k_{3 \perp}} \right. \nonumber \\
&&    \left. \frac{}{} \left.  +
      \left( 1  + {q_\perp^*  x_1 - k_{ 3\perp}^*  x_2  \over 
\sqrt{x_1 x_2 }\spb1.2} \right)
       {\beta  (k_1,k_2,k_3)  s_{3 b b'}\over \sqrt{x_2}}
      \,  \right]  \frac{}{} \right\} \, .\label{qggnpt6}
\end{eqnarray}

\subsection{The NNLO impact factor $q\, g^*\; \to q\,\bar{Q}\, Q$}
\label{sec:nnloqqq}

We consider the amplitude $q\,g\to q\bar{Q}\, Q\,g$ for the production of 
three quarks in the forward-rapidity region of quark $a$ 
(\fig{fig:NNLO}d) in the kinematics (\ref{nnloreg}). Using
\eqn{FourQuarkGluonDecomp} and the
subamplitudes of non-PT type, with two gluons of opposite 
helicities~\cite{4quark}, we obtain
\begin{eqnarray}
&&\A{q g  \to q \bar{Q} Q  g }
(p_a^{-\nu_1}, k_1^{\nu_1}, k_2^{\nu_2},
k_3^{-\nu_2} \sep p_{b'}^{\nu_{b'}}, p_b^{\nu_b}) = \nonumber
\\ && 2 s\, I^{\bar q ; q \bar{Q} Q}(p_a^{-\nu_1};k_1^{\nu_1},k_2^{\nu_2},
k_3^{-\nu_2}) {1\over t}\, \left[ig\, f^{b b' c} 
C^{g; g}(p_b^{\nu_b};p_{b'}^{\nu_{b'}})
\right]\, ,\label{3forwqqq}
\end{eqnarray}
with NNLO impact factor $q\, g^*  \rightarrow q \, \bar{Q}\, Q $
\begin{eqnarray}
&& I^{\bar q ; q \bar{Q} Q}(p_a^{-\nu_1};k_1^{\nu_1},k_2^{\nu_2},
k_3^{-\nu_2}) = \label{qqqqif}\\ && g^3 \left[
\lambda^{c}_{d_3 \bar{a}} \delta_{d_1 \bar{d}_2}   
A_1^{\bar q ; q \bar{Q} Q}(p_a^{-\nu_1};k_1^{\nu_1},k_2^{\nu_2},
k_3^{-\nu_2}) 
-\frac{1}{N_c} \lambda^{c}_{d_1 \bar{a}} \delta_{d_3 \bar{d}_2}   
A_2^{\bar q ; q \bar{Q} Q}(p_a^{-\nu_1};k_1^{\nu_1},k_2^{\nu_2},
k_3^{-\nu_2}) \right. \nonumber\\  && \quad +
\lambda^{c}_{d_1 \bar{d}_2} \delta_{d_3 \bar{a}}  
B_1^{\bar q ; q \bar{Q} Q}(p_a^{-\nu_1};k_1^{\nu_1},k_2^{\nu_2},k_3^{-\nu_2})
-\frac{1}{N_c} \lambda^{c}_{d_3 \bar{d}_2} \delta_{d_1 \bar{a}}   
B_2^{\bar q ; q \bar{Q} Q}(p_a^{-\nu_1};k_1^{\nu_1},k_2^{\nu_2},
k_3^{-\nu_2}) \left.  \frac{}{} \right] \nonumber\\  &&
- \delta_{qQ} \;(1 \leftrightarrow 3 )\, .\nonumber
\end{eqnarray}
The term proportional to $\delta_{qQ}$ is due to the interference
of identical quarks (i.e. with the same flavour and helicity )
in the final state.  
The NNLO impact factor allows for
four helicity configurations. From the non-PT subamplitudes~\cite{4quark} 
we obtain,
\begin{eqnarray}
A_1^{\bar q ; q \bar{Q} Q}(p_a^{-};k_1^{+},k_2^{-},k_3^{+})  &=& i \left\{ 
\sqrt{\frac{x_1 x_3}{x_2}} \frac{\alpha(k_1,k_3,k_2)}{|k_{1\perp}|^2}+
\frac{\sqrt{ x_1^3 x_2 x_3} }{1-x_1}
\frac{|q_\perp|^2}{s_{23}|k_{1\perp}|^2}+
 \frac{\gamma(k_1,k_3,k_2)}{\sqrt{x_1 x_2 x_3} s_{23}}\right.\nonumber\\
&+&\left.
\frac{x_1}{\sqrt{x_2 x_3} s_{23} |k_{1\perp}|^2} 
\left[ \sqrt{{x_1}} \alpha(k_3,k_1,k_2) \,s_{a23}-
\frac{\gamma(k_1,k_3,k_2) s_{123} }{\sqrt{x_1}} \right]\right\}
\nonumber\\
A_2^{\bar q ; q \bar{Q} Q}(p_a^{-};k_1^{+},k_2^{-},k_3^{+})  &=& 
\frac{i}{\sqrt{x_2 x_3} s_{23}} 
\left(\frac{\gamma(k_1,k_3,k_2)}{\sqrt{x_1}}+ \sqrt{x_1}\alpha(k_3,k_1,k_2) \right)
\nonumber\\
B_1^{\bar q;  q \bar{Q} Q}(p_a^{-};k_1^{+},k_2^{-},k_3^{+})  &=& i \left\{
\sqrt{\frac{x_1 x_2}{x_3}} \frac{\beta(k_1,k_3,k_2)}{|k_{1\perp}|^2}-
\frac{\sqrt{ x_1^3 x_2 x_3} }{1-x_1}
\frac{|q_\perp|^2}{s_{23}|k_{1\perp}|^2}+
\sqrt{\frac{x_1}{x_2 x_3}} \frac{\alpha(k_3,k_1,k_2)}{s_{23}}\right.\nonumber\\
&-&\left.
\frac{x_1}{\sqrt{x_2 x_3} s_{23} |k_{1\perp}|^2} 
\left[ \sqrt{x_1} \alpha(k_3,k_1,k_2)\, s_{a23}-
\frac{\gamma(k_1,k_3,k_2) s_{123} }{\sqrt{x_1}} \right]\right\}
 \label{qqqcoeff}\\
B_2^{\bar q ; q \bar{Q} Q}(p_a^{-};k_1^{+},k_2^{-},k_3^{+})  &=& 
\frac{i \sqrt{x_1}}{|k_{1\perp}|^2} 
\left(\sqrt{\frac{x_2}{x_3} } \beta(k_1,k_3,k_2) + 
      \sqrt{\frac{x_3}{x_2} } \alpha(k_1,k_3,k_2)\right)
\nonumber
\end{eqnarray}
with $\alpha,\beta,\gamma$ defined in \eqn{alfabetagamma} and
\begin{eqnarray}
&&A_i^{\bar q ; q \bar{Q} Q}(p_a^{+};k_1^{-},k_2^{-},k_3^{+})  = 
  A_i^{\bar q ; q \bar{Q} Q}(p_a^{-};k_1^{+},k_3^{-},k_2^{+}) 
\qquad i=1,2 \nonumber\\
&&B_i^{\bar q ; q \bar{Q} Q}(p_a^{+};k_1^{-},k_2^{-},k_3^{+})  = 
  B_i^{\bar q ; q \bar{Q} Q}(p_a^{-};k_1^{+},k_3^{-},k_2^{+}) 
\qquad i=1,2\, .\label{qqqqid}
\end{eqnarray}
Note that for each helicity configuration, we have the following relation
between the  functions $A$ and $B$
\begin{equation} 
A_1^{\bar q ; q \bar{Q} Q} + B_1^{\bar q ; q \bar{Q} Q} =
A_2^{\bar q ; q \bar{Q} Q} + B_2^{\bar q ; q \bar{Q} Q}\, ,\label{qqqqlin}
\end{equation}

\subsection{NNLO impact factors in the high-energy limit}
\label{sec:nnlohigh}

The amplitudes (\ref{nnloggg}), (\ref{nnlogqq}), (\ref{nnloqgg}) and
(\ref{3forwqqq}) have been computed in the kinematic limit (\ref{nnloreg}),
in which they factorize into an effective amplitude with a ladder structure,
made of a three-parton forward cluster
and a LO impact factor connected by a gluon exchanged
in the crossed channel (\fig{fig:NNLO}). In the limits
$y_1\simeq y_2\gg y_3$ or $y_1\gg y_2\simeq y_3$, the amplitudes must 
factorize further into NLO impact factors or into NLO Lipatov vertices
for the production of two partons along the ladder. Such limits constitute
then necessary consistency checks, and we display them in this section.
\begin{figure}[t]
\begin{center}
\vspace*{-0.2cm}
\hspace*{0cm}
\epsfxsize=14cm \epsfbox{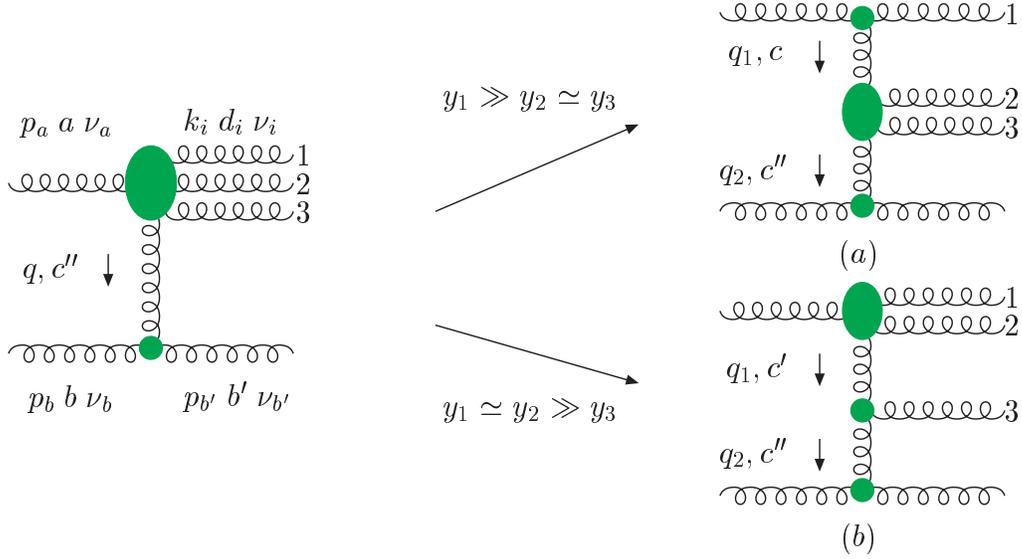}
\vspace*{-0.2cm}
\caption{Limits of the amplitude for the production of three gluons
in the forward-rapidity region of gluon $p_a$, for 
$y_1 \gg y_2 \simeq y_3$ (a) and $y_1 \simeq y_2 \gg y_3$ (b).}
\label{fig:ggglim}
\end{center}
\end{figure}

In the limit, $y_1\gg y_2\simeq y_3$, the NNLO impact factor,
$g^*\, g\, \to g\, g\, g$, \eqn{nnloggg},
factorizes into a NLO Lipatov vertex for the 
production of two gluons convoluted
with a multi-Regge ladder (\fig{fig:ggglim}a)
\begin{eqnarray}
& & \lim_{y_1\gg y_2\simeq y_3}
\left\{ (ig)^3\, \sum_{\sigma\in S_3}
f^{a d_{\sigma_1} c } f^{c d_{\sigma_2} c'} f^{c' d_{\sigma_3} c''}  
A^{g;3g}(p_a^{\nu_a}; k_{\sigma_1}^{\nu_{\sigma_1}}, 
k_{\sigma_2}^{\nu_{\sigma_2}}, k_{\sigma_3}^{\nu_{\sigma_3}}) \right\} 
\label{centrlim} \\ & & = 
\left[i g\, f^{ad_1c}\, C^{g;g}
(p_a^{\nu_a};k_1^{\nu_1}) \right] {1\over t_1}
\left\{ (ig)^2\, \sum_{\sigma\in S_2} f^{c d_{\sigma_2} c' } 
f^{c' d_{\sigma_3} c''} 
A^{gg}(q_1, k_{\sigma_2}^{\nu_{\sigma_2}}, k_{\sigma_3}^{\nu_{\sigma_3}},q_2) 
\right\}
%& & \A{g\,g \to 4g}(p_a^{\nu_a}, k_1^{\nu_1} \sep k_2^{\nu_2}, 
%k_3^{\nu_3} \sep  p_{b'}^{\nu_{b'}}, p_b^{\nu_b}) 
%\lim_{y_1\gg y_2\simeq y_3}
%\A{g\,g \to 4g}(p_a^{\nu_a}, k_1^{\nu_1}, k_2^{\nu_2}, 
%k_3^{\nu_3} \sep  p_{b'}^{\nu_{b'}}, p_b^{\nu_b})
%\nonumber = 2\, s\,
%\, {1\over t_2}\, 
%\left[i g\, f^{bb'c''}\, C^{g;g}(p_b^{\nu_b};p_{b'}^{\nu_{b'}}) \right]\,
,\nonumber
\end{eqnarray}
with the NLO Lipatov vertex, $g^*\, g^* \rightarrow g\, g$, for the 
production of two gluons $k_2$ and $k_3$ given by \cite{fl,ptlipnl,flqq} 
\begin{eqnarray}
& & A^{g\, g}(q_1,k_2^+,k_3^+,q_2) = 2\, {q_{1\perp}^* q_{2\perp}\over 
k_{2\perp}} \sqrt{x_2\over x_3}\, {1\over \langle 2 3\rangle} \label{kosc}\\
& & A^{g\, g}(q_1,k_2^+,k_3^-,q_2) = -2\, {k_{2\perp}^* \over k_{2\perp}} 
\left\{ - {1\over s_{23}} \left[{k_{3\perp}^2 |q_{1\perp}|^2 \over 
(k_2^-+k_3^-)k_3^+} +{k_{2\perp}^2 |q_{2\perp}|^2 \over (k_2^++k_3^+)k_2^-} 
+ {s_{3bb'}\, k_{2\perp}k_{3\perp}\over k_2^-k_3^+} \right]\right. 
\nonumber \\ && + \left. {(q_{2\perp}+k_{3\perp})^2 \over s_{3bb'}} -
{q_{2\perp}+k_{3\perp} \over s_{23}}
\left[{k_2^-+k_3^-\over k_2^-} k_{2\perp} - {k_2^++k_3^+\over k_3^+} 
k_{3\perp} \right]\right\} \nonumber
\end{eqnarray}
with exchanged momenta in the $t$ channel $q_1 = -(p_{a'} + p_a)$, 
$q_2 = p_{b'} + p_b$, three-particle invariant 
$s_{3bb'}= (k_3+q_2)^2 \simeq - \left(
|q_{2\perp}+k_{3\perp}|^2 + k_2^-k_3^+ \right)$,
and with the mass-shell conditions $k_i^-=|k_{i\perp}|^2/k_i^+$ for $i=2,3$.

\noindent
In the collinear limit, $k_2 = zP$ and $k_3 = (1-z)P$, the NLO Lipatov 
vertex (\ref{kosc}) reduces to the splitting factor (\ref{split}), and
amplitude (\ref{nnloggg}) factorizes into a multi-Regge amplitude
(\ref{three}) times a collinear factor (\ref{colfac})
\begin{eqnarray} 
& & \lim_{k_2 || k_3} \A{g\,g \to 4g}(p_a^{\nu_a}, k_1^{\nu_1} \sep
k_2^{\nu_2}, k_3^{\nu_3} \sep p_{b'}^{\nu_{b'}}, p_b^{\nu_b}) = 
\nonumber \\ &&
\sum_\nu \A{g g \to 3g}(p_a^{\nu_a}, k_1^{\nu_1} \sep P^{\nu} \sep
p_{b'}^{\nu_{b'}}, p_b^{\nu_b}) \cdot
{\rm Split}_{-\nu}^{g \to g g}(k_2^{\nu_2}, k_3^{\nu_3})
,\nonumber
\end{eqnarray}

\noindent
In the limit, $y_1\simeq y_2\gg y_3$, the NNLO impact factor
in \eqn{nnloggg}
factorizes into a NLO impact factor, $g^*\, g \rightarrow g\, g$, 
\eqn{nllfg}, convoluted with a multi-Regge ladder (\fig{fig:ggglim}b)
\begin{eqnarray}
& & \lim_{y_1\simeq y_2\gg y_3}
\left\{ (ig)^3\, \sum_{\sigma\in S_3}
f^{a d_{\sigma_1} c } f^{c d_{\sigma_2} c'} f^{c' d_{\sigma_3} c''}  
A^{g;3g}(p_a^{\nu_a}; k_{\sigma_1}^{\nu_{\sigma_1}}, 
k_{\sigma_2}^{\nu_{\sigma_2}}, k_{\sigma_3}^{\nu_{\sigma_3}}) \right\} 
\\ & & = \left\{ 
(ig)^2\, \sum_{\sigma\in S_2} f^{a d_{\sigma_1} c } f^{c d_{\sigma_2} c'} 
A^{g;2g}(p_a^{\nu_a}; k_{\sigma_1}^{\nu_{\sigma_1}}, 
k_{\sigma_2}^{\nu_{\sigma_2}}) \right\} {1\over t_1} 
\left[i g\, f^{c'd_3c''}\, C^g(q_1,k_3^{\nu_3},q_2)\right]\, ,\nonumber
%& &\A{g\,g \to 4g}(p_a^{\nu_a}, k_1^{\nu_1}, k_2^{\nu_2} \sep 
%k_3^{\nu_3} \sep p_{b'}^{\nu_{b'}}, p_b^{\nu_b}) =
% \lim_{y_1\simeq y_2\gg y_3}
%\A{g\,g \to 4g}(p_a^{\nu_a}, k_1^{\nu_1}, k_2^{\nu_2}, 
%k_3^{\nu_3} \sep p_{b'}^{\nu_{b'}}, p_b^{\nu_b}) \nonumber\\ 
%& & = 2\, s\, \, {1\over t_2}\, 
%\left[i g\, f^{bb'c''}\, C^{g;g}(p_b^{\nu_b};p_{b'}^{\nu_{b'}}) \right]\, ,
\end{eqnarray}
with $q_1=-(p_a+k_1+k_2)$,  and with LO Lipatov vertex 
$C^g(q_1,k_3^{\nu_3},q_2)$, \eqn{lip}.

\begin{figure}[tb]
\begin{center}
\vspace*{-0.2cm}
\hspace*{0cm}
\epsfxsize=14cm \epsfbox{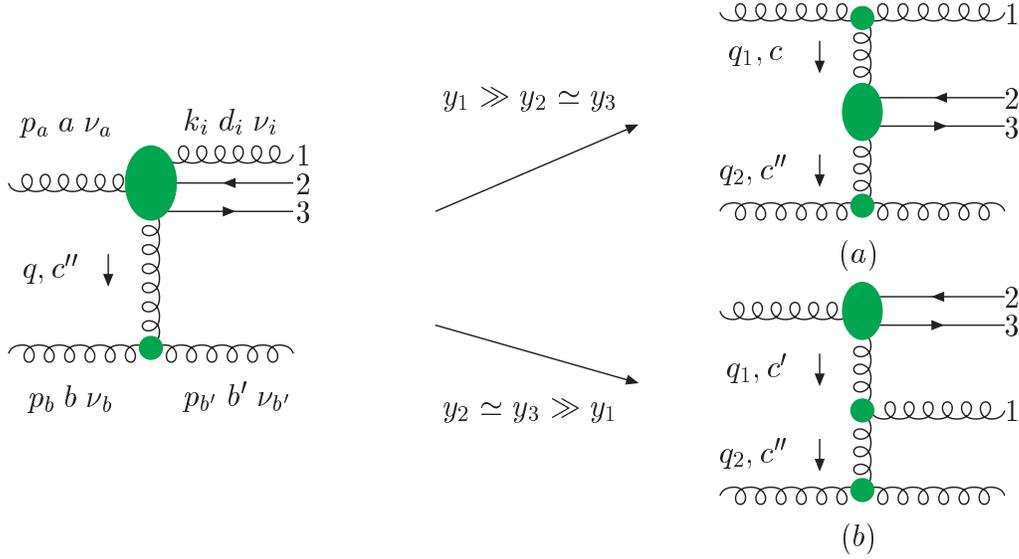}
\vspace*{-0.2cm}
\caption{Same as Fig.~\ref{fig:ggglim} for the
production of a quark-antiquark pair and a gluon
in the forward-rapidity region of gluon $p_a$.}
\label{fig:gaqlim}
\end{center}
\end{figure}

In the limit, $y_1\gg y_2\simeq y_3$, the functions $A$ and $B$ in
\eqn{qqgcoeff}-(\ref{qqgab}) fulfill the relations $A_2^{g; g\bar{q} q}=
A_3^{g; g\bar{q} q}=0$, $B_2^{g; g\bar{q} q}=-A_1^{g; g\bar{q} q}$, and 
$B_1^{g; g\bar{q} q}=-A_4^{g; g\bar{q} q}$,
thus the NNLO impact factor,
$g^*\, g\, \to g\, \bar q\, q$, \eqn{ggqqif},
factorizes into a NLO Lipatov vertex for the 
production of a $\bar q q$ pair convoluted
with a multi-Regge ladder (\fig{fig:gaqlim}a),
\begin{eqnarray}
& & \lim_{y_1\gg y_2\simeq y_3} 
I^{g;g\bar{q}q}(p_a^{\nu_a};k_1^{\nu_1},k_2^{\nu_2},k_3^{-\nu_2})
\nonumber\\ & & = \left[i g\, f^{ad_1c}\, C^{g;g}
(p_a^{\nu_a};k_1^{\nu_1}) \right] \label{qqcentlim}\\ & & \times {1\over t_1}
\left\{g^2 \left[ (\lambda^{c'} \lambda^c)_{d_3\bar d_2}
A^{\bar{q}\, q}(q_1,k_2^{\nu_2}, k_3^{-\nu_2},q_2) +
(\lambda^c \lambda^{c'})_{d_3\bar d_2}
A^{\bar{q}\, q}(q_1,k_3^{-\nu_2}, k_2^{\nu_2},q_2) \right] \right\}\, 
,\nonumber
\end{eqnarray}
with the NLO Lipatov vertex, $g^*\, g^* \rightarrow \bar q\, q$, for the 
production of a $\bar q q$ pair \cite{flqq,ptlipqq} 
\begin{eqnarray}
A^{\bar{q}\, q}(q_1,k_2^+,k_3^-,q_2) &=& - 2 \sqrt{k_2^+\over k_3^+}
\left\{ {k_3^+ |q_{2\perp}|^2 \over (k_2^++k_3^+) s_{23}} +
{k_3^- k_{3\perp} |q_{1\perp}|^2 \over k_{2\perp} (k_2^-+k_3^-) s_{23}}
+ {k_3^+ k_{2\perp}^* (q_{2\perp}+k_{3\perp}) \over k_2^+ s_{3bb'}} \right.
\nonumber\\ &+& \left. {(q_{2\perp}+k_{3\perp}) [k_2^-k_3^+ - k_{2\perp}^* 
k_{3\perp} - (q_{2\perp}^*+k_{3\perp}^*) k_{3\perp}] \over k_{2\perp} 
s_{23}} - {|k_{3\perp}|^2 \over s_{23}} \right\}\, ,\label{centqq}
\end{eqnarray}
with $q_1$, $q_2$, and $s_{3bb'}$ as in \eqn{kosc}.

\noindent
In the collinear limit, $k_2 = zP$ and $k_3 = (1-z)P$, the 
NLO Lipatov vertex (\ref{centqq}) reduces to the splitting factor 
(\ref{split}), and amplitude (\ref{nnlogqq})
factorizes into a multi-Regge amplitude
(\ref{three}) times a collinear factor (\ref{colfac})
\begin{eqnarray} 
& & \lim_{k_2 || k_3} 
\A{g g \to g\,\bar{q} q\, g}
(p_a^{\nu_a}, k_1^{\nu_1} \sep k_2^{\nu_2}, k_3^{-\nu_2} \sep 
p_{b'}^{\nu_{b'}}, p_b^{\nu_b}) \nonumber\\ &&=
\sum_\nu \A{g g \to 3g}(p_a^{\nu_a}, k_1^{\nu_1} \sep P^{\nu} \sep
p_{b'}^{\nu_{b'}}, p_b^{\nu_b}) \cdot 
{\rm Split}_{-\nu}^{g \to \bar q q}(k_2^{\nu_2}, k_3^{-\nu_2})\,
.\nonumber
\end{eqnarray}

\noindent
In the limit, $y_2\simeq y_3\gg y_1$, the functions $A$ and $B$ in
\eqn{qqgcoeff}-(\ref{qqgab}) fulfill the relations $A_1^{g; g\bar{q} q}=
A_4^{g; g\bar{q} q}=0$, $B_2^{g; g\bar{q} q}=-A_3^{g; g\bar{q} q}$, and 
$B_1^{g; g\bar{q} q}=-A_2^{g; g\bar{q} q}$ 
thus the NNLO impact factor, \eqn{ggqqif},
factorizes into a NLO impact factor, $g^*\, g \rightarrow \bar q\, q$, 
\eqn{forwqq}, convoluted with a multi-Regge ladder (\fig{fig:gaqlim}b),
\begin{eqnarray}
&& \lim_{y_2\simeq y_3\gg y_1}
I^{g;g\bar{q}q}(p_a^{\nu_a};k_1^{\nu_1},k_2^{\nu_2},k_3^{-\nu_2})
\nonumber\\ & & = \left\{g^2\, 
\left[\left(\lambda^{c} \lambda^a\right)_{d_3\bar{d_2}} 
A^{g;\bar{q}q}(p_a^{\nu_a}; k_2^{\nu_2}, k_3^{-\nu_2})
+ \left(\lambda^a \lambda^{c}\right)_{d_3\bar{d_2}} 
A^{g;\bar{q}q}(p_a^{\nu_a}; k_3^{-\nu_2}, k_2^{\nu_2})
\right] \right\} \nonumber\\ & & \times {1\over t_1} 
\left[i g\, f^{cd_1c'}\, C^g(q_1,k_1^{\nu_1},q_2)\right]\, 
,\label{qqforwlim}
\end{eqnarray}
with $q_1=-(p_a+k_2+k_3)$.

\begin{figure}[tb]
\begin{center}
\vspace*{-0.2cm}
\hspace*{0cm}
\epsfxsize=14cm \epsfbox{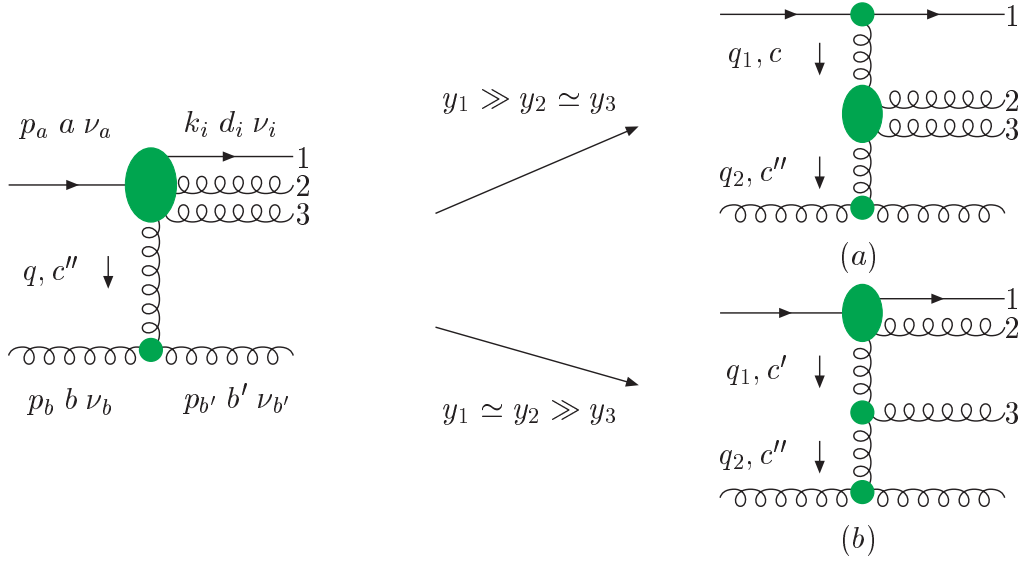}
\vspace*{-0.2cm}
\caption{Same as Fig.~\ref{fig:ggglim} for the
production of a quark  and two gluons
in the forward-rapidity region of quark $p_a$.}
\label{fig:qgglim}
\end{center}
\end{figure}

In the limit, $y_1\gg y_2\simeq y_3$, the functions $A$ and $B$ in
(\ref{qggcoeff})-(\ref{qggnpt6}) fulfill the relations,
\begin{eqnarray}
&& B_1^{\bar q;qgg}(p_a^{-\nu_1};k_1^{\nu_1},k_{\sigma_2}^{\nu_{\sigma_2}},
k_{\sigma_3}^{\nu_{\sigma_3}}) =
A^{\bar q;qgg}(p_a^{-\nu_1};k_1^{\nu_1},k_{\sigma_3}^{\nu_{\sigma_3}},
k_{\sigma_2}^{\nu_{\sigma_2}}) \nonumber\\ 
&& B_2^{\bar q;qgg}(p_a^{-\nu_1};k_1^{\nu_1},k_2^{\nu_2},
k_3^{\nu_3}) = B_2^{\bar q;qgg}(p_a^{-\nu_1};k_1^{\nu_1},k_3^{\nu_3},
k_2^{\nu_2}) \nonumber\\
&& \qquad = -\left( A^{\bar q;qgg}(p_a^{-\nu_1};k_1^{\nu_1},k_2^{\nu_2},
k_3^{\nu_3}) + A^{\bar q;qgg}(p_a^{-\nu_1};k_1^{\nu_1},k_3^{\nu_3},
k_2^{\nu_2}) \right)\, ,\nonumber
\end{eqnarray}
thus the NNLO impact factor, 
$q\,g^*\to q\,g\,g$, \eqn{qqggif},
factorizes into a NLO Lipatov vertex for the 
production of two gluons (\ref{kosc}) convoluted
with a multi-Regge ladder (\fig{fig:qgglim}a)
\begin{eqnarray}
& & \lim_{y_1\gg y_2\simeq y_3}
I^{\bar q;qgg}(p_a^{-\nu_1};k_1^{\nu_1},k_2^{\nu_2},k_3^{\nu_3})
\label{qggcentrlim}\\ & & = \left[g\, \lambda^c_{d_1 \bar a}\,
C^{\bar q;q}(p_a^{-\nu_1}; k_1^{\nu_1}) \right] {1\over t_1}
\left\{ (ig)^2\, \sum_{\sigma\in S_2} f^{c d_{\sigma_2} c' } 
f^{c' d_{\sigma_3} c''} 
A^{gg}(q_1, k_{\sigma_2}^{\nu_{\sigma_2}}, k_{\sigma_3}^{\nu_{\sigma_3}},q_2) 
\right\}\, .\nonumber
\end{eqnarray}

\noindent
In the limit, $y_1\simeq y_2\gg y_3$, the functions $A$ and $B$ in
(\ref{qggcoeff})-(\ref{qggnpt6}) fulfill the relations,
\begin{eqnarray}
A^{\bar q;qgg}(p_a^{-\nu_1};k_1^{\nu_1},k_3^{\nu_3},
k_2^{\nu_2}) &=& B_1^{\bar q;qgg}(p_a^{-\nu_1};k_1^{\nu_1},k_2^{\nu_2},
k_3^{\nu_3}) = 0 \nonumber\\
B_2^{\bar q;qgg}(p_a^{-\nu_1};k_1^{\nu_1},k_2^{\nu_2},
k_3^{\nu_3}) &=& - A^{\bar q;qgg}(p_a^{-\nu_1};k_1^{\nu_1},k_2^{\nu_2},
k_3^{\nu_3}) \nonumber\\
B_2^{\bar q;qgg}(p_a^{-\nu_1};k_1^{\nu_1},k_3^{\nu_3},
k_2^{\nu_2}) &=& B_1^{\bar q;qgg}(p_a^{-\nu_1};k_1^{\nu_1},k_3^{\nu_3},
k_2^{\nu_2})\, ,\nonumber
\end{eqnarray}
thus the NNLO impact factor, \eqn{qqggif},
factorizes into a NLO impact factor, $q\, g^* \rightarrow q\, g$, 
\eqn{forwqg}, convoluted with a multi-Regge ladder (\fig{fig:qgglim}b)
\begin{eqnarray}
&& \lim_{y_1\simeq y_2\gg y_3}
I^{\bar q;qgg}(p_a^{-\nu_1};k_1^{\nu_1},k_2^{\nu_2},k_3^{\nu_3})
\label{forwqqgg}\\ & & = 
\left\{ g^2\, \left[\left(\lambda^{d_2}
\lambda^{c'}\right)_{d_1\bar{a}} 
A^{\bar{q};qg}(p_a^{-\nu_1};k_1^{\nu_1},k_2^{\nu_2}) 
+ \left(\lambda^{c'}\lambda^{d_2}\right)_{d_1\bar{a}} 
B^{\bar{q};qg}(p_a^{-\nu_1};k_1^{\nu_1},k_2^{\nu_2}) \right] \right\} 
\nonumber\\ & & \times {1\over t_1} 
\left[i g\, f^{c'd_3c''}\, C^g(q_1,k_3^{\nu_3},q_2)\right]\, ,\nonumber
\end{eqnarray}
with $q_1=-(p_a+k_1+k_2)$.

\begin{figure}[t]
\begin{center}
\vspace*{-0.2cm}
\hspace*{0cm}
\epsfxsize=14cm \epsfbox{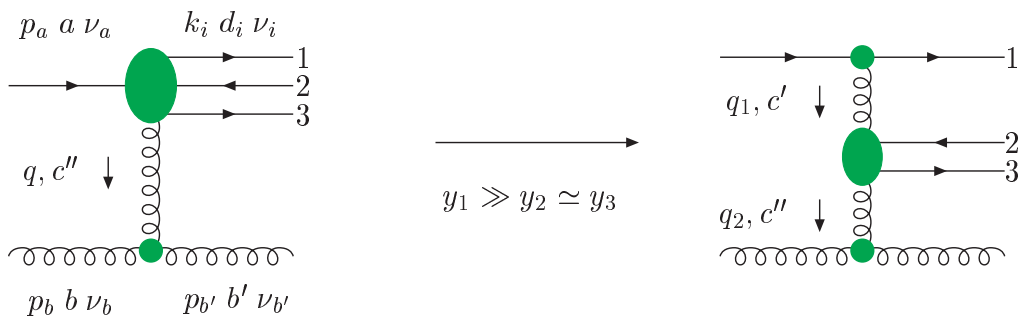}
\vspace*{-0.2cm}
\caption{Limit of the amplitude for the production of a quark and a quark-antiquark pair in the forward-rapidity region of quark $p_a$, for 
$y_1 \gg y_2 \simeq y_3$.}
\label{fig:qAQlim}
\end{center}
\end{figure}

In the limit, $y_1\gg y_2\simeq y_3$, the function $A_2$ in 
\eqn{qqqcoeff} vanishes, $A_2^{\bar q ; q \bar{Q} Q} = 0$, and
using \eqnss{qqqcoeff}{qqqqlin} the NNLO impact factor, 
$q\,g^* \to q\bar{Q}\, Q$,
\eqn{qqqqif}, factorizes into a NLO Lipatov vertex for the 
production of a $\bar q q$ pair (\ref{centqq}) convoluted
with a multi-Regge ladder (\fig{fig:qAQlim})
\begin{eqnarray}
&& \lim_{y_1\gg y_2\simeq y_3}
I^{\bar q ; q \bar{Q} Q}(p_a^{-\nu_1};k_1^{\nu_1},k_2^{\nu_2},
k_3^{-\nu_2}) \nonumber\\ && =
\left[g\, \lambda^c_{d_1 \bar a}\,
C^{\bar q;q}(p_a^{-\nu_1};k_1^{\nu_1}) \right] 
\label{qqqcentrlim}\\ & & \times {1\over t_1}
\left\{g^2 \left[ (\lambda^{c'} \lambda^c)_{d_3\bar d_2}
A^{\bar{q}\, q}(q_1,k_2^{\nu_2}, k_3^{-\nu_2},q_2) +
(\lambda^c \lambda^{c'})_{d_3\bar d_2}
A^{\bar{q} q}(q_1,k_3^{-\nu_2}, k_2^{\nu_2},q_2) \right] \right\}\, .\nonumber
\end{eqnarray}

\subsection{NNLO impact factors in the triple collinear limit}
\label{sec:nnlocoll}

In the triple collinear limit, $k_i = z_i P$, with $z_1+z_2+z_3=1$
a generic amplitude must factorize as~\cite{glover,cat} 
\begin{equation}
\lim_{k_1 || k_2 || k_3} \A{... d_1 d_2 d_3 ...}(..., k_1^{\nu_1}, 
k_2^{\nu_2}, k_3^{\nu_3}, ...)
= \sum_\nu \A{... c ...}(..., P^{\nu}, ...) \cdot 
{\rm Split}_{-\nu}^{f \to f_1 f_2 f_3}(k_1^{\nu_1}, 
k_2^{\nu_2},k_3^{\nu_3})\, .\label{tricoll}
\end{equation}
Accordingly, we must show that taking the triple collinear limit of the 
NNLO impact factors, we can write the amplitudes 
(\ref{nnloggg}), (\ref{nnlogqq}), (\ref{nnloqgg}) and
(\ref{3forwqqq}) as 
\begin{eqnarray}
& &\lim_{k_1 || k_2 || k_3} \A{f g \to f_1 f_2 f_3 g}(p_a^{\nu_a}, 
k_1^{\nu_1}, k_2^{\nu_2}, k_3^{\nu_3} \sep p_{b'}^{-\nu_b}, p_b^{\nu_b}) 
\nonumber\\ & & = 
\A{f g \to f g}(p_a^{\nu_a}, P^{-\nu_a} \sep p_{b'}^{-\nu_b}, 
p_b^{\nu_b}) \cdot {\rm Split}_{\nu_a}^{f \to f_1 f_2 f_3}
(k_1^{\nu_1}, k_2^{\nu_2}, k_3^{\nu_3})\, ,\label{hetricoll}
\end{eqnarray}
with $f$ denoting the parton species, $\A{f g \to f g}$ given in 
Eqs.~(\ref{elas}), (\ref{elasqa}) and (\ref{elasaqa}), and
with ${\rm Split}_{-\nu}^{f \to f_1 f_2 f_3}$
the polarized double-splitting functions. 

In the triple collinear limit, the functions $A$ of 
Sect.~\ref{sec:nnlogggg}, \ref{sec:nnloqqgg}, \ref{sec:nnloqgg} and
\ref{sec:nnloqqq} yield a quadratic divergence as $s_{123}\to 0$
or $s_{ij}\to 0$ with $i,j=1,2,3$. In the same limit, the functions $B$ 
have a single collinear divergence since only
two out of the three partons are color adjacent.
However, terms with a single divergence when integrated over the
triple collinear region of phase space yield a negligible 
contribution~\cite{glover}, thus we ignore them.

It is easy to show that a function $A^{g;3g}$, \eqnss{pt3g}{fournonpt},
differs from its reflection by a term which contains only a single divergence.
Using this property and \eqn{baaa}, we obtain a dual Ward identity
and a reflection identity for the  functions $A^{g;3g}$,
up to singly divergent terms,
\begin{eqnarray}
&&A^{g;3g}(p_a^{\nu_a}; k_1^{\nu_1}, k_2^{\nu_2}, k_3^{\nu_3}) +
A^{g;3g}(p_a^{\nu_a}; k_1^{\nu_1}, k_3^{\nu_3}, k_2^{\nu_2}) +
A^{g;3g}(p_a^{\nu_a}; k_3^{\nu_3}, k_1^{\nu_1}, k_2^{\nu_2})=0 \nonumber\\ 
&&A^{g;3g}(p_a^{\nu_a};k_1^{\nu_1},k_2^{\nu_2},k_3^{\nu_3})=
A^{g;3g}(p_a^{\nu_a};k_3^{\nu_3},k_2^{\nu_2},k_1^{\nu_1})\, .\label{wardcoll}
\end{eqnarray}
Using the identities (\ref{wardcoll}) in the impact factor
in \eqn{nnloggg}, we can factorize the color structure  on a leg
\begin{eqnarray}
&& (ig)^3 \sum_{\sigma \in S_3}
f^{a d_{\sigma_1} c } f^{c d_{\sigma_2} c'} f^{c' d_{\sigma_3} c''} 
A^{g;3g}(p_a^{\nu_a};k_{\sigma_1}^{\nu_{\sigma_1}},
k_{\sigma_2}^{\nu_{\sigma_2}},k_{\sigma_3}^{\nu_{\sigma_3}}) \nonumber\\
&&\hspace{2cm}=\frac{(ig)^3}{3}\, f^{a c c''}\,\sum_{\sigma \in S_3}
  f^{c d_{\sigma_1} c'} f^{c' d_{\sigma_2} d_{\sigma_3} } 
A^{g;3g}(p_a^{\nu_a};k_{\sigma_1}^{\nu_{\sigma_1}},
k_{\sigma_2}^{\nu_{\sigma_2}},k_{\sigma_3}^{\nu_{\sigma_3}})\, 
\nonumber\\
&&\hspace{2cm} = i g f^{a c c''}\; \left\{ g^2  \sum_{\sigma \in S_2}
 \left( F^{ d_{\sigma_1} } F^{ d_{\sigma_2} }\right)_{c d_3}
% f^{c d_{\sigma_1} c'} f^{c' d_{\sigma_2} d_{3} } 
A^{g;3g}(p_a^{\nu_a};k_{\sigma_1}^{\nu_{\sigma_1}},
k_{\sigma_2}^{\nu_{\sigma_2}},k_{3}^{\nu_{3}}) \right\}\, 
,\label{colorfac}
\end{eqnarray}
where $(F^a)_{bc}\equiv i f^{bac}$.
Thus amplitude (\ref{nnloggg}) can be put in the form of \eqn{hetricoll}
with collinear factor
\begin{eqnarray}
{\rm Split}_{-\nu}^{g \to 3g}(k_1^{\nu_1}, k_2^{\nu_2}, k_3^{\nu_3})
&=& {g^2} \sum_{\sigma \in S_2}
 \left( F^{ d_{\sigma_1} } F^{ d_{\sigma_2} }\right)_{c d_3}
%  f^{c d_{\sigma_1} c'} f^{c' d_{\sigma_2} d_{3} } 
{\rm split}_{-\nu}^{g \to 3g}(k_{\sigma_1}^{\nu_{\sigma_1}},
k_{\sigma_2}^{\nu_{\sigma_2}},k_{3}^{\nu_{3}})\, .
\label{trisplit}
\end{eqnarray}
The splitting factors ${\rm split}_{-\nu}^{g \to 3g}$ are the
functions $A$, \eqnss{pt3g}{fournonpt}, in the triple collinear limit,
up to singly divergent terms, and thus they fulfill the identities,
\eqn{wardcoll}. The splitting factors of PT type can be soon read off 
from \eqnss{pt3g}{fourpt}, while for the ones of 
non-PT type we note that the coefficients of \eqn{alfabetagamma}
reduce to
\begin{eqnarray}
&& \alpha (k_1,k_2,k_3) \rightarrow \frac{z_1 z_3}{z_1+z_3} \nonumber\\
&& \beta (k_1,k_2,k_3) \rightarrow 
-\frac{\sqrt{z_1 z_2}}{P_{\perp}^{*}} [1\, 2]
\label{abcd}\\
&& \gamma (k_1,k_2,k_3) \rightarrow \frac{\sqrt{z_1 z_2 z_3}}{s_{123}} 
\delta(1,2,3) \nonumber
\end{eqnarray}
with
\begin{equation}
\delta(1,2,3) \equiv \spb1.2
(\sqrt{z_1}\, \langle 1 \,3\rangle+\sqrt{z_2}\,\langle 2 \,3 \rangle  )\,
.\label{delta123}
\end{equation}
Thus we obtain
\begin{eqnarray}
{\rm split}_{-}^{g \to 3g}(k_1^+, k_2^+, k_3^+) &=&
2 {1\over \sqrt{z_1z_3}} {1\over \langle 1 \,2\rangle
\langle 2 \,3\rangle} \nonumber\\
{\rm split}_{+}^{g \to 3g}(k_1^-, k_2^+, k_3^+) &=&
2 {z_1^2\over \sqrt{z_1z_3}} {1\over \langle 1 \,2\rangle
\langle 2 \,3\rangle} \nonumber\\
{\rm split}_{+}^{g \to 3g}(k_1^+, k_2^-, k_3^+) &=&
2 {z_2^2\over \sqrt{z_1z_3}} {1\over \langle 1 \,2\rangle
\langle 2 \,3\rangle} \nonumber\\
{\rm split}_{+}^{g \to 3g}(k_1^+, k_2^+, k_3^-) &=&
2 {z_3^2\over \sqrt{z_1z_3}} {1\over \langle 1 \,2\rangle
\langle 2 \,3\rangle} \label{triplea}\\
{\rm split}_{-}^{g \to 3g}(k_{1}^{+}, k_{2}^{+}, k_{3}^{-}) &=&
\frac{2}{s_{12} s_{23}}
\left[\frac{s_{12} z_2}{(1-z_1)} + 
        \frac{\delta(1,2,3)^2}{s_{123}} +
        \sqrt{\frac{z_2}{z_1 z_3} } (1-z_3) \delta(1,2,3)
\right]\nonumber\\
{\rm split}_{-}^{g \to 3g}(k_1^-,k_2^+,k_3^+) &=& 
{\rm split}_{-}^{g \to 3g}(k_3^+,k_2^+,k_1^-)\nonumber\\
{\rm split}_{-}^{g \to 3g}(k_1^+,k_2^-,k_3^+) &=&
- {\rm split}_{-}^{g \to 3g}(k_2^-,k_1^+,k_3^+) -
{\rm split}_{-}^{g \to 3g}(k_1^+,k_3^+,k_2^-)\, .\nonumber
\end{eqnarray}

In the triple collinear limit of 
the NNLO impact factor $g\, g^* \rightarrow \bar q\, q\, g$,
the functions $A^{g\,g\bar{q} q}$, 
\eqnss{qqgcoeff}{qqgab}, fulfill the relations
$A_2^{g; g\bar{q} q}=-A_1^{g; g\bar{q} q}$ and 
$A_4^{g; g\bar{q} q}=-A_3^{g; g\bar{q} q}$, and
$A_3^{g; g\bar{q} q}(p_a^{\nu_a}; k_1^{\nu_1}, k_2^{\nu_2}, k_3^{-\nu_2}) 
= A_1^{g; g\bar{q} q}(p_a^{\nu_a}; k_1^{\nu_1}, k_3^{-\nu_2}, k_2^{\nu_2})$.
Thus amplitude (\ref{nnlogqq}) can be put in the form of \eqn{hetricoll}
with collinear factor
\begin{eqnarray}
&& {\rm Split}_{-\nu}^{g \to g \bar{q} q}(k_1^{\nu_1}, k_2^{\nu_2}, 
k_3^{-\nu_2}) \label{triqqsplit}\\ &&
= g^2 \left[ \left(\lambda^{c} \lambda^{d_1}\right)_{d_3\bar{d_2}} 
{\rm split}_{-\nu}^{g \to g \bar{q} q}(k_1^{\nu_1}, k_2^{\nu_2}, k_3^{-\nu_2})
+ \left(\lambda^{d_1} \lambda^{c}\right)_{d_3\bar{d_2}} 
{\rm split}_{-\nu}^{g \to g \bar{q} q}(k_1^{\nu_1}, k_3^{-\nu_2}, k_2^{\nu_2})
\right]\, ,\nonumber
\end{eqnarray}
with
\begin{eqnarray}
&& {\rm split}_{+}^{g \to g \bar{q} q}(k_1^+, k_2^-, k_3^+) =
2 \sqrt{z_2^3\over z_1} {1\over \langle 1 \,2\rangle
\langle 2 \,3\rangle} \nonumber\\
&& {\rm split}_{+}^{g \to g \bar{q} q}(k_1^+, k_2^+, k_3^-) =
2 z_3 \sqrt{z_2\over z_1} {1\over \langle 1 \,2\rangle
\langle 2 \,3\rangle} \label{qqgsplit}\\
&&{\rm split}_{-}^{g \to g \bar{q} q}(k_1^+,k_2^-,k_3^+) = 
-\frac{2}{s_{12} s_{23}}
\, \left[ \frac{\delta(1,3,2)^2 \spb1.2 }{\spb1.3  s_{123} } 
\right. \nonumber\\ && \left. \qquad +
\frac{\delta(1,3,2)}{z_3}
\left( \frac{z_2 (z_1 - z_3)}{\sqrt{z_1}}- 
       \frac{z_1 \sqrt{z_2} \spb2.3}{\spb1.3 z_3 }\right)   +
\frac{\sqrt{z_2} (- z_2 s_{13} + z_3 s_{23} + z_1 z_2
s_{123})}{\sqrt{z_3} (1-z_1) } \right]\nonumber\\
&&{\rm split}_{-}^{g \to g \bar{q} q}(k_1^+,k_2^+,k_3^-)=
- \frac{2}{s_{12} s_{23} }
\left[\frac{\delta(1,2,3)^2 \spb1.3 }{\spb1.2  s_{123}}
  + \frac{\delta(1,2,3) (1 - z_3)}{\sqrt{z_1}}
  + \frac{ \sqrt{z_2 z_3} s_{12}}{ (1 - z_1) }
\right]\, .\nonumber
\end{eqnarray}

Writing the functions $A$, \eqnss{qggcoeff}{qggnpt6},
in the triple collinear limit of
the NNLO impact factor $q\, g^* \rightarrow q\, g\, g$, 
the amplitude (\ref{nnloqgg}) can be put in the form of \eqn{hetricoll}
with collinear factor
\begin{equation}
{\rm Split}_{-\nu}^{q \to q g g}(k_1^{\nu}, k_2^{\nu_2}, 
k_3^{\nu_3}) = g^2 \sum_{\sigma \in S_2}
\left(\lambda^{d_{\sigma_2}} \lambda^{d_{\sigma_3}}\right)_{d_1\bar{c}} 
{\rm split}_{-\nu}^{q \to q g g}(k_1^\nu,
k_{\sigma_2}^{\nu_{\sigma_2}},k_{\sigma_3}^{\nu_{\sigma_3}})\, 
.\label{triqggsplit} 
\end{equation}
with 
\begin{eqnarray}
&& {\rm split}_{-}^{q \to q g g}(k_1^+, k_2^+, k_3^+) =
 -{2 i \over \sqrt{z_3}} {1\over \langle 1 \,2\rangle
\langle 2 \,3\rangle} \nonumber\\
&& {\rm split}_{+}^{q \to q g g}(k_1^-, k_2^+, k_3^+) =
{2 i z_1\over \sqrt{z_3}} {1\over \langle 1 \,2\rangle
\langle 2 \,3\rangle} \label{qggsplit}\\
&& {\rm split}_{-}^{q \to q g g}(k_1^+,k_2^-,k_3^+)=\frac{2i}{s_{12} s_{23}}
\times\nonumber\\
&& \left[
     \frac{ \delta(1,3,2)^2 (\sqrt{z_1} \spb1.3 + \sqrt{z_2} \spb2.3)}
        { \spb1.3 s_{123} }+
     \frac{ \sqrt{z_2}\, (1 - z_3) \,\delta(1,3,2)}{\sqrt{z_3} } + 
     \frac{\sqrt{z_1} z_2 s_{12} }{ 1 - z_1 } + 
     \sqrt{z_2} \spa2.3 \spb1.3 
    \right]\nonumber\\
&& {\rm split}_{-}^{q \to q g g}(k_1^+,k_2^+,k_3^-)=\frac{2i}{s_{12} s_{23}}
\times\nonumber\\
&&  \left[\frac{ \delta(1,2,3)^2 (\sqrt{z_1} \spb1.2 - \sqrt{z_3} \spb2.3
) }
    {\spb1.2 s_{123} }+
     \frac{\sqrt{z_2} \,(1 - z_3)  \,\delta(1,2,3) }{\sqrt{z_3} } + 
    \frac{ \sqrt{z_1} z_2 s_{23} }{1 - z_1}  
    \right]\nonumber
\end{eqnarray}

In the triple collinear limit of
the NNLO impact factor $q\, g^*\; \to q\,\bar{Q}\, Q$,
the functions $A$ (\ref{qqqcoeff}) fulfill the relation 
$A_1^{\bar{q};q \bar{Q} Q}=A_2^{\bar{q};q \bar{Q} Q}$.
Thus the amplitude (\ref{3forwqqq}) can be put in the form of 
Eq.~(\ref{hetricoll}), with  collinear factor
\begin{eqnarray}
{\rm Split}_{-\nu}^{q \to q \bar{Q}Q}(k_1^{\nu },k_2^{\nu _2},k_3^{\nu _3}) &=&
2 g^2 \left\{ \lambda^a_{d_1\bar{c}} \lambda^a_{d_3\bar{d_2}} \cdot 
{\rm split}_{-\nu}^{q \to q \bar{Q}Q}(k_1^{\nu },k_2^{\nu _2},k_3^{\nu _3})
\right. \nonumber \\
&& \left. -\delta _{qQ}
\lambda^a_{d_3\bar{c}} \lambda^a_{d_1\bar{d_2}} \cdot 
{\rm split}_{-\nu}^{q \to q \bar{Q}Q}(k_3^{\nu _3},k_2^{\nu _2},k_1^{\nu _1})
 \right\}\, ,
\end{eqnarray}
where the second term occurs for the case of identical quarks, and
$\bar{c}$ is the color index of the parent quark. The splitting 
factors are
\begin{eqnarray}
{\rm split}_{-}^{q \to q \bar{Q} Q}(k_1^+,k_2^-,k_3^+)&=&
\frac{i}{s_{23}  }
\left( \frac{\sqrt{z_1 z_2 z_3}}{1-z_1}+
\frac{\delta(1,3,2)}{s_{123}}\right)\label{qqqsplit}\\
{\rm split}_{+}^{q \to q \bar{Q} Q}(k_1^-,k_2^-,k_3^+)&=& 
{\rm split}_{-}^{q \to q \bar{Q} Q}(k_1^+,k_3^-,k_2^+)\, .\nonumber
\end{eqnarray}
The factor ${\rm split}_{\nu}^{f \to f_1 f_2 f_3}(k_1^{-\nu_1}, k_2^{-\nu_2}, 
k_3^{-\nu_3})$ can be obtained from ${\rm split}_{-\nu}
^{f \to f_1 f_2 f_3}(k_1^{\nu_1}, k_2^{\nu_2}, k_3^{\nu_3})$ 
in Eqs.~(\ref{triplea}), (\ref{qqgsplit}), (\ref{qggsplit}) and 
(\ref{qqqsplit}) by exchanging $\langle i
j\rangle$ with $[j i]$, and multiplying by a coefficient $S$, \eqn{qparity},
for each quark pair the splitting factor includes. 

Using \eqn{trisplit} and \eqnss{triplea}{qqqsplit}, and summing over the 
two helicity states of partons 1, 2 and 3, we obtain, as in \sec{sec:nlocoll},
the two-dimensional Altarelli-Parisi
polarization matrix at fixed color and helicity of the parent parton,
\begin{equation}  
\sum_{\nu_1\nu_2\nu_3}\, 
{\rm Split}_{\lambda}^{f \to f_1 f_2 f_3} (k_1^{\nu_1}, k_2^{\nu_2},
k_3^{\nu_3}) [{\rm Split}_{\rho}^{f \to f_1 f_2 f_3} (k_1^{\nu_1}, k_2^{\nu_2},
k_3^{\nu_3})]^* = \delta^{cc'} {4g^4\over s_{123}^2} 
P^{f\to f_1f_2f_3}_{\lambda\rho}\, ,\label{ap3matrix}
\end{equation}
where  $P^{f\to f_1f_2f_3}_{++} = 
P^{f\to f_1f_2f_3}_{--}$, and
$P^{f\to f_1f_2f_3}_{-+} = (P^{f\to f_1f_2f_3}_{+-})^*$.
For splitting functions of type $P^{q \to q\, f_2f_3}$, namely for
$P^{q \to q\, g\, g}$, $P^{q \to q\, \bar Q\, Q}$ and
$P^{q_1 \to q_1\, \bar q_2\, q_2}$, where the last splitting function
is for identical quarks, helicity conservation on the quark line
sets the off-diagonal elements equal to zero. 

Averaging over the trace of matrix (\ref{ap3matrix}), i.e. over
color and helicity of the parent parton, we obtain
the unpolarized Altarelli-Parisi splitting functions~\cite{glover}
\begin{equation}
{1\over 2{\cal C}}  \sum_{\nu\nu_1\nu_2\nu_3}\, 
|{\rm Split}_{-\nu}^{f \to f_1 f_2 f_3} (k_1^{\nu_1}, k_2^{\nu_2},
k_3^{\nu_3})|^2 = \frac{4g^4}{s_{123}^2}\, \langle P^{f\to f_1f_2f_3} \rangle\,
,\label{unpolap3}
\end{equation}
with ${\cal C}$ defined below \eqn{unpolap}. For 
$\langle P^{g \to g_1 g_2 g_3} \rangle$, 
the sum over colors can be immediately done using \eqn{square2}, and it
yields
\begin{equation}
|{\rm Split}_{-\nu}^{g \to g_1 g_2 g_3} 
(k_1^{\nu_1}, k_2^{\nu_2}, k_3^{\nu_3})|^2=
4 {\cal C}_4(N_c) \sum_{\sigma \in S_{3}} 
|{\rm split}_{-\nu}^{g \to 3g} 
(k_{\sigma_1}^{\nu_{\sigma_1}},
 k_{\sigma_2}^{\nu_{\sigma_2}},
 k_{\sigma_3}^{\nu_{\sigma_3}})|^2\, ,\label{colorggg}
\end{equation}
with ${\cal C}_n(N_c)$ as in \eqn{calc}.
\eqn{colorggg} shows that for the purely gluonic unpolarized
splitting function the color factorizes.

\noindent
Since the averaged trace of $P^{f\to f_1f_2f_3}$ 
is $\langle P^{f\to f_1f_2f_3} \rangle
= \tr P^{f\to f_1f_2f_3}/2 = P^{f\to f_1f_2f_3}_{++}$,
we have checked that for the diagonal elements, $P^{f\to f_1f_2f_3}_{++}$,
our expressions agree with the unpolarized splitting functions of
Ref.~\cite{glover} by setting there the RS parameter $\epsilon=0$.
Finally, for the off-diagonal elements of the splitting functions of type 
$P^{g \to g\, f_2f_3}$ we obtain
\begin{eqnarray}
\lefteqn{P_{+-}^{g \to g_1 g_2 g_3 } = 
C_A^2 \sum_{\sigma \in S_3}\frac{s_{123}}{s_{{{\sigma _1}} {{\sigma _2}}  }}\left\{ 
-\frac{2 \left[{{\sigma _1}}\,{{\sigma _2}}\right] ^2 
 z_{{\sigma _1}}  z_{{\sigma _2}}   
  }{   s_{{{\sigma _1}} {{\sigma _2}}  } (1 - z_{{\sigma _3}}   ) z_{{\sigma _3}}   }
 \right. }\nonumber \\
&& -  \frac{\sqrt{z_{{\sigma _2}}   z_{{\sigma _3}}   }   }{    s_{{{\sigma _1}} {{\sigma _3}}   }} 
  \left[   D_{{\sigma _2}} D_{{\sigma _3}}   
          \left( -3 + \frac{2 (1 - z_{{\sigma _2}}  ) z_{{\sigma _2}}     }{   (1 - z_{{\sigma _3}}   ) z_{{\sigma _3}}   } \right)
             \right. \label{offap} \\
&& \left. \left.      -  \frac{z_{{\sigma _1}}  z_{{\sigma _2}}   D_{{\sigma _2}}  ^2 
               (1 - 2 z_{{\sigma _3}}   )   }{    (1 - z_{{\sigma _3}}   ) z_{{\sigma _3}}   } 
    - \frac{z_{{\sigma _1}}  z_{{\sigma _3}}    D_{{\sigma _3}}   ^2 (1 - 2 z_{{\sigma _2}}  ) 
                  }{    (1 - z_{{\sigma _2}}  ) z_{{\sigma _2}}  } \right] \right\} \, ,\nonumber
\end{eqnarray}
with
\begin{equation}
D_i=\left[i\,j\right] \sqrt{z_j}+
\left[i\,k\right] \sqrt{z_k } \qquad \mbox{\rm with $i,j,k=1,2,3$ and $j,k \neq i$}\, ,
\end{equation}
and
\begin{equation}
P_{+-}^{g \to g_1 \bar{q}_2 q_3 } = {1\over 2} \left(C_F P_{+-}^{g \to g_1 
\bar{q}_2 q_3 \, ({\rm ab})} + C_A P_{+-}^{g \to g_1 \bar{q}_2 q_3 \, 
({\rm nab})} \right)
\end{equation}
where the abelian and non-abelian terms are,
\begin{eqnarray}  
\lefteqn{P_{+-}^{g \to g_1 \bar{q}_2 q_3 \, ({\rm ab})} = 
\frac{2 s_{123}}{s_{12}  s_{13}} \left\{ z_1  D_1^2
   - 2 \sqrt{z_2   z_3   }  D_2   D_3     \right\} }\nonumber \\
&& \nonumber \\
\lefteqn{P_{+-}^{g \to g_1 \bar{q}_2 q_3 \, ({\rm nab})} =
\sum_{\sigma \in S_2} \frac{s_{123}}{  s_{1 {{\sigma _2}}  } s_{1 {{\sigma _3}}   }} }\nonumber \\
&& 
\left\{  -\frac{1}{2} z_1  D_1^2    + 
       \sqrt{z_{{\sigma _2}}   z_{{\sigma _3}}   }  D_{{\sigma _2}}   D_{{\sigma _3}} 
       + \frac{2 \left[{{\sigma _2}}\,{{\sigma _3}}\right] ^2  z_{{\sigma _2}}   z_{{\sigma _3}} 
           s_{1 {{\sigma _2}}  }
  s_{1  {{\sigma _3}}   } }
 {  s_{{{\sigma _2}}  {{\sigma _3}}   }^2 (1 - z_1  )
  z_1  }       \right. \label{offapgqq} \\ 
 &&  \left.    + {s_{1\sigma_3}\over s_{\sigma_2\sigma_3}} \left[
     - z_{{\sigma _2}}   D_{{\sigma _2}}  ^2 + 
        \frac{ 2 D_{{\sigma _3}}   ^2 z_{{\sigma _2}}  ^2 z_{{\sigma _3}}      }
        {  (1 - z_1  ) z_1  } 
        - \sqrt{z_{{\sigma _2}}   z_{{\sigma _3}}   } D_{{\sigma _2}}   D_{{\sigma _3}}   
          \left( 1 + \frac{2 z_{{\sigma _2}}   (z_{{\sigma _3}}    - z_1  )  }
          {   (1 - z_1  ) z_1  }\right) \right]   \right\}\, .\nonumber 
\end{eqnarray}

We have checked that \eqnss{offap}{offapgqq} agree with the corresponding 
spin-correlated splitting functions of Ref.~\cite{cat}
after contracting them with a parent-gluon polarization as in \app{sec:appe},
and after setting the RS parameter $\epsilon=0$.

\section{Four-Parton Forward Clusters}
\label{sec:cluster}

The procedure of Sects.~\ref{sec:nloif} and \ref{sec:nnloif} can be clearly
extended to $n$-parton forward clusters.
In a forward cluster there are one incoming and $n$ outgoing 
partons. Thus,
for purely gluonic clusters there are $2^{n+1}$ helicity configurations.
However, in the high-energy limit two of these are subleading, thus an
$n$-gluon forward cluster  contains $2(2^n-1)$ helicity configurations.
For $n$-parton forward clusters including $\bar q\, q$ pairs,
all the helicity configurations are leading; then an easy counting
yields $2^n$ helicity configurations for the one
including a $\bar q\, q$ pair, $2^{n-1}$ for the one including two
$\bar q\, q$ pairs, and so on. For $n=3$, we obtain the helicity 
configurations dealt with in \sec{sec:nnloif}.

\subsection{The NNNLO impact factor $g\, g^* \rightarrow g\, g\, g\, g$}
\label{sec:nnnloif}

\begin{figure}[t]
\begin{center}
\vspace*{-0.2cm}
\hspace*{0cm}
\epsfxsize=14cm \epsfbox{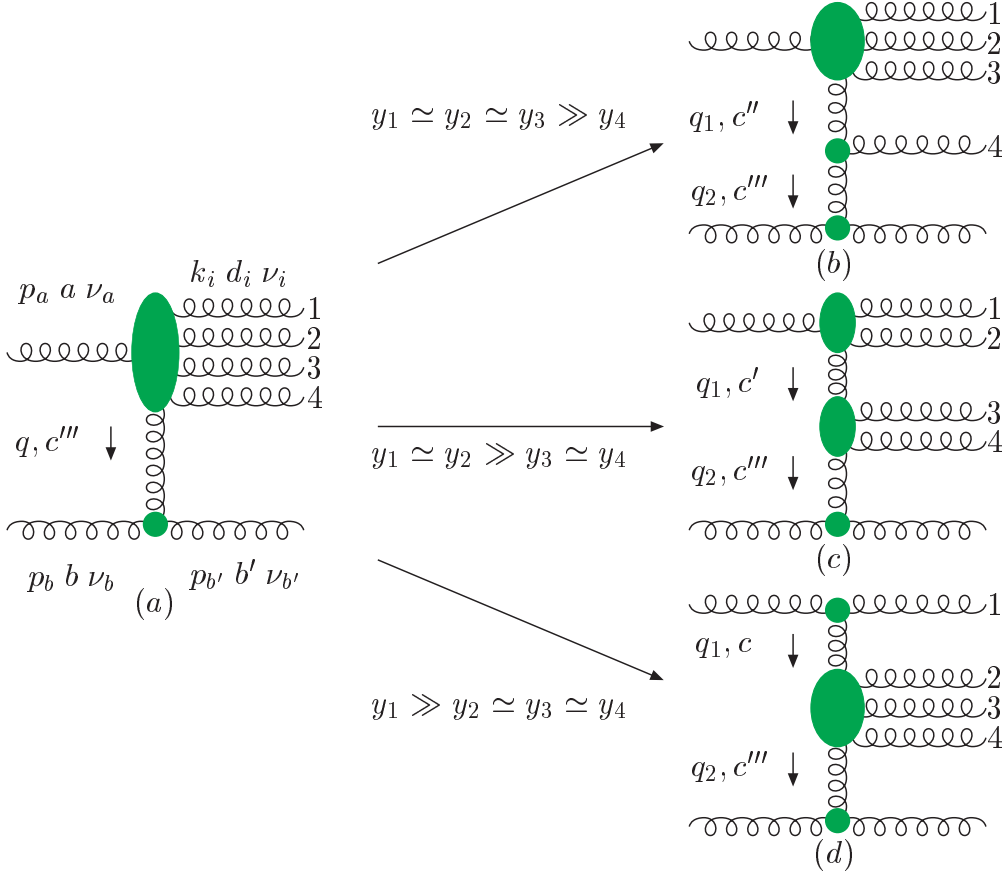}
\vspace*{-0.2cm}
\caption{Amplitude for the production of five gluons, with gluons
$k_1$, $k_2$, $k_3$ and $k_4$ in the forward-rapidity region of gluon $p_a$.}
\label{fig:gggglim}
\end{center}
\end{figure}

Here we analyse in detail the four-gluon forward cluster.
We take the production of five gluons
with momenta $k_1$, $k_2$, $k_3$, $k_4$ and $p_{b'}$ 
in the scattering between two partons of momenta $p_a$ and $p_b$,
and we take partons $k_1$, $k_2$, $k_3$ and $k_4$ in the forward-rapidity
region of parton $p_a$ (\fig{fig:gggglim}a),
\begin{equation}
y_1 \simeq y_2 \simeq y_3 \simeq y_4 \gg y_{b'}\,;\qquad |k_{1\perp}|
\simeq|k_{2\perp}| \simeq|k_{3\perp}| \simeq|k_{4\perp}| \simeq|p_{b'\perp}|\, 
.\label{nnnloreg}
\end{equation}
Using Eqs.~(\ref{GluonDecomp}), (\ref{relations}) and (\ref{ptgluon})
and the subamplitudes of non-PT type,
with four gluons of $+$ helicity and three gluons of $-$ 
helicity~\cite{BeGiKu90}, we obtain
\begin{eqnarray}
& & \A{g\,g \to 5g}(p_a^{\nu_a}, k_1^{\nu_1}, k_2^{\nu_2}, 
k_3^{\nu_3}, k_4^{\nu_4} \sep p_{b'}^{\nu_{b'}}, p_b^{\nu_b}) = 
\label{fourgggg}\\ & & = 
4\, g^5\, {s\over |q_{\perp}|^2}\, 
C^{g;g}(p_b^{\nu_b};p_{b'}^{\nu_{b'}})\, \sum_{\sigma\in S_4} \left[
A^{g;4g}(p_a^{\nu_a}; k_{\sigma_1}^{\nu_{\sigma_1}}, 
k_{\sigma_2}^{\nu_{\sigma_2}}, k_{\sigma_3}^{\nu_{\sigma_3}},
k_{\sigma_4}^{\nu_{\sigma_4}}) \right. \nonumber \\ & &
\tr \left( \lambda^a \lambda^{d_{\sigma_1}} \lambda^{d_{\sigma_2}} 
\lambda^{d_{\sigma_3}} \lambda^{d_{\sigma_4}} \lambda^{b'} \lambda^b 
- \lambda^a \lambda^{d_{\sigma_1}} \lambda^{d_{\sigma_2}} 
\lambda^{d_{\sigma_3}} \lambda^{d_{\sigma_4}} \lambda^b \lambda^{b'}
\right. \nonumber \\ & & \left. \qquad
- \lambda^b \lambda^{b'} \lambda^{d_{\sigma_4}} \lambda^{d_{\sigma_3}} 
\lambda^{d_{\sigma_2}} \lambda^{d_{\sigma_1}} \lambda^a
+ \lambda^{b'} \lambda^b \lambda^{d_{\sigma_4}} \lambda^{d_{\sigma_3}} 
\lambda^{d_{\sigma_2}} \lambda^{d_{\sigma_1}} \lambda^a \right) \nonumber\\ 
& & + B^{g;4g}(p_a^{\nu_a}; k_{\sigma_1}^{\nu_{\sigma_1}}, 
k_{\sigma_2}^{\nu_{\sigma_2}}, k_{\sigma_3}^{\nu_{\sigma_3}},
k_{\sigma_4}^{\nu_{\sigma_4}}) \nonumber\\ & &
\tr \left( \lambda^a \lambda^{d_{\sigma_1}} \lambda^{d_{\sigma_2}} 
\lambda^{d_{\sigma_3}} \lambda^{b'} \lambda^b \lambda^{d_{\sigma_4}} 
- \lambda^a \lambda^{d_{\sigma_1}} \lambda^{d_{\sigma_2}} 
\lambda^{d_{\sigma_3}} \lambda^b \lambda^{b'} \lambda^{d_{\sigma_4}}
\right. \nonumber\\ & & \left. \qquad
- \lambda^b \lambda^{b'} \lambda^{d_{\sigma_3}} \lambda^{d_{\sigma_2}} 
\lambda^{d_{\sigma_1}} \lambda^a \lambda^{d_{\sigma_4}}
+ \lambda^{b'} \lambda^b \lambda^{d_{\sigma_3}} \lambda^{d_{\sigma_2}} 
\lambda^{d_{\sigma_1}} \lambda^a \lambda^{d_{\sigma_4}}
\right) \nonumber\\
& & + D^{g;4g}(p_a^{\nu_a}; k_{\sigma_1}^{\nu_{\sigma_1}}, 
k_{\sigma_2}^{\nu_{\sigma_2}}, k_{\sigma_3}^{\nu_{\sigma_3}},
k_{\sigma_4}^{\nu_{\sigma_4}}) \nonumber\\ & & \left. 
\tr \left( \lambda^a \lambda^{d_{\sigma_1}} \lambda^{d_{\sigma_2}} 
\lambda^{b'} \lambda^b \lambda^{d_{\sigma_4}} \lambda^{d_{\sigma_3}}
- \lambda^a \lambda^{d_{\sigma_3}} \lambda^{d_{\sigma_4}}
\lambda^{b'} \lambda^b \lambda^{d_{\sigma_2}} \lambda^{d_{\sigma_1}} 
\right) \right]\, ,\nonumber
\end{eqnarray}
with the sum over the permutations of the four gluons 1, 2, 3 and 4.
From the PT subamplitudes (\ref{ptgluon}) we obtain the functions of
$(-++++)$ helicities
\begin{eqnarray}
A^{g;4g}(p_a^{\nu_a}; k_1^{\nu_1}, k_2^{\nu_2}, k_3^{\nu_3}, k_4^{\nu_4}) 
&=& C^{g;4g}(p_a^{\nu_a}; k_1^{\nu_1}, k_2^{\nu_2}, k_3^{\nu_3}, k_4^{\nu_4})
A^{\bar\nu}(k_1, k_2, k_3, k_4)\, \label{pt4g}, 
\end{eqnarray}
where $\bar\nu={\rm sign}(\nu_a+\nu_1+\nu_2+\nu_3+\nu_4)$ and
\begin{eqnarray}
A^+(k_1,k_2,k_3,k_4) &=& - 2\sqrt{2}\, {q_\perp\over k_{1\perp}}
\sqrt{x_1\over x_4}\, {1\over \langle 1 2\rangle
\langle 2 3\rangle \langle 3 4\rangle} \nonumber,
\end{eqnarray}
and
\begin{equation}
x_i= \frac{k_i^+}{k_1^+ + k_2^+ + k_3^+ + k_4^+}   
\qquad i=1,2,3,4 \quad (x_1+x_2+x_3+x_4=1)\, .\label{fourfrac}
\end{equation}
As in \eqn{fourpt}, the functions $C^{g;4g}$ are 
\begin{eqnarray}
C^{g;4g}(p_a^{\nu_a}; k_1^{\nu_1}, k_2^{\nu_2},k_3^{\nu_3},k_4^{\nu_4}) &=& 
\left\{
\begin{array}{l}
1{~}\quad \nu_a=- \\
x_i^2 \quad \nu_i=- \quad i=1,2,3,4\\
\end{array}\right.
\quad {\rm with}\; \bar\nu=+, \label{fivept}
\end{eqnarray}
%C^{g;4g}(p_a^-,k_1^+,k_2^+,k_3^+,k_4^+) &=& 1 \nonumber\\
%C^{g;4g}(p_a^+,k_1^-,k_2^+,k_3^+,k_4^+) &=& x_1^2 \label{fivept}\\
%C^{g;4g}(p_a^+,k_1^+,k_2^-,k_3^+,k_4^+) &=& x_2^2 \nonumber\\
%C^{g;4g}(p_a^+,k_1^+,k_2^+,k_3^-,k_4^+) &=& x_3^2 \nonumber\\
%C^{g;4g}(p_a^+,k_1^+,k_2^+,k_3^+,k_4^-) &=& x_4^2 \nonumber\\

From the non-PT subamplitudes~\cite{BeGiKu90} we have obtained the functions of
$(--+++)$ helicities. We do not reproduce them here 
because they are quite lengthy. They are available from the authors upon
request.

Using the U(1) decoupling equations for one and two photons, 
the functions $B$ and $D$ in \eqn{fourgggg} can be written as
\begin{eqnarray}
&& B^{g;4g}(p_a^{\nu_a}; k_1^{\nu_1}, k_2^{\nu_2}, k_3^{\nu_3}, 
k_4^{\nu_4}) = \nonumber\\ &&  \quad- \left[A^{g;4g}(p_a^{\nu_a}; k_1^{\nu_1}, 
k_2^{\nu_2}, k_3^{\nu_3}, k_4^{\nu_4}) 
+ A^{g;4g}(p_a^{\nu_a}; k_1^{\nu_1}, k_2^{\nu_2}, k_4^{\nu_4}, 
k_3^{\nu_3}) \right. \nonumber\\ && \left. \quad
+ \;\;A^{g;4g}(p_a^{\nu_a}; k_1^{\nu_1}, k_4^{\nu_4}, 
k_2^{\nu_2}, k_3^{\nu_3}) + A^{g;4g}(p_a^{\nu_a}; k_4^{\nu_4}, 
k_1^{\nu_1}, k_2^{\nu_2}, k_3^{\nu_3})\right] \label{baaaa}\\ 
& & \nonumber\\ &&
D^{g;4g}(p_a^{\nu_a}; k_1^{\nu_1}, k_2^{\nu_2}, k_3^{\nu_3}, 
k_4^{\nu_4}) = \nonumber\\ &&\quad
\left[A^{g;4g}(p_a^{\nu_a}; k_1^{\nu_1}, k_2^{\nu_2}, 
k_3^{\nu_3}, k_4^{\nu_4}) + A^{g;4g}(p_a^{\nu_a}; k_1^{\nu_1},
k_3^{\nu_3}, k_2^{\nu_2}, k_4^{\nu_4}) \right. \nonumber\\ && \quad 
+ \;\;A^{g;4g}(p_a^{\nu_a}; k_3^{\nu_3}, k_1^{\nu_1}, k_2^{\nu_2}, k_4^{\nu_4})
+ A^{g;4g}(p_a^{\nu_a}; k_1^{\nu_1}, k_3^{\nu_3}, k_4^{\nu_4}, k_2^{\nu_2})
\label{daaaa}\\ && \left. \quad
+ \;\;A^{g;4g}(p_a^{\nu_a}; k_3^{\nu_3}, k_1^{\nu_1}, k_4^{\nu_4}, k_2^{\nu_2})
+ A^{g;4g}(p_a^{\nu_a}; k_3^{\nu_3}, k_4^{\nu_4}, k_1^{\nu_1}, k_2^{\nu_2})
\right]\, .\nonumber
\end{eqnarray}

In the quadruple collinear limit, $k_1 || k_2 || k_3 || k_4$, 
Sect.~\ref{sec:nnnlocoll}, the function $A$ has a triple collinear divergence;
the function $B$, whose gluon 4 is not
color adjacent to gluons 1, 2 and 3, has only a double collinear divergence;
the function $D$, where gluon 1 is adjacent to 2 and gluon 3 is adjacent to 4
but the pairs are not adjacent one to another, has two single
collinear divergences.

Using Eqs.~(\ref{baaaa}) and (\ref{daaaa}), we can rewrite \eqn{fourgggg} as
\begin{eqnarray}
& & \A{g\,g \to 5g}(p_a^{\nu_a}, k_1^{\nu_1}, k_2^{\nu_2}, 
k_3^{\nu_3}, k_4^{\nu_4}  \sep p_{b'}^{\nu_{b'}}, p_b^{\nu_b}) = 
\label{nnnlogggg}\\ & & 
2\, s\, \left\{ (ig)^4\, \sum_{\sigma\in S_4}
f^{a d_{\sigma_1} c } f^{c d_{\sigma_2} c'} f^{c' d_{\sigma_3} c''}
f^{c'' d_{\sigma_4} c'''}  
A^{g;4g}(p_a^{\nu_a}; k_{\sigma_1}^{\nu_{\sigma_1}}, 
k_{\sigma_2}^{\nu_{\sigma_2}}, k_{\sigma_3}^{\nu_{\sigma_3}}, 
k_{\sigma_4}^{\nu_{\sigma_4}}) \right\}
\nonumber \\& & \times
{1\over t}\, \left[ig\, f^{b b' c'''} 
C^{g;g}(p_b^{\nu_b};p_{b'}^{\nu_{b'}})
\right]\, ,\nonumber
\end{eqnarray}
where the NNLO impact factor $g^*\, g \rightarrow g\, g\, g\, g$
is enclosed in curly brackets, and includes
30 helicity configurations, in agreement with the counting above.

\subsection{NNNLO impact factors in the high-energy limit}
\label{sec:nnnlohigh}

The amplitude (\ref{nnnlogggg})
has been computed in the kinematic limit (\ref{nnnloreg}), in which it
factorizes into a four-gluon cluster and
a LO impact factor connected by a gluon exchanged
in the cross channel. In the limits
$y_1\simeq y_2 \simeq y_3\gg y_4$ or $y_1\simeq y_2\gg y_3\simeq y_4$, 
or $y_1\gg y_2 \simeq y_3\simeq y_4$, \eqn{nnnlogggg} must
factorize further into a NNLO impact factor 
or into a NLO impact factor times a NLO Lipatov vertex, or into a
NNLO Lipatov vertex (\fig{fig:gggglim}), respectively.
While the first two limits constitute necessary consistency checks, the
last one allows us to derive the so far unknown NNLO Lipatov vertex for 
the production of three gluons along the ladder.

In the limit, $y_1\simeq y_2\simeq y_3\gg y_4$, the NNNLO impact factor, 
$g^*\, g \rightarrow g\, g\, g\, g$, in \eqn{nnnlogggg}
factorizes into a NNLO impact factor, $g^*\, g \rightarrow g\, g\, g$, 
\eqn{nnloggg}, convoluted with a multi-Regge ladder (\fig{fig:gggglim}a)
\begin{eqnarray}
& &  \lim_{y_1\simeq y_2 \simeq y_3\gg y_4}
\left\{ (ig)^4\, \sum_{\sigma\in S_4}
f^{a d_{\sigma_1} c } f^{c d_{\sigma_2} c'} f^{c' d_{\sigma_3} c''}
f^{c'' d_{\sigma_4} c'''}  
A^{g;4g}(p_a^{\nu_a}; k_{\sigma_1}^{\nu_{\sigma_1}}, 
k_{\sigma_2}^{\nu_{\sigma_2}}, k_{\sigma_3}^{\nu_{\sigma_3}}, 
k_{\sigma_4}^{\nu_{\sigma_4}}) \right\} \nonumber \\
& & = \left\{ (ig)^3\, \sum_{\sigma\in S_3}
f^{a d_{\sigma_1} c } f^{c d_{\sigma_2} c'} f^{c' d_{\sigma_3} c''}  
A^{g;3g}(p_a^{\nu_a}; k_{\sigma_1}^{\nu_{\sigma_1}}, 
k_{\sigma_2}^{\nu_{\sigma_2}}, k_{\sigma_3}^{\nu_{\sigma_3}}) \right\}
\nonumber \\ & & \times {1\over t_1} 
\left[i g\, f^{c''d_4c'''}\, C^g(q_1,k_4^{\nu_4},q_2)\right]
%\, {1\over t_2}\, 
%\left[i g\, f^{bb'c'''}\, C^{g;g}(p_b^{\nu_b};p_{b'}^{\nu_{b'}}) \right]\,
%& &\A{g\,g \to 5g}(p_a^{\nu_a}, k_1^{\nu_1}, k_2^{\nu_2} \sep
%k_3^{\nu_3}, k_4^{\nu_4}  \sep p_{b'}^{\nu_{b'}}, p_b^{\nu_b})=
%\A{g\,g \to 5g}(p_a^{\nu_a}, k_1^{\nu_1}, k_2^{\nu_2}, 
%k_3^{\nu_3}, k_4^{\nu_4}  \sep p_{b'}^{\nu_{b'}}, p_b^{\nu_b})  
,\label{4forwlim}
\end{eqnarray}
with $q_1=-(p_a+k_1+k_2+k_3)$, $q_2=p_{b'}+p_b$, 
and with LO Lipatov vertex $C^g(q_1,k_3^{\nu_3},q_2)$, \eqn{lip}.

\noindent
In the limit $y_1\simeq y_2\gg y_3\simeq y_4$, the NNNLO impact factor in
\eqn{nnnlogggg}
factorizes into a NLO impact factor, $g^*\, g \rightarrow g\, g$,
\eqn{nllfg}, times a NLO Lipatov vertex for
production of two gluons $g^*\, g^* \rightarrow g\, g$ (\ref{kosc}), convoluted
with a multi-Regge ladder (\fig{fig:gggglim}b)
\begin{eqnarray}
& & \lim_{y_1\simeq y_2\gg y_3\simeq y_4}
\left\{ (ig)^4\, \sum_{\sigma\in S_4}
f^{a d_{\sigma_1} c } f^{c d_{\sigma_2} c'} f^{c' d_{\sigma_3} c''}
f^{c'' d_{\sigma_4} c'''}  
A^{g;4g}(p_a^{\nu_a}; k_{\sigma_1}^{\nu_{\sigma_1}}, 
k_{\sigma_2}^{\nu_{\sigma_2}}, k_{\sigma_3}^{\nu_{\sigma_3}}, 
k_{\sigma_4}^{\nu_{\sigma_4}}) \right\} \nonumber\\
& & = \left\{ 
(ig)^2\, \sum_{\sigma\in S_2} f^{a d_{\sigma_1} c } f^{c d_{\sigma_2} c'} 
A^{g;gg}(p_a^{\nu_a}; k_{\sigma_1}^{\nu_{\sigma_1}}, 
k_{\sigma_2}^{\nu_{\sigma_2}}) \right\} \nonumber\\ & & 
\times {1\over t_1} 
\left\{ (ig)^2\, \sum_{\sigma\in S_2} f^{c' d_{\sigma_3} c'' } 
f^{c'' d_{\sigma_4} c'''} 
A^{gg}(q_1, k_{\sigma_3}^{\nu_{\sigma_3}}, k_{\sigma_4}^{\nu_{\sigma_4}},q_2) 
\right\}\, ,\label{2centrlim}
%& &\hspace{-1cm} \A{g\,g \to 5g}(p_a^{\nu_a}, k_1^{\nu_1}, k_2^{\nu_2} \sep
%k_3^{\nu_3}, k_4^{\nu_4} \sep p_{b'}^{\nu_{b'}}, p_b^{\nu_b}) 
%\A{g\,g \to 5g}(p_a^{\nu_a}, k_1^{\nu_1}, k_2^{\nu_2},
%k_3^{\nu_3}, k_4^{\nu_4} \sep p_{b'}^{\nu_{b'}}, p_b^{\nu_b})   \nonumber\\
%\\ & & \times {1\over t_2}\, 
%\left[i g\, f^{bb'c'''}\, C^{g;g}(p_b^{\nu_b};p_{b'}^{\nu_{b'}}) \right]\,
%,\nonumber
\end{eqnarray}
with $q_1=-(p_a+k_1+k_2)$.

\subsection{The NNLO Lipatov vertex}
\label{sec:nnlolip}

In the limit $y_1\gg y_2 \simeq y_3\simeq y_4$, the NNNLO impact factor in
\eqn{nnnlogggg} factorizes into a NNLO Lipatov vertex convoluted
with a multi-Regge ladder (\fig{fig:gggglim}c)
\begin{eqnarray}
& & \lim_{y_1\gg y_2 \simeq y_3\simeq y_4}
\left\{ (ig)^4\, \sum_{\sigma\in S_4}
f^{a d_{\sigma_1} c } f^{c d_{\sigma_2} c'} f^{c' d_{\sigma_3} c''}
f^{c'' d_{\sigma_4} c'''}  
A^{g;4g}(p_a^{\nu_a}; k_{\sigma_1}^{\nu_{\sigma_1}}, 
k_{\sigma_2}^{\nu_{\sigma_2}}, k_{\sigma_3}^{\nu_{\sigma_3}}, 
k_{\sigma_4}^{\nu_{\sigma_4}}) \right\} \nonumber\\
& & =  \left[i g\, f^{ad_1c}\, C^{g;g}
(p_a^{\nu_a};k_1^{\nu_1}) \right] \label{3centrlim}\\ & & \times {1\over t_1}
\left\{ (ig)^3\, \sum_{\sigma\in S_3} f^{c d_{\sigma_2} c' } 
f^{c' d_{\sigma_3} c''} f^{c'' d_{\sigma_4} c'''} 
A^{3g}(q_1, k_{\sigma_2}^{\nu_{\sigma_2}}, k_{\sigma_3}^{\nu_{\sigma_3}},
k_{\sigma_4}^{\nu_{\sigma_4}},q_2) \right\}\nonumber
%& &\hspace{-1cm}  \A{g\,g \to 5g}(p_a^{\nu_a}, k_1^{\nu_1}\sep k_2^{\nu_2}, 
%k_3^{\nu_3}, k_4^{\nu_4} \sep p_{b'}^{\nu_{b'}}, p_b^{\nu_b})   \nonumber\\
%\A{g\,g \to 5g}(p_a^{\nu_a}, k_1^{\nu_1}, k_2^{\nu_2},
%k_3^{\nu_3}, k_4^{\nu_4} \sep p_{b'}^{\nu_{b'}}, p_b^{\nu_b})   \nonumber\\
% \\ & & \times {1\over t_2}\, 
%\left[i g\, f^{bb'c''}\, C^{g;g}(p_b^{\nu_b};p_{b'}^{\nu_{b'}}) \right]\,
,\nonumber
\end{eqnarray}
with the NNLO Lipatov vertex, $g^*\, g^* \rightarrow g\, g\, g$, for the 
production of three gluons $k_2$, $k_3$ and $k_4$ enclosed in curly brackets
in the right hand side, with
\begin{eqnarray}
&& A^{3g}(q_1, k_2^+, k_3^+, k_4^+,q_2) = 
- 2  \sqrt{2} \sqrt{\frac{\x(2) }{\x(4) }}
\frac{1}{\br(2,3)   \br(3,4)  } 
\frac{  \qoperpStar  \qtperp}{  k_{2\perp} } \label{nnlolip1} \\[20pt]
%%%%%%%%%%%%%%%%%%%%%%%%%%%%%%%%%%%%%%%
&& A^{3g}(q_1, k_2^-, k_3^+, k_4^+,q_2) = 2 \sqrt{2}   \left\{
-{\frac{\qoperpt  k_{2\perp}   
      {(\PperpStar(3) \!)^2}  
      \left( \qtperpStar + \PperpStar(4) \right)   
      \x(2) }{s_{4bb'}  \PperpStar(2)   \s(2,3)   
      \left( \Pperpt(3)   \x(2) + 
        \Pperpt(2)   \x(3) \right) }} \right.
\nonumber\\[7pt]&&+
\frac{k_{2\perp} }{s_{4bb'}  \br(3,4)   
     \PperpStar(2)   \s(2,3)   {\sqrt{\x(3) }}}      
     \left[ \left( \qoperpStar  \qtperpStar  
           \PperpStar(3)    \br(3,4)  + 
          \br(2,4)   
           \left( \qoperpStar - 
             \PperpStar(2) \right)   
           \left( \qtperpStar + 
             \PperpStar(4) \right)   \sq(2,3)
           \right)   {\sqrt{\x(3) }}
\right.\nonumber\\[5pt]&& \left. 
\hspace*{1cm} -  \PperpStar(3)   
        \left( \left( \qoperpStar - 
             \PperpStar(2) \right)   \s(2,3) + 
          \qoperpStar  \s(3,4) \right)   {\sqrt{\x(4) }} \right]
\nonumber\\[7pt]&&+
{\frac{\qoperpt  k_{2\perp}   
     {(\PperpStar(3) \!)^2}  \x(2)   {\sqrt{\x(3) }}}{
\br(3,4)   \PperpStar(2)   \s(2,3)   
     \left( \Pperpt(3)   \x(2) + 
       \Pperpt(2)   \x(3) \right)   {\sqrt{\x(4) }}}}
\nonumber\\[7pt]&&+
\frac{1}{\br(3,4)   \PperpStar(2)   \s(2,3)   \sqrt{\x(3) \x(4) }} 
\left( - \Pperpt(2)   \PperpStar(3) + {\sqrt{\x(2) }}  
\left( - \qoperp  
           \left( \qoperpStar - \PperpStar(2)
             \right)   \PperpStar(3)   {\sqrt{\x(2) }} \right. \right. 
\nonumber\\[5pt]&&\hspace*{1cm} - \qoperpStar  
         \left( - k_{2\perp}    \PperpStar(3)   {\sqrt{\x(2) }}  + 
           \br(2,3)   \PperpStar(3)   
            {\sqrt{\x(3) }} + k_{2\perp}   
            \sq(2,3)   {\sqrt{\x(3) }} + 
           \br(2,4)   \PperpStar(3)   
            {\sqrt{\x(4) }} \right)
\nonumber\\[5pt]&& \hspace*{1cm} + \left.\left.
        \PperpStar(2)   
         \left( \left( \br(2,3)   
               \PperpStar(3) + 
              k_{2\perp}   \sq(2,3) \right)   
            {\sqrt{\x(3) }} + \br(2,4)   
            \PperpStar(3)   {\sqrt{\x(4) }} \right)  \right) \right)
\nonumber\\[7pt]&&+
{\frac{\qtperpt  \x(2)   
     \left( \br(2,3)   {\sqrt{\x(3) }} + 
       \br(2,4)   {\sqrt{\x(4) }} \right) }{\br(2,
      3)   \br(3,4)   \t(2,3,4)   {\sqrt{\x(4) }}}}+
{\frac{\qtperp  k_{2\perp}   
     {{\left( \qoperpStar - \PperpStar(2)
          \right) }^2}  \left( \qtperpStar  {\sqrt{\x(3) }} + 
       \sq(3,4)   {\sqrt{\x(4) }} \right) }{s_{2aa'}  
     s_{4bb'}  \br(3,4)   
     \PperpStar(2)   {\sqrt{\x(4) }}}}
\nonumber\\[7pt]&&+
\frac{\qoperpt  (k_{2\perp} \!)^2  
     \PperpStar(3)   \PperpStar(4) 
            {\sqrt{ x_3 }}}{\br(3,4)   
     \s(2,3)   \left( \qomqtperpt + \t(2,3,4) \right)   
     \left( \Pperpt(3)   \x(2) + 
       \Pperpt(2)   \x(3) \right) \sqrt{x_4} }
\nonumber\\[7pt]&&+
{\frac{\qoperpt  k_{2\perp}   
     \sq(3,4)     
     \left( \br(2,4)   \PperpStar(4)   
        {\sqrt{\x(3) }} + \br(2,3)   
        \PperpStar(3)   {\sqrt{\x(4) }} \right) }
     { {\sqrt{\x(2) \x(3) \x(4) }}  \br(3,4)   \s(2,3)   \t(2,3,4)   
     \left( \qomqtperpt + \t(2,3,4) \right) }}
\nonumber\\[7pt]&&+ 
\frac{\sq(3,4) }{\br(3,4)   \s(2,3)   \t(2,3,4)   
     {\sqrt{\x(2) }}  \x(3)    \x(4) } 
\left( - \qoperpStar  k_{2\perp} \x(3) \x(4)   
\left( \br(2,3)   {\sqrt{\x(3) }} + 
            \br(2,4)   {\sqrt{\x(4) }} \right) \right.
\nonumber\\[5pt]&& 
%\hspace*{1cm} 
\left.
    + \qtperp  {\sqrt{x_2}}  
        \left( \Pperp(4)   
           \left( \qtperpStar + 
             \PperpStar(4) \right)   \x(2)   \x(3) + 
          \left( \Pperpt(3)   \x(2) + 
             k_{2\perp}   
           (\PperpStar(2) -\qoperpStar)   \x(3) \right)   \x(4)
           \right)  \right)  \left. \frac{}{} \!\! \right\}
\label{nnlolip2} 
\end{eqnarray}
%%%%%%%%%%%%%%%%%%%%%%%%%%%%%%%%%%%
\vspace*{0.5cm}
%%%%%%%%%%%%%%%%%%%%%%%%%%%%%%%%%%%
\begin{eqnarray}
&& A^{3g}(q_1, k_2^+, k_3^-, k_4^+,q_2) = 2 \sqrt{2} \times \nonumber\\[7pt]
&&\left[
{\frac{\qtperpt  \Pperp(3)   {\sqrt{\x(2) }}  
      {\sqrt{\x(3) }}}{\br(2,3)   k_{2\perp}   
      \s(3,4)   \left( 1 - \x(2) \right) }} \right.
+{\frac{\qtperpt  \left( \qoperp - 
        k_{2\perp} \right)   \PperpStar(2)   \x(3) }
      {s_{2aa'}  k_{2\perp}   \s(3,4)   
      \left(1 - \x(2) \right) }}-
{\frac{\qtperpt  {\sqrt{\x(2) x_3^3 }}}
   {\br(2,3)   \s(3,4)   \left( 1 - \x(2) \right) }}
\nonumber\\[7pt]&&-{\frac{\qoperpt  {(\Pperp(3) \!)^2}  
      \PperpStar(2) \left( \qtperpStar + \PperpStar(4) \right)
          \x(2) }{s_{4bb'}  k_{2\perp}   \s(2,3)   
      \left( \Pperpt(3)   \x(2) + 
        \Pperpt(2)   \x(3) \right) }}
\nonumber\\[7pt]&&+
{\frac{ (\Pperp(3) \!)^2   {\sqrt{\x(2) }}  
     \left( - \left( \qtperpStar + 
            \PperpStar(4) \right)   \sq(2,3)   
          {\sqrt{\x(3) }}   + 
       \qoperpStar  
        \left( \qtperpStar  {\sqrt{\x(2) }} + 
          \sq(2,4)   {\sqrt{\x(4) }} \right)  \right) }{
     s_{4bb'}  k_{2\perp}   \s(2,3)   \x(3) }}
\nonumber\\[7pt]&&+
{\frac{\qoperpStar  \Pperp(3)   
     {{\sq(2,4) }^2}  
     \left( \br(2,3)   
        \left( \qoperp - k_{2\perp} \right)   
        {\sqrt{\x(2) }} + \br(3,4)   k_{2\perp}   
        {\sqrt{\x(4) }} \right) }{k_{2\perp}   \s(2,3)   \s(3,4)   
     \t(2,3,4)   {\sqrt{\x(3) }}}}
\nonumber\\[7pt]&&
-{\frac{\Pperp(3)   \PperpStar(2)   
      \left(  \qtperpStar k_{2\perp} {\sqrt{\x(2) }}  - 
\qoperp   \left( \qtperpStar + \PperpStar(4) \right)   {\sqrt{\x(2) }}  
            + k_{2\perp}   \sq(2,4)   
         {\sqrt{\x(4) }} \right) }{s_{2aa'}  s_{4bb'}  
      k_{2\perp}   {\sqrt{\x(2) }}}}
\nonumber\\[7pt]&&
-{\frac{\PperpStar(2)   {\sqrt{\x(3) }}  
      \left( - \qoperp  \qtperp  
           \PperpStar(4)   {\sqrt{\x(2) \x(3) }} 
         - \left( \qoperp - 
                k_{2\perp} \right)   \Pperp(3)   
              \sq(3,4)   {\sqrt{\x(2) \x(4) }}  + 
           \qtperp  k_{2\perp}   
            \sq(2,4)   {\sqrt{\x(3) \x(4) }}  
         \right) }{s_{2aa'}  k_{2\perp}   \s(3,4)   
      {\sqrt{\x(2) }}  \x(4) }}
\nonumber\\[7pt]&&+
{\frac{\qoperpt  {(\Pperp(3) \!)^2}  
     \PperpStar(2)   \x(2)   {\sqrt{\x(3) }}}{\br(3,
      4)   k_{2\perp}   \s(2,3)   
     \left( \Pperpt(3)   \x(2) + 
       \Pperpt(2)   \x(3) \right)   {\sqrt{\x(4) }}}}
+
{\frac{\qtperpt  \sq(2,4)   \x(3)   
     \left( -\br(2,3)   {\sqrt{\x(2) }}   + 
       \br(3,4)   {\sqrt{\x(4) }} \right) }{\br(2,
      3)   \s(3,4)   \t(2,3,4)   {\sqrt{\x(4) }}}}
\nonumber\\[7pt]&&+
\frac{\qtperp  {{\sq(2,4) }^2}  {\sqrt{\x(3) }}  
     \left( \br(3,4)   
        \left( \qtperpStar + \PperpStar(4)
           \right)   {\sqrt{\x(2) }} - 
       \br(2,3)   \PperpStar(2)   {\sqrt{\x(4) }}
        \right) }{\s(2,3)   \s(3,4)   \t(2,3,4)   {\sqrt{\x(2) \x(4) }}}
\nonumber\\[7pt]&&
-{\frac{\qoperpt  {(\Pperp(3) \!)^3}  
      \PperpStar(2)   \PperpStar(4) x_2
      }{ {\sqrt{\x(3) \x(4) }}  \br(3,4)   
      k_{2\perp}   \s(2,3)   
      \left( \qomqtperpt + \t(2,3,4) \right)   
      \left( \Pperpt(3)   \x(2) + 
        \Pperpt(2)   \x(3) \right) }}
\nonumber\\[7pt]&&
-{\frac{\qoperpt  {(\Pperp(3) \!)^2}  
      \sq(2,4)      
      \left( \br(3,4)   \PperpStar(4)   
         {\sqrt{\x(2) }} - \br(2,3)   
         \PperpStar(2)   {\sqrt{\x(4) }} \right)   }
      { x_3 {\sqrt{\x(4) }} \br(3,4)   k_{2\perp}   \s(2,3)   \t(2,3,4)   
      \left( \qomqtperpt + \t(2,3,4) \right) }}
\nonumber\\[7pt]&&+
\frac{1}{k_{2\perp} \s(2,3) \s(3,4) \sqrt{\x(2) \x(4) } \x(3) }
\left( \x(3)   \left( -  \x(2) 
\left( \Pperp(3)  \qtperp \PperpStar(3) \sq(2,4) + 
             \Pperp(3) \sq(2,3) \sq(3,4) \right) 
\right. \right.  \nonumber\\[5pt]&&\hspace*{1cm}+ \left.
        \qtperp  \Pperp(3) \PperpStar(2)   \sq(3,4)   
         {\sqrt{\x(2) \x(3) }} - 
        \qtperp  \Pperpt(2)   
         \sq(2,4)   \x(3) \right)
  \nonumber\\[5pt]&&\hspace*{1cm} +\left.
     \qoperpStar  \Pperp(3)   \x(2)   
      \left( \qtperp  \sq(2,4)   \x(3) + 
        \Pperp(3)   
          \sq(3,4)   {\sqrt{\x(2) \x(3) }} + \Pperp(3)
           \sq(2,4)   \x(4)  \right) \right) 
\left. \frac{}{} \right] \label{nnlolip3}
\end{eqnarray}
%%%%%%%%%%%%%%%%%%%%%%%%%%%%%%%%%%%
\vspace*{0.5cm}
%%%%%%%%%%%%%%%%%%%%%%%%%%%%%%%%%%
\begin{eqnarray}
&& A^{3g}(q_1, k_2^+, k_3^+, k_4^-,q_2) = 2 \sqrt{2}   \left[
\frac{\PperpStar(2) }{s_{2aa'}  
      \br(2,3)   k_{2\perp}   \s(3,4)   {\sqrt{\x(2) }}}
\left( \qoperp  \qtperp  
              \br(2,3)   \PperpStar(3) {\sqrt{\x(2) }} 
\right. \right. \nonumber\\[5pt]
&&\hspace*{1cm}- \left. \br(2,4)   \sq(3,4) {\sqrt{\x(2) }}   
              \left( \qoperp - k_{2\perp} \right)
                  \left( \qtperp + 
                \Pperp(4) \right)  + 
        k_{2\perp}   {\sqrt{\x(3) }}   
         \left(  \qtperp  \s(2,3) + 
           \left( \qtperp + \Pperp(4) \right)   
            \s(3,4) \right)  \right)
\nonumber\\[7pt]&&+
{\frac{\qoperpStar  {{\left( \qtperp + 
          \Pperp(4) \right) }^2}  
     \left( \qoperp  \PperpStar(3)   {\sqrt{\x(2) }} - 
       k_{2\perp}   \sq(2,3)   {\sqrt{\x(3) }}
       \right) }{s_{2aa'}  s_{4bb'}  
     \br(2,3)   k_{2\perp}   {\sqrt{\x(3) }}}}
+{\frac{\qtperpt  \Pperp(3)   {\sqrt{\x(2) \x(3) }}}{\br(2,3)   k_{2\perp}   
      \s(3,4)   \left( 1 - \x(2) \right) }}
\nonumber\\[7pt]&&+
{\frac{\qoperpt  {(\Pperp(4) \!)^2}    
     \left( \br(3,4)   \PperpStar(3)   
        {\sqrt{\x(2) }} + \br(2,4)   
        \PperpStar(2)   {\sqrt{\x(3) }} \right)   }
     { x_4 {\sqrt{\x(3) }} \br(2,3)   \br(3,4)   
     k_{2\perp}   \t(2,3,4)   
     \left( \qomqtperpt + \t(2,3,4) \right) }}
+{\frac{\qtperpt  \left( \qoperp - 
        k_{2\perp} \right)   \PperpStar(2)   \x(3) }
      {s_{2aa'}  k_{2\perp}   \s(3,4)   
      \left( 1 - \x(2) \right) }}
\nonumber\\[7pt]&&+
{\frac{s_{4bb'}  \Pperp(4)   {\sqrt{\x(2) }}  
      {\sqrt{\x(3) }} + \left( \qtperp + 
        \Pperp(4) \right)   
      \left( \left( \br(3,4)   \PperpStar(3) + 
           \Pperp(4)   \sq(3,4) \right)   
         {\sqrt{\x(2) }} + \br(2,4)   
         \PperpStar(2)   {\sqrt{\x(3) }} \right)   {\sqrt{\x(4) }}}
     {\br(2,3)   k_{2\perp}   \s(3,4)   \x(4) }}
\nonumber\\[7pt]&&+
{\frac{\qtperpt  \sq(2,3)   
     \left( \br(2,4)   {\sqrt{\x(2) }} + 
       \br(3,4)   {\sqrt{\x(3) }} \right)   {\sqrt{\x(4) }}}{
     \br(2,3)   \s(3,4)   \t(2,3,4) }}
+{\frac{\qtperpt  {\sqrt{\x(2) \x(3) }}  \x(4) }
    {\br(2,3)   \s(3,4)   \left( 1 - \x(2) \right) }}
\nonumber\\[7pt]&&+
\frac{\sq(2,3) }{\br(2,3)   k_{2\perp}   
     \s(3,4)   \t(2,3,4)   \x(4) }
     \left( \qoperpt  {(\Pperp(4) \!)^2}  \x(2) + 
       \qtperp  \left( {\frac{\br(2,4)   
              \Pperpt(2) }{{\sqrt{\x(2) }}}} + 
          {\frac{\br(3,4)   k_{2\perp}   
              \PperpStar(3) }{{\sqrt{\x(3) }}}} \right)   
        \sqrt{ x_4^3 } \right.
\nonumber\\[7pt]&&\hspace*{1cm} - \left.
       \qoperpStar  k_{2\perp}    
        \Pperp(4)   
        \left( \Pperp(4) + \qtperp  \x(4) \right)
           \right) \left. \frac{}{} \right]\, ,\label{nnlolip4}
\end{eqnarray}
where in \eqnss{nnlolip2}{nnlolip4} we have used the
three-particle invariants, $s_{2aa'} = (k_2-q_1)^2$ and 
$s_{4bb'} = (k_4+q_2)^2$.

\eqn{3centrlim} must not diverge more rapidly than $1/|q_{i\perp}|$ 
for $|q_{i\perp}|\rightarrow 0$, with $i=1,2$, in order for the related 
cross section not to diverge more than logarithmically.
Since \eqn{3centrlim} is proportional to $1/|q_{i\perp}|^2$,
the NNLO Lipatov vertex must be at least linear in $|q_{i\perp}|$,
\begin{equation}
\lim_{|q_{i\perp}|\rightarrow 0} A^{3g}(q_1,k_1^{\nu_1},k_2^{\nu_2},
k_3^{\nu_3},q_2) = O(|q_{i\perp}|)\, ,\label{aqlim}
\end{equation}
which is fulfilled by \eqnss{nnlolip1}{nnlolip4}.

As a consistency check on \eqn{3centrlim}, in the further limits 
$y_2 \gg y_3\simeq y_4$ or $y_2 \simeq y_3\gg y_4$, the NNLO Lipatov
vertex in \eqn{3centrlim}
must factorize into a NLO Lipatov vertex convoluted
with a multi-Regge ladder,
\begin{eqnarray}
& & \lim_{y_2 \gg y_3\simeq y_4}
\left\{ (ig)^3\, \sum_{\sigma\in S_3} f^{c d_{\sigma_2} c' } 
f^{c' d_{\sigma_3} c''} f^{c'' d_{\sigma_4} c'''} 
A^{3g}(q_1, k_{\sigma_2}^{\nu_{\sigma_2}}, k_{\sigma_3}^{\nu_{\sigma_3}},
k_{\sigma_4}^{\nu_{\sigma_4}},q_2) \right\} \label{forwcentrlim}\\& & =
\left[i g\, f^{cd_2c'}\, C^g(q_1,k_2^{\nu_2},q_{12})\right]  
{1\over t_{12}}\, \left\{ (ig)^2\, \sum_{\sigma\in S_2} 
f^{c' d_{\sigma_3} c''} f^{c'' d_{\sigma_4} c'''} 
A^{gg}(q_{12}, k_{\sigma_3}^{\nu_{\sigma_3}},
k_{\sigma_4}^{\nu_{\sigma_4}},q_2) \right\},\nonumber
\end{eqnarray}
with $q_{12}=q_1-k_2$, 
and
\begin{eqnarray}
& & \lim_{y_2 \simeq y_3\gg y_4}
\left\{ (ig)^3\, \sum_{\sigma\in S_3} f^{c d_{\sigma_2} c' } 
f^{c' d_{\sigma_3} c''} f^{c'' d_{\sigma_4} c'''} 
A^{3g}(q_1, k_{\sigma_2}^{\nu_{\sigma_2}}, k_{\sigma_3}^{\nu_{\sigma_3}},
k_{\sigma_4}^{\nu_{\sigma_4}},q_2) \right\}\\& & =
 \left\{ (ig)^2\, \sum_{\sigma\in S_2} f^{c d_{\sigma_2} c' } 
f^{c' d_{\sigma_3} c''} 
A^{gg}(q_1, k_{\sigma_2}^{\nu_{\sigma_2}}, k_{\sigma_3}^{\nu_{\sigma_3}},
q_{12}) \right\}\, \label{2forwcentrlim} {1\over t_{12}}\,
\left[i g\, f^{c''d_4c'''}\, C^g(q_{12},k_4^{\nu_4},q_2)\right]
,\nonumber
\end{eqnarray}
with $q_{12}=q_2+k_4$.

In the triple collinear limit, $k_2 = z_2P$, $k_3 = z_3P$ and $k_4 = z_4P$, 
with $z_2+z_3+z_4=1$, the coefficients of the NNLO Lipatov 
vertex (\ref{nnlolip1})-(\ref{nnlolip4}) reduce to the splitting functions 
(\ref{triplea}), and
amplitude (\ref{3centrlim}) factorizes into a multi-Regge amplitude
(\ref{three}) times a double-collinear factor (\ref{trisplit})
\begin{eqnarray} 
& & \lim_{k_2 || k_3 || k_4} \A{g\,g \to 5g}
(p_a^{\nu_a}, k_1^{\nu_1} \sep
k_2^{\nu_2}, k_3^{\nu_3}, k_4^{\nu_4} \sep p_{b'}^{\nu_{b'}}, p_b^{\nu_b}) = 
\nonumber \\ &&
\sum_\nu \A{g g \to 3g}(p_a^{\nu_a}, k_1^{\nu_1} \sep P^{\nu} \sep
p_{b'}^{\nu_{b'}}, p_b^{\nu_b}) \cdot
{\rm Split}_{-\nu}^{g \to 3g}(k_1^{\nu_1}, k_2^{\nu_2}, k_3^{\nu_3})\,
.\nonumber
\end{eqnarray}

\subsection{NNNLO impact factors in the quadruple collinear limit}
\label{sec:nnnlocoll}

In the quadruple collinear limit, $k_i = z_i P$, with $z_1+z_2+z_3+z_4=1$
a generic amplitude is expected to factorize as
\begin{eqnarray}
\lefteqn{
\lim_{k_1 || k_2 || k_3 || k_4} \A{... d_1 d_2 d_3 d_4...}(..., k_1^{\nu_1}, 
k_2^{\nu_2}, k_3^{\nu_3}, k_4^{\nu_4}, ...) } \nonumber\\ & &
= \sum_\nu \A{... c ...}(..., P^{\nu}, ...) \cdot
{\rm Split}_{-\nu}^{f\to f_1f_2f_3f_4}(k_1^{\nu_1}, 
k_2^{\nu_2},k_3^{\nu_3},k_4^{\nu_4})\, .\label{quadricoll}
\end{eqnarray}
Accordingly, we show that we can write \eqn{nnnlogggg} as
\begin{eqnarray}
\lefteqn{\lim_{k_1 || k_2 || k_3 || k_4} \A{g g \to 5g}(p_a^{\nu_a}, 
k_1^{\nu_1}, k_2^{\nu_2}, k_3^{\nu_3}, k_4^{\nu_4} \sep p_{b'}^{-\nu_b}, 
p_b^{\nu_b}) } \nonumber\\ & & = 
\A{g g \to g g}(p_a^{\nu_a}, P^{-\nu_a} \sep p_{b'}^{-\nu_b}, 
p_b^{\nu_b}) \cdot {\rm Split}_{\nu_a}^{g \to 4g}
(k_1^{\nu_1}, k_2^{\nu_2}, k_3^{\nu_3}, k_4^{\nu_4})\, ,\label{hequadricoll}
\end{eqnarray}
by taking the quadruple collinear limit of the NNNLO impact factor.

In the quadruple collinear limit, the functions $A^{g;4g}$ of \eqn{fivept}
yield a cubic divergence as $s_{1234}=(k_1+k_2+k_3+k_4)^2\to 0$
or $s_{ijk}\to 0$, or $s_{ij}\to 0$ with $i,j,k=1,2,3,4$. 
Analogously to Sect.~\ref{sec:nnlocoll}, a function $A^{g;4g}$ differs from its 
reflection by a term which contains only a quadratic divergence in the
vanishing invariants. 
Using this property and Eqs.~(\ref{baaaa}) and (\ref{daaaa}), we obtain a 
reflection identity and dual Ward identities,
up to quadratically divergent terms,
\begin{eqnarray}
&&A^{g;4g}(p_a^{\nu_a};k_1^{\nu_1},k_2^{\nu_2},k_3^{\nu_3},k_4^{\nu_4})=
-A^{g;4g}(p_a^{\nu_a};k_4^{\nu_4},k_3^{\nu_3},k_2^{\nu_2},k_1^{\nu_1}),
\label{reflcoll}\\[10pt] && A^{g;4g}(p_a^{\nu_a}; k_1^{\nu_1}, 
k_2^{\nu_2}, k_3^{\nu_3}, k_4^{\nu_4}) 
+ A^{g;4g}(p_a^{\nu_a}; k_1^{\nu_1}, k_2^{\nu_2}, k_4^{\nu_4}, 
k_3^{\nu_3}) \nonumber\\ && 
+ A^{g;4g}(p_a^{\nu_a}; k_1^{\nu_1}, k_4^{\nu_4}, 
k_2^{\nu_2}, k_3^{\nu_3}) + A^{g;4g}(p_a^{\nu_a}; k_4^{\nu_4}, 
k_1^{\nu_1}, k_2^{\nu_2}, k_3^{\nu_3}) = 0 ,\label{ward1photcoll}\\[10pt] &&
A^{g;4g}(p_a^{\nu_a}; k_1^{\nu_1}, k_2^{\nu_2}, 
k_3^{\nu_3}, k_4^{\nu_4}) + A^{g;4g}(p_a^{\nu_a}; k_1^{\nu_1},
k_3^{\nu_3}, k_2^{\nu_2}, k_4^{\nu_4}) \nonumber\\ && 
+ A^{g;4g}(p_a^{\nu_a}; k_3^{\nu_3}, k_1^{\nu_1}, k_2^{\nu_2}, k_4^{\nu_4})
+ A^{g;4g}(p_a^{\nu_a}; k_1^{\nu_1}, k_3^{\nu_3}, k_4^{\nu_4}, k_2^{\nu_2})
\nonumber\\ && 
+ A^{g;4g}(p_a^{\nu_a}; k_3^{\nu_3}, k_1^{\nu_1}, k_4^{\nu_4}, k_2^{\nu_2})
+ A^{g;4g}(p_a^{\nu_a}; k_3^{\nu_3}, k_4^{\nu_4}, k_1^{\nu_1}, k_2^{\nu_2})
= 0\, .\label{ward2photcoll}
\end{eqnarray}
We note, however that the last identity is not independent from
the first two.
Using the identities (\ref{reflcoll})-(\ref{ward2photcoll})
in \eqn{nnnlogggg}, we can factorize the color structure on a leg 
\begin{eqnarray}
&& (ig)^4 \sum_{\sigma \in S_4}
f^{a d_{\sigma_1} c } f^{c d_{\sigma_2} c'} f^{c' d_{\sigma_3} c''} 
f^{c'' d_{\sigma_4} c'''} 
A^{g;4g}(p_a^{\nu_a};k_{\sigma_1}^{\nu_{\sigma_1}},
k_{\sigma_2}^{\nu_{\sigma_2}},k_{\sigma_3}^{\nu_{\sigma_3}},
k_{\sigma_4}^{\nu_{\sigma_4}}) \nonumber\\
%&&\hspace{1cm}=\frac{1}{4}\, f^{a c c'''}\,\sum_{\sigma \in S_4}
%  f^{c d_{\sigma_1} c'} f^{c' d_{\sigma_2} c''} 
%f^{c'' d_{\sigma_3} d_{\sigma_4} } 
%A^{g;3g}(p_a^{\nu_a};k_{\sigma_1}^{\nu_{\sigma_1}},
%k_{\sigma_2}^{\nu_{\sigma_2}},k_{\sigma_3}^{\nu_{\sigma_3}},
%k_{\sigma_4}^{\nu_{\sigma_4}})\nonumber\\
&&\hspace{1cm}=
\, i g f^{a c c'''}\,\left\{ g^3 \sum_{\sigma \in S_3}
 \left( F^{ d_{\sigma_1} } F^{ d_{\sigma_2}} F^{ d_{\sigma_3} }\right)_{c d_4}
%  f^{c d_{\sigma_1} c'} f^{c' d_{\sigma_2} c''} 
%f^{c'' d_{\sigma_3} d_{4} } 
A^{g;4g}(p_a^{\nu_a};k_{\sigma_1}^{\nu_{\sigma_1}},
k_{\sigma_2}^{\nu_{\sigma_2}},k_{\sigma_3}^{\nu_{\sigma_3}},
k_{4}^{\nu_{4}})\right\}\, \label{color4fac}
\end{eqnarray}
thus amplitude (\ref{nnnlogggg}) can be put in the form of \eqn{hequadricoll}
with collinear factor
\begin{eqnarray}
&& {\rm Split}_{-\nu}^{g \to 4g}(k_1^{\nu_1}, k_2^{\nu_2}, k_3^{\nu_3},
k_4^{\nu_4}) \nonumber\\ & &=  
{g^3} \sum_{\sigma \in S_3}
 \left( F^{ d_{\sigma_1} } F^{ d_{\sigma_2}} F^{ d_{\sigma_3} }\right)_{c d_4}
%  f^{c d_{\sigma_1} c'} f^{c' d_{\sigma_2} c''} 
%f^{c'' d_{\sigma_3} d_{4} }
{\rm split}_{-\nu}^{g \to 4g}(k_{\sigma_1}^{\nu_{\sigma_1}},
k_{\sigma_2}^{\nu_{\sigma_2}},k_{\sigma_3}^{\nu_{\sigma_3}},
k_{4}^{\nu_{4}})\,  .\label{quadrisplit}
\end{eqnarray}
The splitting factors ${\rm split}_{-\nu}^{g \to 4g}$ are the
functions $A$ of \sec{sec:nnnloif} in the quadruple collinear limit, 
up to quadratically divergent terms, and thus they fulfill the identities,
\eqnss{reflcoll}{ward2photcoll}.
The splitting factors of PT type can be soon read off from 
\eqnss{pt4g}{fivept},
while the ones of non-PT type can be given in terms of three
functions of the collinear momenta,
\begin{eqnarray}
{\rm split}_{-}^{g \to 4g}(k_1^+, k_2^+, k_3^+, k_4^+) &=&
2\sqrt{2} {1\over \sqrt{z_1z_4}} {1\over \langle 1 \,2\rangle
\langle 2 \,3\rangle \langle 3 \,4\rangle} \nonumber\\
{\rm split}_{+}^{g \to 4g}(k_1^-, k_2^+, k_3^+, k_4^+) &=&
2\sqrt{2} {z_1^2\over \sqrt{z_1z_4}} {1\over \langle 1 \,2\rangle
\langle 2 \,3\rangle \langle 3 \,4\rangle} \nonumber\\
{\rm split}_{+}^{g \to 4g}(k_1^+, k_2^-, k_3^+, k_4^+) &=&
2\sqrt{2} {z_2^2\over \sqrt{z_1z_4}} {1\over \langle 1 \,2\rangle
\langle 2 \,3\rangle \langle 3 \,4\rangle} \nonumber\\
{\rm split}_{+}^{g \to 4g}(k_1^+, k_2^+, k_3^-, k_4^+) &=&
2\sqrt{2} {z_3^2\over \sqrt{z_1z_4}} {1\over \langle 1 \,2\rangle
\langle 2 \,3\rangle \langle 3 \,4\rangle} \nonumber\\
{\rm split}_{+}^{g \to 4g}(k_1^+, k_2^+, k_3^+, k_4^-) &=&
2\sqrt{2} {z_4^2\over \sqrt{z_1z_4}} {1\over \langle 1 \,2\rangle
\langle 2 \,3\rangle \langle 3 \,4\rangle} \label{foursplit}\\
{\rm split}^{g\to 4g}_-(k_1^-, k_2^+, k_3^+, k_4^+) &=&  
-{\cal B}_1(4, 3, 2, 1)
\nonumber\\
%G
{\rm split}^{g\to 4g}_-(k_1^+, k_2^-, k_3^+, k_4^+) &=& ~~
{\cal B}_1(4, 3, 1, 2) + {\cal B}_1(4, 1, 3, 2) + {\cal B}_1(1, 4, 3, 2)
\nonumber\\
%H
{\rm split}^{g\to 4g}_-(k_1^+, k_2^+, k_3^-, k_4^+) &=& 
-{\cal B}_1(1, 2, 4, 3) - {\cal B}_1(1, 4, 2, 3) - {\cal B}_1(4, 1, 2, 3)
\nonumber\\
%I
{\rm split}^{g\to 4g}_-(k_1^+, k_2^+, k_3^+, k_4^-) &=& ~~{\cal B}_1(1, 2, 3, 4)
\nonumber\\
%J
{\rm split}^{g\to 4g}_-(k_1^-, k_2^-, k_3^+, k_4^+) &=& -{\cal B}_2(4, 3, 2, 1)
\nonumber\\
%K  
{\rm split}^{g\to 4g}_-(k_1^-, k_2^+, k_3^-, k_4^+) &=& ~~{\cal B}_3(1, 2, 3, 4) 
\nonumber\\
%L
{\rm split}^{g\to 4g}_-(k_1^-, k_2^+, k_3^+, k_4^-) &=& 
-{\cal B}_3(1, 2, 4, 3) + {\cal B}_2(3, 2, 4, 1) + {\cal B}_2(3, 2, 1, 4)   
\nonumber\\
%M
{\rm split}^{g\to 4g}_-(k_1^+, k_2^-, k_3^-, k_4^+) &=& 
-{\cal B}_2(1, 4, 2, 3) + {\cal B}_3(3, 4, 2, 1) - {\cal B}_2(4, 1, 2, 3)   
\nonumber\\
%N
{\rm split}^{g\to 4g}_-(k_1^+, k_2^-, k_3^+, k_4^-) &=& -{\cal B}_3(4, 3, 2, 1)
\nonumber\\
%O 
{\rm split}^{g\to 4g}_-(k_1^+, k_2^+, k_3^-, k_4^-) &=& ~~{\cal B}_2(1, 2, 3, 4)\nonumber 
\end{eqnarray}
with
\begin{eqnarray}
%&&{\cal B}_1(1, 2, 3, 4)=
%2\sqrt{2} {1\over \sqrt{z_1z_4}} {1\over \langle 1 \,2\rangle
%\langle 2 \,3\rangle \langle 3 \,4\rangle} \nonumber\\
%
%%%%%%%%%%%%%%%%%%%%%%%%%%%%%%%%%%%%%%%%%%%%%%%%%%%%%%%%%%%%%%%%%%%%%%%%%%%
%
%&&{\cal B}_2(1, 2, 3, 4)=
%2\sqrt{2} {z_1^2\over \sqrt{z_1z_4}} {1\over \langle 1 \,2\rangle
%\langle 2 \,3\rangle \langle 3 \,4\rangle} \nonumber\\
%
%%%%%%%%%%%%%%%%%%%%%%%%%%%%%%%%%%%%%%%%%%%%%%%%%%%%%%%%%%%%%%%%%%%%%%%%%%%
%
&&{\cal B}_1(1, 2, 3, 4)=\nonumber\\
&&\frac{2\,\sqrt{2} } {\br(1,2) \,\br(2,3) \,\s(3,4) }
\left[
-\frac{z_3 \delta(1,2,3)}{\sqrt{ z_1} ( z_3 + z_4 ) \sq(1,2) }
-\frac{\sqrt{z_2 z_3} z_4 \br(1,2)   }{(1 - z_1) \,(z_3 + z_4)}
+\sqrt{\frac{z_3}{z_1 z_4} } \, (1 - z_4)\,\eps(1,2,3,4) 
\right.\nonumber\\
&& - {\frac{  {{\eps(1,2,3,4) }^2}   \left(
      \br(2,4)   \sq(1,2) + \br(3,4)   \sq(1,3) \right)   
      \sq(3,4) }{{s_{1234}}     \t(2,3,4) }}+   \sqrt{\frac{z_2}{z_1}}   
      \frac{{\eps(1,2,3,4) }^2 \sq(2,3) }{\t(2,3,4) }\nonumber\\
&&\left. +\frac{ \sqrt{z_4} \br(1,2) \sq(2,3) \left(
       \sqrt{z_1} \br(1,4) - \left( 2 - z_1 \right)   \eps(1,2,3,4)\right) }
       { (1 - z_1) \t(2,3,4) } \right]\\[10pt]
%
%%%%%%%%%%%%%%%%%%%%%%%%%%%%%%%%%%%%%%%%%%%%%%%%%%%%%%%%%%%%%%%%%%%%%%%%%%%
%
&&{\cal B}_2(1, 2, 3, 4)=\nonumber\\
&&\frac{2  \sqrt{2}}{\sq(3,4) \s(2,3) } \left[
  \frac{ z_2  \sqrt{z_3} } {(1 - z_1) \sqrt{z_4} }
+ \frac{ \sqrt{z_2} (z_1 + z_2) \delta(1,2,3) }{ \sqrt{z_1 z_4} \s(1,2) } 
- \frac{ {\delta(1,2,3) }^2  \eps(1,2,3,4) \sq(3,4) } 
      {s_{1234} \sqrt{z_4} \s(1,2)  \t(1,2,3) } \right.
\nonumber\\
&& +  \frac{\delta(1,2,3) \left( 
      \sqrt{z_1} \delta(3,4,1) + 
      \sqrt{z_2} \delta(3,4,2) \right) + 
      \sqrt{z_1 z_2 z_4}  \s(1,2) \s(3,4) }
      { s_{1234} \sqrt{z_4} \br(1,2) \sq(3,4) } 
\nonumber\\
&&\left.
   +  {\frac{ z_1  \sqrt{z_2}  \delta^*(3,4,2) }
      {(1 - z_1) \t(2,3,4) }} 
   -  {\frac{ z_1 \br(3,4)  \sq(1,2)   \left(
      \br(1,3)  \sq(2,3) + \br(1,4)   \sq(2,4) \right) }{
      {s_{1234}} \t(2,3,4) }} \right]\\[10pt]
%
%%%%%%%%%%%%%%%%%%%%%%%%%%%%%%%%%%%%%%%%%%%%%%%%%%%%%%%%%%%%%%%%%%%%%%%%%%%
%
&&{\cal B}_3(1, 2, 3, 4)=
\frac{2  \sqrt{2} }{\s(1,2) \s(2,3) \s(3,4) } \left[ 
     \frac{z_2 \sqrt{z_2 z_3} }{\sqrt{z_1}  ( z_1+ z_2)} 
        \left( \frac{ \sqrt{z_1} \br(2,3)   \s(3,4) }{1-z_4} - 
      \frac{\sqrt{z_3} \br(1,2)   \s(2,3) }{z_3+z_4} \right) \right.
\nonumber\\
&& -  \sqrt{\frac{z_2}{z_1 z_4}}    
      \left({\sqrt{z_2 z_3}} \br(1,2) - z_4  \br(1,3) \right)
        \delta(2,4,3) -
      {\frac{ z_2   
      \left ( \sqrt{z_3} \eps(1,2,4,3)   \s(2,3) - \br(2,3)   \br(3,4)   
      \sq(2,4) \right) }{\left( z_1 + z_2 \right) }}
\nonumber\\
&& -  \frac{  \left(
      {\sqrt{z_2}}   \br(1,2)   {{\delta(2,4,3) }^2} +
      {\sqrt{z_1}}  \br(1,3)    \left( z_2  \br(1,2)    \br(2,3) +
      {\sqrt{z_3}}  {\sqrt{z_4}}   \br(1,3)   \br(3,4) +
      z_4  \br(1,4)    \br(3,4) \right)   {{\sq(2,4) }^2} \right)
      }{\sqrt{z_1}   s_{1234} }
\nonumber\\
&& +  \frac{z_2  \s(2,3) \s(3,4) \delta^*(1,3,2) }{ \sqrt{z_1}   
      \left( 1 - z_4 \right)    \sq(2,3)    \t(1,2,3) }
   +  \frac{ \s(1,2)    {\delta(2,4,3) }^2  
      \eps(2,3,4,1) }{s_{1234}  {\sqrt{z_1}}  \t(2,3,4) }
   +  \sqrt{\frac{z_3}{z_1}}   z_2   \br(1,3)   \t(2,3,4) 
\nonumber\\
&& +  \frac{\s(3,4) }{ s_{1234}   \t(1,2,3) }
      \sqrt{z_4}    {\br(1,3) }^2   \sq(2,4)    
      \left( \eps(1,2,4,3)   \sq(2,3) -
             \eps(2,3,4,1)   \sq(1,2) \right)    
\nonumber\\
&&\left. 
   - \sqrt{\frac{z_2}{z_1}}   \frac{ {\br(1,3) }^2   \s(3,4) }{\t(1,2,3) }
      \left(- \sqrt{z_2} \eps^*(2,3,4,1) + {\sqrt{z_1 z_4}}   \sq(2,4) \right) 
    \right]
\end{eqnarray}
with $\delta(1,2,3)$ as in \eqn{delta123}, and
\begin{equation}
\eps(1,2,3,4) =
\sqrt{z_1} \br(1,4) +\sqrt{z_2} \br(2,4) +\sqrt{z_3} \br(3,4) \, 
.\label{eps1234}
\end{equation}

As in \sec{sec:nnlocoll}, summing over the 
helicities of gluons 1, 2, 3 and 4, one can obtain the two-dimensional
polarization matrix,
\begin{equation}  
\sum_{\nu_1\nu_2\nu_3\nu_4}\,
{\rm Split}_{-\nu}^{g \to 4g} (k_1^{\nu_1}, k_2^{\nu_2},
k_3^{\nu_3}, k_4^{\nu_4}) 
[{\rm Split}_{-\nu}^{g \to 4g} (k_1^{\nu_1}, k_2^{\nu_2},
k_3^{\nu_3}, k_4^{\nu_4})]^* = \delta^{cc'} \frac{8g^6}{s_{1234}^3}
P^{g \to 4g}_{\lambda\rho}\, ,\label{ap4matrix}
\end{equation}
where $P^{g \to 4g}_{++} = P^{g \to 4g}_{--}$, and
$P^{g \to 4g}_{-+} = (P^{g \to 4g}_{+-})^*$.
Averaging then over the trace of matrix (\ref{ap4matrix}), i.e. over
color and helicity of the parent gluon, one can obtain
the unpolarized Altarelli-Parisi gluon triple-splitting function 
\begin{equation}
{1\over 2(N_c^2-1)} \sum_{\nu\nu_1\nu_2\nu_3\nu_4}\, 
|{\rm Split}_{-\nu}^{g \to 4g} (k_1^{\nu_1}, k_2^{\nu_2},
k_3^{\nu_3}, k_4^{\nu_4})|^2 = \frac{8g^6}{s_{1234}^3} 
\langle P^{g\to 4g} \rangle\, ,\label{dglap4}
\end{equation}
with $\langle P^{g\to 4g} \rangle = P^{g \to 4g}_{++}$. As in 
\sec{sec:nnlocoll}, the sum over colors can be done using \eqn{square2}, 
and we obtain,
\begin{equation}
|{\rm Split}_{-\nu}^{g \to g_1 g_2 g_3 g_4} 
(k_1^{\nu_1}, k_2^{\nu_2}, k_3^{\nu_3} ,k_4^{\nu_4})|^2=
4 {\cal C}_5(N_c) \sum_{\sigma \in S_{4}} 
|{\rm split}_{-\nu}^{g \to 4g} 
(k_{\sigma_1}^{\nu_{\sigma_1}},
 k_{\sigma_2}^{\nu_{\sigma_2}},
 k_{\sigma_3}^{\nu_{\sigma_3}},
 k_{\sigma_4}^{\nu_{\sigma_4}})|^2\, ,
\end{equation}
with ${\cal C}_5(N_c)$ as in \eqn{calc}.
It is then clear that for the splitting functions $P^{g\to n g}$, 
with $n>4$, the color will not factorize since LCA, \eqn{square2}, 
is not exact any more. We do not compute 
here $P^{g \to 4g}_{++}$ and $P^{g \to 4g}_{+-}$, all the information
about them being already contained in \eqnss{foursplit}{eps1234}.

\section{Conclusions}
\label{sec:conc}

In this paper, the structure of QCD amplitudes in the high-energy limit
and in the collinear limit has been explored beyond NLO.
We have computed forward clusters of three partons and four gluons, 
which in the
BFKL theory constitute the tree parts of NNLO and NNNLO impact factors 
for jet production. In the BFKL theory the NNLO impact factors could be used
to compute jet rates at NNLL accuracy. In Sect.~\ref{sec:nnlogggg},
\ref{sec:nnloqqgg}, \ref{sec:nnloqgg} and \ref{sec:nnloqqq}, we have computed
the tree parts of the NNLO impact factors for all the parton flavors.
On these we have performed in Sect.~\ref{sec:nnlohigh} a set of consistency
checks in the high-energy limit, and we have obtained in the triple collinear 
limit (Sect.~\ref{sec:nnlocoll}) the polarized, the spin-correlated
and the unpolarized double-splitting functions. The last two agree with
previous calculations by Catani-Grazzini and Campbell-Glover, respectively.
They can be used to set up general algorithms to compute jet rates at NNLO.

From the four-gluon forward cluster we have obtained in 
Sect.~\ref{sec:nnnloif} the 
tree part of the purely gluonic NNNLO impact factor. 
In the quadruple collinear limit, this yields 
(Sect.~\ref{sec:nnnlocoll}) the 
purely gluonic unpolarized triple-splitting functions. They could be used 
to compute the three-loop Altarelli-Parisi evolution, or 
to compute jet rates at NNNLO.
In addition, by separating a central cluster of three gluons out of
the four-gluon forward cluster, we have computed the emission of three gluons 
along the ladder, \eqnss{3centrlim}{nnlolip4}, 
which contributes to the NNLO Lipatov vertex.
This constitutes one of the universal building blocks in an eventual
construction of a BFKL resummation at NNLL accuracy.

Finally,
inspired by the color structure in the
high-energy limit, we have found a compact color decomposition
of the tree multigluon amplitudes 
in terms of the linearly independent subamplitudes only, \eqn{GluonDecompNew}.
It would be interesting to analyse whether this structure generalizes to
multigluon amplitudes at one loop, and beyond.

The decomposition in rapidity of amplitudes in terms of
gauge-invariant parton clusters performed in this work
suggests naturally a modular decomposition of a generic multiparton 
amplitude, where each module is an 
$n$-parton cluster. Such an approximation could be tested against
existing approximations of multiparton amplitudes~\cite{ksti,max}. 
In the high-energy limit, the cluster decomposition seems superior, in that
it does not use only PT-type subamplitudes, like the Kunszt-Stirling
approximation~\cite{ksti}, and within a cluster it is not limited to collinear
kinematics, like the Maxwell approximation~\cite{max}.

\appendix

\section{Multiparton kinematics}
\label{sec:appa}

We consider the production of $n$ partons of momentum $p_i$, with 
$i=1,...,n$ and $n\ge 2$, in the scattering between two partons of  
momenta $p_a$ and $p_b$\footnote{By convention
we consider the scattering in the unphysical region where all momenta 
are taken as outgoing, and then we analitically continue to the
physical region where $p_a^0<0$ and $p_b^0<0$. Thus
partons are ingoing or outgoing depending on the sign
of their energy. Since the helicity of a positive-energy 
(negative-energy) massless spinor has the same (opposite) sign as its
chirality, the helicities assigned to the partons 
depend on whether they are incoming or outgoing.
Our convention is to label outgoing (positive-energy) particles 
with their helicity; so if they are incoming the 
actual helicity and charge is reversed.}.

Using light-cone coordinates $p^{\pm}= p_0\pm p_z $, and
complex transverse coordinates $p_{\perp} = p^x + i p^y$, with scalar
product $2 p\cdot q = p^+q^- + p^-q^+ - p_{\perp} q^*_{\perp} - p^*_{\perp} 
q_{\perp}$, the 4-momenta are,
\begin{eqnarray}
p_a &=& \left(p_a^+/2, 0, 0,p_a^+/2 \right) 
     \equiv  \left(p_a^+ , 0; 0, 0 \right)\, ,\nonumber \\
p_b &=& \left(p_b^-/2, 0, 0,-p_b^-/2 \right) 
     \equiv  \left(0, p_b^-; 0, 0\right)\, ,\label{in}\\
p_i &=& \left( (p_i^+ + p_i^- )/2, 
                {\rm Re}[p_{i\perp}],
                {\rm Im}[p_{i\perp}], 
                (p_i^+ - p_i^- )/2 \right)\nonumber\\
    &\equiv& \left(|p_{i\perp}| e^{y_i}, |p_{i\perp}| e^{-y_i}; 
|p_{i\perp}|\cos{\phi_i}, |p_{i\perp}|\sin{\phi_i}\right)\, \,,\nonumber
\end{eqnarray}
where $y$ is the rapidity. The first notation in \eqn{in} is the 
standard representation 
$p^\mu =(p^0,p^x,p^y,p^z)$, while in the second we have the + and -
components on the left of the semicolon, 
and on the right the transverse components.
In the following, if not  differently stated, $p_i$ and $p_j$ are always 
understood for $1\le i,j \le n$.
From the momentum conservation,
\begin{eqnarray}
0 &=& \sum_{i=1}^n p_{i\perp}\, ,\nonumber \\
p_a^+ &=& -\sum_{i=1}^n p_i^+\, ,\label{nkin}\\ 
p_b^- &=& -\sum_{i=1}^n p_i^-\, ,\nonumber
\end{eqnarray}
the Mandelstam invariants may be written as,
\begin{eqnarray}
s_{ij} &=& 2 p_i\cdot p_j = p_i^+ p_j^- + p_i^- p_j^+
- p_{i\perp} p_{j\perp}^* - p_{i\perp}^* p_{j\perp}\, .\nonumber
\end{eqnarray}
so that 
\begin{eqnarray}
s &=& 2 p_a\cdot p_b = \sum_{i,j=1}^n p_i^+ p_j^- \nonumber\\ 
s_{ai} &=& 2 p_a\cdot p_i = -\sum_{j=1}^n p_i^- p_j^+ \label{inv}\\ 
s_{bi} &=& 2 p_b\cdot p_i = -\sum_{j=1}^n p_i^+ p_j^- \nonumber.
\end{eqnarray}

Massless Dirac spinors $\psi_{\pm}(p)$ of fixed helicity are
defined by the projection,
\begin{equation}
\psi_{\pm}(p) = {1\pm \gamma_5\over 2} \psi(p)\, ,\label{spi}
\end{equation}
with the shorthand notation,
\begin{eqnarray}
\psi_{\pm}(p) &=& |p\pm\rangle, \qquad \overline{\psi_{\pm}(p)} = 
\langle p\pm|\, ,\nonumber\\
\langle p k\rangle &=& \langle p- | k+ \rangle = \overline{\psi_-(p)}
\psi_+(k)\, ,\label{cpro}\\ 
\left[pk\right] &=& \langle p+ | k- \rangle = \overline{\psi_+(p)}\psi_-(k)\, 
.\nonumber
\end{eqnarray}
Using the chiral representation of the $\gamma$-matrices,
\begin{equation}
\gamma^0 = \left( \begin{array}{cc} 0 & I\\ I & 0\end{array} \right),
\qquad \gamma^i = \left( \begin{array}{cc} 0 & -\sigma^i\\ \sigma^i & 0
\end{array} \right)\, ,\label{gam}
\end{equation}
and the normalization condition:
\begin{equation}
\langle p\pm| \gamma_{\mu} |p\pm\rangle = 2p_{\mu}\, ,\label{norm}
\end{equation}
and the complex notation $p_\perp=|p_\perp| e^{i\phi}$, the spinors
for the momenta (\ref{in}) are\footnote{The spinors of the 
incoming partons must be continued
to negative energy after the complex conjugation. For instance,
$\overline{\psi_{+}(p_a)}= i \left( \sqrt{-p_a^+}, 0, 0, 0 \right)$.}
\begin{equation} \begin{array}{cc}
\psi_+(p_i) = \left( \begin{array}{c} \sqrt{p_i^+}\\ \sqrt{p_i^-} e^{i\phi_i}\\ 0\\ 0\end{array}\right) 
& \psi_-(p_i) = \left(
\begin{array}{c} 0\\ 0\\ \sqrt{p_i^-} e^{-i\phi_i}\\ 
-\sqrt{p_i^+}\end{array}\right)
\\ \\ \psi_+(p_a) = i \left( \begin{array}{c} \sqrt{-p_a^+}\\ 0\\
0\\ 0\end{array} \right) & \psi_-(p_a) = i \left( \begin{array}{c}
0\\ 0\\ 0\\ -\sqrt{-p_a^+} \end{array}\right)\\ \\ 
\psi_+(p_b) = -i
\left( \begin{array}{c} 0\\ \sqrt{-p_b^-} \\ 0\\ 0\end{array}\right) &
\psi_-(p_b) = -i \, \left( \begin{array}{c} 0\\ 0\\
\sqrt{-p_b^-}\\ 0\end{array}\right) \end{array}\label{spin}.
\end{equation}
Using the above spinor representation, the spinor products for the
momenta (\ref{in}) are
\begin{eqnarray}
\langle p_i p_j\rangle &=& p_{i\perp}\sqrt{p_j^+\over p_i^+} - p_{j\perp}
\sqrt{p_i^+\over p_j^+}\, , \nonumber\\ 
\langle p_a p_i\rangle &=& - i \sqrt{-p_a^+
\over p_i^+}\, p_{i\perp}\, ,\label{spro}\\ 
\langle p_i p_b\rangle &=&
i \sqrt{-p_b^- p_i^+}\, ,\nonumber\\ 
\langle p_a p_b\rangle 
&=& -\sqrt{\hat s}\, ,\nonumber
\end{eqnarray}
where we have used the mass-shell condition 
$|p_{i\perp}|^2 = p_i^+ p_i^-$. Note that in the present convention
the spinors (\ref{spin}) and the spinor products (\ref{spro})
differ by phases with respect to the same in Ref.~\cite{ptlip}.

\noindent
We consider also the spinor products
$\langle p_i+| \gamma\cdot p_k |p_j+\rangle$, which in the 
spinor representation (\ref{spin}) take the form,
\begin{eqnarray}
\langle p_i+| \gamma\cdot p_k |p_j+\rangle &=& {1\over\sqrt{p_i^+ p_j^+}}
\left(p_i^+ p_j^+ p_k^- - p_i^+ p_{j\perp} p_{k\perp}^* - p_{i\perp}^* p_j^+
p_{k\perp} + p_{i\perp}^* p_{j\perp} p_k^+ \right) ,\; \forall k \nonumber\\
\langle p_i+| \gamma\cdot p_j |p_a+\rangle &=& i \sqrt{-p_a^+\over p_i^+}
\left(p_i^+ p_j^- - p_{i\perp}^* p_{j\perp}\right) \label{compspi}\;,\qquad\forall j \\
\langle p_i+| \gamma\cdot p_j |p_b+\rangle &=& -i \sqrt{-p_b^-\over p_i^+}
\left(-p_i^+ p_{j\perp}^* + p_{i\perp}^* p_j^+\right)\;,\qquad \forall j .\nonumber
\end{eqnarray}
The spinor products fulfill the identities ($ i\equiv  p_i, j\equiv p_j$),
\begin{eqnarray}
\langle i j\rangle &=& - \langle j i\rangle \nonumber\\
           \left[ i j \right] &=& - \left[ j i \right] \nonumber\\
\langle i j\rangle^* &=& {\rm sign}(p^0_i p^0_j) \left[ j i\right] 
\nonumber\\
\left( \langle i+| \gamma^{\mu}  |j+\rangle \right)^* &=&
{\rm sign}(p^0_i p^0_j)
\langle j+| \gamma^{\mu}  |i+\rangle \nonumber\\
\langle i j \rangle \left[ji\right] &=& 
2p_i\cdot p_j = \hat{s}_{ij}\,  \label{flip0}\\
\langle i+| \slash \!\!\! k  |j+\rangle  &=&
\left[ i k \right] \langle k j \rangle \nonumber \\
\langle i-| \slash  \!\!\! k  |j-\rangle  &=& 
\langle i k \rangle \left[k j\right]\,\nonumber \\
\langle ij \rangle\langle kl \rangle   &=& 
\langle ik \rangle\langle jl \rangle + 
\langle il \rangle\langle kj \rangle\,\nonumber\\
\left[ ij \right]\left[ kl \right]   &=& \left[ ik \right]\left[ jl \right] + \left[ il \right]\left[ kj \right]\,\nonumber
\end{eqnarray}
and if $\sum_{i=1}^{n} p_i=0$ then
\begin{eqnarray}
\sum_{i=1}^{n} \left[ji\right] \langle ik \rangle =0\,.
\end{eqnarray}

Throughout the paper the following representation for the gluon 
polarization is used,
\begin {equation}
\epsilon_{\mu}^{\pm}(p,k) = \pm {\langle p\pm |\gamma_{\mu}| k\pm\rangle\over
\sqrt{2} \langle k\mp | p\pm \rangle}\, ,\label{hpol}
\end{equation}
which enjoys the properties
\begin {eqnarray}
\epsilon_{\mu}^{\pm *}(p,k) &=& \epsilon_{\mu}^{\mp}(p,k)\, ,\nonumber\\
\epsilon_{\mu}^{\pm}(p,k)\cdot p &=& \epsilon_{\mu}^{\pm}(p,k)\cdot k = 0\,
,\label{polc}\\
\sum_{\nu=\pm} \epsilon_{\mu}^{\nu}(p,k) \epsilon_{\rho}^{\nu *}(p,k) &=&
- g_{\mu\rho} + {p_{\mu} k_{\rho} + p_{\rho} k_{\mu}\over p\cdot k}\, 
,\nonumber
\end{eqnarray}
where $k$ is an arbitrary light-like momentum. The sum in Eq.~(\ref{polc}) is
equivalent to use an axial, or physical, gauge.

\section{Multi-Regge kinematics}
\label{sec:appb}

In the multi-Regge kinematics, we require that the gluons
are strongly ordered in rapidity and have comparable transverse momentum,
\begin{equation}
y_1 \gg ...\gg y_n;\qquad |p_{1\perp}| \simeq ...\simeq|p_{n\perp}|\, 
.\nonumber
\end{equation}
Momentum conservation (\ref{nkin}) then becomes
\begin{eqnarray}
0 &=& \sum_{i=1}^n p_{i\perp}\, ,\nonumber \\
p_a^+ &\simeq& -p_1^+\, ,\label{mrkin}\\ 
p_b^- &\simeq& -p_n^-\, .\nonumber
\end{eqnarray}
The Mandelstam invariants (\ref{inv}) are reduced to,
\begin{eqnarray}
s &=& 2 p_a\cdot p_b \simeq p_1^+ p_n^- \nonumber\\ 
s_{ai} &=& 2 p_a\cdot p_i \simeq - p_1^+ p_i^- \label{mrinv}\\ 
s_{bi} &=& 2 p_b\cdot p_i \simeq - p_i^+ p_n^- \nonumber\\ 
s_{ij} &=& 2 p_i\cdot p_j \simeq |p_{i\perp}| |p_{j\perp}| e^{|y_i-y_j|}\,
\nonumber
\end{eqnarray}
to leading accuracy. The spinor products (\ref{spro}) become,
\begin{eqnarray}
\langle p_i p_j\rangle &\simeq& -\sqrt{p_i^+\over p_j^+}\,
p_{j\perp}\, \qquad {\rm for}\, y_i>y_j \;, \nonumber\\
\langle p_a p_i\rangle &\simeq& - i\sqrt{p_a^+\over p_i^+}\,
p_{i\perp}\, ,\label{mrpro}\\ \langle p_i p_b\rangle 
&\simeq& i\sqrt{p_i^+ p_n^-}\, ,\nonumber\\ 
\langle p_a p_b\rangle &\simeq& -\sqrt{p_1^+ p_n^-}\, .\nonumber
\end{eqnarray}

\section{NLO Multi-Regge kinematics}
\label{sec:appc}

We consider the production of $n$ partons of momenta $p_1,...,p_n$,
with partons 1 and 2 in the forward-rapidity region of parton $p_a$,
\begin{equation}
y_1 \simeq y_2 \gg y_3\gg ...\gg y_n\,;\qquad |p_{1\perp}|
\simeq |p_{2\perp}| \simeq ...\simeq |p_{n\perp}|\, .\label{qmrapp}
\end{equation}
Momentum conservation (\ref{nkin}) becomes
\begin{eqnarray}
0 &=& \sum_{i=1}^n p_{i\perp}\, ,\nonumber \\
p_a^+ &\simeq& -(p_1^+ + p_2^+)\, ,\label{frkapp}\\ 
p_b^- &\simeq& -p_n^-\, .\nonumber
\end{eqnarray}
The spinor products (\ref{spro}) become
\begin{eqnarray}
\langle p_a p_b\rangle &=& 
-\sqrt{s} \simeq - \sqrt{(p_1^+ + p_2^+) p_n^-}\,
,\nonumber\\ 
\langle p_a p_n\rangle &=& 
-i \sqrt{-p_a^+\over p_n^+}\, p_{n\perp} \simeq i
{p_{n\perp}\over |p_{n\perp}|} \langle p_a p_b\rangle\, ,\nonumber\\
\langle p_a p_k\rangle &=& -i \sqrt{-p_a^+\over p_k^+}\, p_{k\perp}
\simeq -i \sqrt{p_1^+ + p_2^+\over p_k^+} p_{k\perp}\, ,
\nonumber \qquad k=1,\dots,n-1\\
\langle p_k p_b\rangle &=& i \sqrt{-p_b^- p_k^+}\, 
\simeq i \sqrt{p_k^+ p_n^-}\, ,\label{frpro}\qquad k=1,\dots,n-1\nonumber\\
\langle p_n p_b\rangle &=& i \sqrt{-p_b^- p_n^+}\,
\simeq i |p_{n\perp}|\, ,\label{qmrkspro}\\
\langle p_k p_n\rangle &=& p_{k\perp}\sqrt{p_n^+\over p_k^+} - p_{n\perp}
\sqrt{p_k^+\over p_n^+} \simeq - p_{n\perp}\, \sqrt{p_k^+\over p_n^+}\, 
,\nonumber \qquad k=1,\dots,n-1\\
\langle p_1 p_2\rangle &=& p_{1\perp}\sqrt{p_2^+\over p_1^+} - 
p_{2\perp}\sqrt{p_1^+\over p_2^+}\, .\nonumber\\
\langle p_k p_i\rangle &=& p_{k\perp}\sqrt{p_i^+\over p_k^+} - p_{i\perp}
\sqrt{p_k^+\over p_i^+} \simeq - p_{i\perp}\, \sqrt{p_k^+\over p_i^+}\, 
,\nonumber\qquad k=1,2 \; ; i=3,\dots,n-1 \,.
\end{eqnarray}
which differ by phases with respect to the same spinor products
in Ref.~\cite{ptlipnl} because of the convention for the spinor
representation we use in Sect.~\ref{sec:appa}.

\section{NNLO Multi-Regge kinematics}
\label{sec:appd}

The extension to the production of $n$ partons of momenta $p_1,...,p_n$,
with partons 1, 2 and 3 in the forward-rapidity region of parton $p_a$,
\begin{equation}
y_1 \simeq y_2 \simeq y_3 \gg y_4\gg ...\gg y_n\,;\qquad |p_{1\perp}|
\simeq |p_{2\perp}| \simeq ...\simeq |p_{n\perp}|\, ,\label{nnloqmr}
\end{equation}
is straightforward. We mention it here because by taking the further
limit $y_1\gg y_2 \simeq y_3$, one obtains the kinematics of the
NLO Lipatov vertex (sect.~\ref{sec:nnlohigh}).

\noindent
With \eqn{nnloqmr}, momentum conservation (\ref{nkin}) becomes
\begin{eqnarray}
0 &=& \sum_{i=1}^n p_{i\perp}\, ,\nonumber \\
p_a^+ &\simeq& -(p_1^+ + p_2^+ + p_3^+)\, ,\label{nnlofrk}\\ 
p_b^- &\simeq& -p_n^-\, .\nonumber
\end{eqnarray}
The spinor products (\ref{spro}) become
\begin{eqnarray}
\langle p_a p_b\rangle &=& 
-\sqrt{s} \simeq - \sqrt{(p_1^+ + p_2^+ + p_3^+) p_n^-}\,
,\nonumber\\ 
\langle p_a p_n\rangle &=& 
-i \sqrt{-p_a^+\over p_n^+}\, p_{n\perp} \simeq i
{p_{n\perp}\over |p_{n\perp}|} \langle p_a p_b\rangle\, ,\nonumber\\
\langle p_a p_k\rangle &=& -i \sqrt{-p_a^+\over p_k^+}\, p_{k\perp}
\simeq -i \sqrt{p_1^+ + p_2^+ + p_3^+\over p_k^+} p_{k\perp}\, ,
\nonumber \qquad k=1,\dots,n-1\\
\langle p_k p_b\rangle &=& i \sqrt{-p_b^- p_k^+}\, 
\simeq i \sqrt{p_k^+ p_n^-}\, ,\label{nnlopro}\qquad k=1,\dots,n-1\nonumber\\
\langle p_n p_b\rangle &=& i \sqrt{-p_b^- p_n^+}\,
\simeq i |p_{n\perp}|\, ,\nonumber\\
\langle p_k p_n\rangle &=& p_{k\perp}\sqrt{p_n^+\over p_k^+} - p_{n\perp}
\sqrt{p_k^+\over p_n^+} \simeq - p_{n\perp}\, \sqrt{p_k^+\over p_n^+}\, 
,\nonumber \qquad k=1,\dots,n-1\\
\langle p_k p_i\rangle &=& p_{k\perp}\sqrt{p_i^+\over p_k^+} - p_{i\perp}
\sqrt{p_k^+\over p_i^+} \simeq - p_{i\perp}\, \sqrt{p_k^+\over p_i^+}\, 
,\nonumber\qquad k=1,2,3 \; ; i=4,\dots,n-1 \, ,
\end{eqnarray}
while the others spinor products remain unchanged. The spinor products
(\ref{nnlopro}) generalize straightforwardly to the
 kinematics (\ref{nnnloreg}).

\section{The Sudakov parametrization}
\label{sec:appe}

We want to elucidate the relationship between our parametrization of
the momenta and the one of Ref.~\cite{cat}.
Recalling the last of Eqs.~(\ref{in}), we can write,
\begin{equation}
p_i=\frac{ x_i P^+}{2} (1,0,0,1)+(0,{\rm Re}[p_{i\perp }],{\rm Im}[p_{i\perp }],0)+
        \frac{ |p_{i\perp }|^2}{2x_i P^+} (1,0,0,-1)
\end{equation}
where $P^\mu $ is the sum of the three  momenta, the $ x_i$ are the 
momentum fractions 
and we used the mass-shell condition $p^+_i p^-_i =|p_{i\perp }|^2$.
This is exactly what is obtained from 
 the general Sudakov parametrization of Ref.~\cite{cat},
\begin{equation}
p_i^\mu =x_i p^\mu +k_{\perp i}^\mu -\frac{k_{\perp i}^2 }{x_i} \frac{n^\mu }{2p\cdot n}\, 
\end{equation}
through the following choices for  the lightlike vectors,
\begin{equation}
p^\mu =\frac{P^+}{2} (1,0,0,1) \qquad \mbox{\rm and} \qquad
n^\mu = (1,0,0,-1)\, ,
\end{equation}
and the identification,
\begin{equation}
k_{\perp i}^\mu =(0, {\rm Re} [p_{i\perp }], {\rm Im} [p_{i\perp }],0)\, .
\end{equation}
The spin-correlated splitting functions of Ref.~\cite{ cat} are 
expressed in terms of
the vectors $\tilde{k}_i$  defined as $\tilde{k}_i^\mu =k_{ \perp i }^\mu - z_i P_{\perp }^\mu  $,
where, as in our case, the $z_i$ variables represent the momentum fractions in the collinear limit. In order to compare \eqn{ap3matrix} with
the spin-correlated splitting functions of Ref.~\cite{ cat}, we must
project the latter onto the helicity basis, namely to contract
them with the  polarization vector, \eqn{hpol},
\begin{equation} 
\epsilon_\mu ^\pm (P, \, n) =\frac{1}{\sqrt{2}} (0,1,\mp i,0)\, .
\end{equation}
The contraction of the $\tilde{k}_i^\mu $  vectors with $\epsilon ^+$ is,
\begin{equation}
\tilde{k}_i \cdot \epsilon ^+ = \sqrt{\frac{z_i}{2}} 
(\left[i\,j\right] \sqrt{z_j}+
\left[i\,\l \right] \sqrt{z_{\l } })\, ,
\end{equation}
with $i,j,\l=1,2,3$ and $j,\l \neq i$, with the analogous
expressions for $\epsilon ^- $ obtained by complex conjugation.

For the off-diagonal terms, $P^{g\to g f_2 f_3}$,
we find a  relative minus sign between the results of Ref.~\cite{ cat}
and ours, which, however, has no physical relevance.

\section*{Acknowledgments}
We should like to thank Stefano Catani, Walter Giele, 
David Kosower and Zoltan Trocsanyi for discussions. 
We are particularly grateful to Lance Dixon for his valuable insight.

\end{document}